\documentclass[12pt]{article}
\pdfoutput=1 

\usepackage{jheppub} 

\usepackage[T1]{fontenc} 
\usepackage{amsmath,physics}
\usepackage{comment}
\usepackage{bbm}
\usepackage{tikz-cd}
\usepackage[toc,page]{appendix}
\usepackage{etoolbox}
\usepackage{tabularx}

\BeforeBeginEnvironment{appendices}{\clearpage}
\AfterEndEnvironment{appendices}{\clearpage}
\AfterEndEnvironment{appendices}{\pretocmd{\section}{\clearpage}{}{}}{}
\AfterEndEnvironment{appendices}{\pretocmd{\subsubsection}{\clearpage}{}{}}{}

\usepackage{graphicx,float,caption,subcaption,xcolor}
\usepackage{hyperref}
\hypersetup{
    colorlinks=true,
    linkcolor=blue,
    filecolor=magenta,      
    urlcolor=cyan,
    pdftitle={Penrose limits and Double Copies},
    pdfpagemode=FullScreen,
    }

\graphicspath{{./img/}} 
\usepackage{multirow}

\usepackage{bbm}

\newcommand{\cA}{\mathcal{A}}

\newcommand{\lenf}{15}
\newcommand{\lenff}{19}
\newcommand{\lenfff}{18}

\title{The Penrose limit of the Weyl double copy}
\author[a]{Samarth Chawla,}
\author[b]{Kwinten Fransen,}
\author[a]{Cynthia Keeler}

\affiliation[a]{Department of Physics, Arizona State University, Tempe, AZ 85281, USA}
\affiliation[b]{Department of Physics, University of California, Santa Barbara, CA 93106, USA}

\emailAdd{samarthc@asu.edu}
\emailAdd{kfransen@ucsb.edu}
\emailAdd{keelerc@asu.edu}

\abstract{We embed the Penrose limit into the Weyl classical double copy. Thereby, we provide a lift of the double copy properties of plane wave spacetimes into black hole geometries and we open a novel avenue towards taking the classical double copy beyond statements about algebraically special backgrounds. In particular, the Penrose limit, viewed as the leading order Fermi coordinate expansion around a null geodesic, complements approaches leveraging asymptotic flatness such as the asymptotic Weyl double copy. Along the way, we show how our embedding of the Penrose limit within the Weyl double copy naturally fixes the functional ambiguity in the double copy for Petrov type N spacetimes. We also highlight the utility of a spinorial approach to the Penrose limit. In particular, we use this spinorial approach to derive a simple analytical expression for arbitrary Penrose limits of four-dimensional, vacuum type D spacetimes.}

\begin{document} 
\maketitle
\flushbottom
 
\newpage

\section{Introduction}
Gravity and gauge theory are related in surprising ways. Among these connections is BCJ duality (or
color-kinematics duality), which provides a map between scattering amplitudes of gravitational and Yang-Mills
theories \cite{Bern:2019prr,Bern:2022wqg,Adamo:2022dcm}. This map is usually termed a ``double copy'' map, since it
involves replacing color factors in each term by a second copy of the respective kinematic factor.

On the one hand, the double copy presents a powerful technical tool --- gravitational amplitudes can be expressed
in terms of simpler gauge theory amplitudes, which allows for efficient computations when modeling gravitational
wave observations of black hole binary systems
\cite{Bern:2019nnu,Bern:2019crd,Bern:2020uwk,Bern:2021yeh,Bern:2022jvn}. On the other hand, the double copy
relation appears to persist in UV-complete theories of quantum gravity, for instance manifesting as the KLT
relations between open and closed string amplitudes \cite{Kawai:1985xq}. The double copy is thus arguably of
foundational interest.

The double copy was originally discovered and studied in the context of perturbative scattering amplitudes
formulated on flat spacetimes with trivial background fields. Subsequently, there have been significant efforts
to treat non-trivial backgrounds \cite{Adamo:2017nia,Adamo:2023fbj,Sivaramakrishnan:2021srm,
Herderschee:2022ntr,Cheung:2022pdk,Ilderton:2024oly} (\cite{Adamo:2017qyl} in particular considered plane wave
backgrounds), and of understanding how the double copy manifests at the
level of exact classical solutions, first explored in \cite{Monteiro:2014cda} for Kerr-Schild spacetimes.

The classical double copy has by now an impressive array of examples
\cite{Luna:2016due,Luna:2015paa,Kim:2019jwm,Luna:2020adi,Luna:2018dpt,Godazgar:2021iae,Berman:2018hwd,Lee:2018gxc,Cho:2019ype,Cristofoli:2020hnk,Easson:2021asd,Easson:2022zoh,Easson:2020esh,Mkrtchyan:2022ulc,Ilderton:2018lsf,Barrientos:2024uuq,Ortaggio:2023cdz,Adamo:2022rob,Keeler:2020rcv,Keeler:2024bdt}
under its belt, but such examples are all special in various ways. The Kerr-Schild double copy introduced in
\cite{Monteiro:2014cda} of course requires that the spacetime under study be of the Kerr-Schild type, that is, it
requires the existence of a geodesic null vector $k^{\mu}$ such that the metric is of the form
\begin{equation} \label{} 
    g_{\mu \nu} = \eta_{\mu \nu} + \phi k_{\mu} k_{\nu}.
\end{equation}
A double copy relation in terms of the Weyl tensor was first formulated in \cite{Luna:2018dpt} and has
been studied extensively since
\cite{Godazgar:2020zbv,White:2020sfn,Chacon:2021wbr,Chacon:2021lox,Han:2022ubu,Alawadhi:2019urr,Alawadhi:2020jrv,Alkac:2023glx,Armstrong-Williams:2023ssz,Chawla:2022ogv,Mao:2023yle,Easson:2023dbk}.
In particular, type N spacetimes --- which include the plane wave spacetimes that arise out of a Penrose limit ---
were treated in \cite{Godazgar:2020zbv}.  The Weyl double copy proposal covers a broader class of spacetimes, but
still restricts them to be of Petrov type D or N. Indeed, it has been argued \cite{Luna:2022dxo} that an exact
classical double copy in position space is only possible for these special algebraic types.

One approach to generalizing beyond special algebraic types is to consider that any asymptotically flat spacetime,
to leading order in the radial coordinate $r$, is type N \cite{newman1968new}. This universal property was used in
\cite{Godazgar:2021iae,Adamo:2021dfg} to argue that a double copy map exists near asymptotic infinity for generic
asymptotically flat spacetimes. In the same spirit, Penrose observed that to leading order
around a null geodesic in an aribtrary spacetime, the limiting metric is that of a plane wave \cite{penrose1976any}
and thus type N. The existence of a double copy relation for type N spacetimes \cite{Godazgar:2020zbv} might then
allow us to study the double copy structure of an arbitrary spacetime, in an expansion about a null geodesic.

In this paper, we take a step towards this understanding by first studying how the procedure of taking a Penrose
limit interacts with the classical double copy structure of the full spacetime (where one is available). In section
\ref{sec:background} we provide a quick review of relevant material from the literature on the classical double
copy and on the Penrose limit. Our main results are contained in section \ref{sec:dcandPenrose}, in which we
outline the general procedure for taking a Penrose limit of a gravitational solution, along with the gauge theory
and scalar solutions that are associated to it via its double copy structure. We illustrate our approach in detail for the
Schwarzschild and Kerr solutions in sections \ref{sec:Schwarzschild} and \ref{sec:Kerr}, which has a
straightforward generalization to the generic vacuum type D (Plebanski-Demianski) metric in section \ref{sec:typeD}. We
also examine type N spacetimes in \ref{sec:typeN}. Detailed computations for the examples can be found in appendix
\ref{app:examples}.

\paragraph{Conventions} We will generally follow the conventions of Penrose and Rindler \cite{penrose1984spinors, Penrose:1986ca}. An important exception is the Riemann curvature which differs by a sign from that reference. See also Appendix \ref{app:conventions}.

\section{Background}\label{sec:background}

In this section, we review the classical double copy and the Penrose limit, which we will subsequently combine. We
highlight the main elements we need later. For a broader overview of the double copy and further references to the
literature see \cite{Bern:2022wqg,Kosower:2022yvp}. A more complete introduction to the Penrose limit can be found in \cite{blau2011plane}.  

\subsection{Classical Double Copy}\label{sec:dc}
The double copy of exact classical solutions was first discussed in
\cite{Monteiro:2014cda} for Kerr-Schild solutions. These spacetimes possess a metric that can be written (in
Kerr-Schild coordinates \cite{Kerr:1965vyg}) as 
\begin{equation} \label{eqn:KSmetric} 
    g_{\mu \nu} = \eta_{\mu \nu} + \phi k_{\mu} k_{\nu},
\end{equation}
where $k_{\mu}$ is a null, affinely parametrized geodesic vector field. Such spacetimes map to solutions of Maxwell
electrodynamics\footnote{In keeping with the double copy's amplitude roots, the single copy and zeroth copy
solutions are often interpreted as living on flat spacetime. This is also the view taken in this paper. There is
however a body of work
\cite{Didenko:2022qxq,Liang:2023zxo,Farnsworth:2023mff,Han:2022ubu,Han:2022mze,Alkac:2021bav,Gurses:2018ckx,Prabhu:2020avf,Carrillo-Gonzalez:2017iyj,Bahjat-Abbas:2017htu,Luna:2015paa,Alawadhi:2020jrv} which interprets these fields as living on the curved double copy spacetime itself,
whether as a test field or as coupled self-consistently to the metric.} via
\begin{equation} \label{eqn:KSgauge} 
    A_{\mu} = \phi k_{\mu}.
\end{equation}
The Kerr-Schild form \eqref{eqn:KSmetric}, along with the affineness of $k_{\mu}$, fixes the ``single copy'' gauge
potential $A_{\mu}$ and the ``zeroth copy'' scalar $\phi$ only up to a constant scaling as one can always shift $(\phi, k_{\mu}) \to
(c^{-2} \phi, c k_{\mu})$ without changing the metric for a constant $c$.

Another widely used formulation of the classical double copy is the Weyl double copy
\cite{Luna:2018dpt,Godazgar:2020zbv}, which is formulated in terms of the Weyl spinor $\Psi_{ABCD}$ and the Maxwell
field strength spinor $f_{AB}$, instead of the metric and gauge potential. The Weyl double copy relation is
\begin{equation} \label{WeylDCRelation} 
    \Psi_{ABCD} = \frac{1}{S} f_{(AB} f_{CD)}.
\end{equation}
This version of the classical double copy has the advantage of not manifestly privileging a certain choice of coordinates on the
gravity side, and of working with the gauge-invariant field strength on the gauge side\footnote{However,
    in order to interpret the Maxwell field strength and the scalar as living on a flat background
    (as was initially the case in the double copy literature), we need a well-defined map from the curved
    spacetime to a flat spacetime. In the case of pp-waves, which are type N, and the most general vacuum type D
solution (Plebanski-Demianski), \cite{Luna:2018dpt} uses the fact that a (double-)Kerr-Schild form is available to
obtain such a map.}. Moreover, it was argued in \cite{Luna:2018dpt} that whenever a Kerr-Schild double copy exists in addition to the Weyl double copy, both formulations are compatible. When appropriately scaled, the relation between the Weyl double copy $f_{AB}$, $S$ satisfying \eqref{WeylDCRelation} and the Kerr-Schild double copy $A_{\mu}$, $\phi$ satisfying \eqref{eqn:KSmetric} and \eqref{eqn:KSgauge} is simply
\begin{equation}\label{eqn:WeyltoKSrelation}
	(dA)_{AA' BB'} = f_{AB} \bar{\epsilon}_{A'B'} + \bar{f}_{A'B'} \epsilon_{AB} \, , \quad \phi = S + S^* \, .
\end{equation}

The Weyl spinor for a generic spacetime is constrained to be fully symmetric in its indices. In four dimensions,
fully symmetric spinorial objects can always be written as the symmetrized outer products of single-index spinors
\begin{equation} \label{} 
    \Psi_{ABCD} = \alpha_{(A} \beta_{B} \gamma_{C} \delta_{D)}.
\end{equation}
Clearly, if \eqref{WeylDCRelation} holds, the spinors $\alpha_{A}$, $\beta_{A}$, $\gamma_{A}$ and $\delta_{D}$
cannot be distinct. Indeed, they must either match up in pairs
\begin{equation} \label{eqn:spinortypeD} 
    \Psi_{ABCD} = \alpha_{(A} \alpha_{B} \beta_{C} \beta_{D)},
\end{equation}
known in the Petrov classification as a type D solution, or must in fact be the same,
\begin{equation} \label{eqn:spinortypeN} 
    \Psi_{ABCD} = \beta_{A} \beta_{B} \beta_{C} \beta_{D},
\end{equation}
which is a type N solution.

The black hole solutions of vacuum general relativity fall into the type D class with \eqref{eqn:spinortypeD} while radiative spacetimes, such as the plane waves obtained from the Penrose limit, will be of type N \eqref{eqn:spinortypeN}. Our focus in this paper will thus be on the interplay between the Weyl double copy for type D and type N solutions. However, plane waves are only a very special class of type N solutions, so we will also discuss the relation between the double copy of these plane waves and more generic type N spacetimes. 

For type N spacetimes, neither the Weyl nor the Kerr-Schild classical double copies are unique up to constant scaling. For plane wave spacetimes, the remaining freedom is particularly large but, as was already noted in \cite{Luna:2018dpt}, it is natural to additionally impose that a gravitational plane wave maps to a gauge theory plane wave.  Even with this additional constraint for plane waves, generic type N spacetimes have single copy field strengths and scalars that can be scaled by an arbitrary function of some of the coordinates such that the double copy relation as well as the equations of motion are maintained \cite{Godazgar:2020zbv}. We will discuss this issue further in section \ref{sec:ambiguity}.

While we have just argued that the form \eqref{WeylDCRelation} with identical field strength spinors $f_{AB}$
imposes the double copy spacetime to be Petrov type D or Petrov type N, one can also consider using two different
Maxwell spinors $f^{(1)}_{AB}$ and $f^{(2)}_{AB}$, which would allow extending the Weyl double copy to more general
Petrov type. The possibility of such a ``mixed'' double copy was considered already in \cite{Luna:2018dpt}, and
indeed the amplitudes literature has an established history of multiplying together distinct gauge theories to
produce a variety of gravity theories \cite{Bern:2019prr}. The mixed Weyl double copy was used in
\cite{White:2020sfn,Chacon:2021wbr} to produce examples of Petrov types I, II and III, albeit at the linearized
level using a twistor space formulation. More recently, a type II fluid dual metric was shown to have a Weyl double
copy structure \cite{Keeler:2024bdt} using the mixed prescription when expanded perturbatively in a near-horizon
expansion.

\subsection{Penrose limit}\label{sec:Penrosereview}

The Penrose limit was originally introduced in \cite{penrose1976any} starting from a coordinate system adapted to a twist-free null geodesic congruence. The metric in such coordinates in $d$ dimensions, by definition, takes the form
\begin{equation}\label{eqn:adapted}
ds^2 = 2 dU dV + D(x^{\mu}) dV^2 + B_i(x^{\mu}) dV dX^i - C_{ij}(x^{\mu})dX^i dX^j \, ,
\end{equation}
with $D(x^{\mu})$, $B_i(x^{\mu})$, and $C_{ij}(x^{\mu})$ functions of all the coordinates $x^{\mu} = \left\lbrace U, V, X^i \right\rbrace$ and with indices $i$, $j$ running over the $d-2$  ``transverse'' directions ($i,j, \ldots \in \left\lbrace 1, \ldots, d-2 \right \rbrace$). Note that $C_{ij}(x^{\mu})$ is symmetric and positive definite. The null coordinate vector field $\partial_U$ is geodesic and it is in this sense that the coordinates are ``adapted'' to the null geodesic congruence generated by $\partial_U$. 

The requirement that the geodesic congruence is twist-free arises as an integrability condition for it to be written as a coordinate vector field. On the other hand, any geodesic can locally be embedded in a twist-free geodesic congruence. More precisely, given any reference geodesic $\gamma$, a coordinate system can be found in a neighborhood of a portion of $\gamma$ free of conjugate points\footnote{Conjugate points (or focal points) are a pair of points such that there exists a Jacobi vector field (or equivalently here: a non-trivial solution to the geodesic deviation equation) vanishing at both points. Conjugate points are sometimes referred to as caustics but, more commonly, caustics are the geometric locus of conjugate points for a family of geodesics \cite{Hawking:1973uf}.} such that the metric takes the form \eqref{eqn:adapted} with (the conjugate point-free portion of) $\gamma$ corresponding to the geodesic at $X^i = V = 0$ \cite{penrose1972techniques}.

Having constructed the adapted coordinates  $\left\lbrace U, V, X^i \right\rbrace$ associated to $\gamma$, the
Penrose limit amounts to the leading order of an expansion in $\epsilon \to 0$ where $\left\lbrace U, V, X^i
\right\rbrace \sim \left\lbrace \epsilon^0, \epsilon^2, \epsilon^1 \right\rbrace$. One way to interpret
the scalings is by considering the combination of an ordinary expansion around a point but simultaneously boosting
along the null geodesic direction $\left\lbrace U, V, X^i \right\rbrace \sim \epsilon \times \left\lbrace
\epsilon^{-1}, \epsilon, 1 \right\rbrace$. With these scalings, on the assumption that $D(x^{\mu})$,
$B_i(x^{\mu})$, and $C_{ij}(x^{\mu})$ are smooth,\footnote{The smoothness assumption is made by \cite{penrose1976any} but can be somewhat relaxed. We will not attempt a precise formulation of weaker assumptions here but ideally shockwaves and other weak solutions to the Einstein equations would be included where possible.} the metric scales as $ds^2 = O(\epsilon^2)$ and the Penrose limit line element $ds_{\gamma}^2$ is thus given by only keeping the terms in \eqref{eqn:adapted} scaling as $\epsilon^2$. Explicitly, the Penrose limit of \eqref{eqn:adapted} with respect to the null geodesic $\gamma$ at $V=X^i=0$ yields
\begin{equation}\label{eqn:Rosen}
ds_{\gamma}^2 = 2 dU dV - C_{ij}\left(U\right)dX^i dX^j  \, , \quad C_{ij}\left(U\right) = C_{ij}\left(x^{\mu}\right)|_{(V=0, X^i=0)} \, ,
\end{equation}
which is a plane wave spacetime in Rosen coordinates \cite{rosen1937plane}.

Rosen coordinates are not, in general, globally well-defined. Similarly, the presented construction of the Penrose limit so far only works for a portion of $\gamma$ free of conjugate points. On the other hand, it is well-known \cite{robinson1956report,penrose1965remarkable} that the coordinate singularities of the Rosen coordinates can be removed by going to Brinkmann coordinates $\left\lbrace u, v, x^a \right\rbrace$ with $a, b, \ldots \in \left\lbrace 1, \ldots, d-2 \right \rbrace$, in which the plane wave metric takes the form \cite{brinkmann1925einstein}
\begin{equation}\label{eqn:Brinkmann}
ds_{\gamma}^2 = 2 du dv - H_{ab}(u)x^a x^b du^2 - \delta_{ab} dx^a dx^b  \, .
\end{equation}
In order to explicitly relate $\left\lbrace U, V, X^i \right\rbrace$ to $\left\lbrace u, v, x^a \right\rbrace$, let
$u = U$ and introduce a parallel propagated, transverse frame $E^{(\gamma)}{}^i{}_a(u)$ such that
\begin{equation}\label{eqn:frame}
X^i = E^{(\gamma)}{}^i{}_a x^a \, , \quad  C_{ij} E^{(\gamma)}{}^i{}_a  E^{(\gamma)}{}^j{}_b \equiv E^{(\gamma)}{}^i{}_a  E^{(\gamma)}{}_{ib} = \delta_{ab}\, ,
\end{equation}
and\footnote{For the condition \eqref{eqn:Esymm}, it is important to note the distinction from the identity
	\begin{equation}\label{eqn:notEsymm}
	E^{(\gamma)}{}^i{}_a  \dot{E}^{(\gamma)}{}_{ib} = - \dot{E}^{(\gamma)}{}^i{}_a  E^{(\gamma)}{}_{ib}  \, .
	\end{equation}
	In particular, \eqref{eqn:Esymm} and \eqref{eqn:notEsymm} are not in contradiction because the latin ``Rosen'' indices are raised and lowered by $C_{ij}$, which is time-dependent.}
\begin{equation}\label{eqn:Esymm}
E^{(\gamma)}{}^i{}_a  \dot{E}^{(\gamma)}{}_{ib} = E^{(\gamma)}{}^i{}_b  \dot{E}^{(\gamma)}{}_{ia} \, ,
\end{equation} 
where a dot, such as in $\dot{E}^{(\gamma)}{}^i{}_a$, denotes a derivative with respect to $u=U$. Equation \eqref{eqn:frame} is simply the condition on $E^{(\gamma)}{}^i{}_a(u)$ to form an orthonormal transverse frame while \eqref{eqn:Esymm} ensures parallel transport along $\partial_U$. The full change of coordinates from Rosen to Brinkmann coordinates is then given by
\begin{equation}\label{eqn:RtoB}
\begin{aligned}
U &= u \, , \\
V &= v + \frac{1}{2}  \dot{E}^{(\gamma)}{}^i{}_a  E^{(\gamma)}{}_{ib} x^a x^b \, , \\
X^i &= E^{(\gamma)}{}^i{}_a x^a \, .
\end{aligned}
\end{equation}
In terms of the frame $E^{(\gamma)}{}^i{}_a$, the Brinkmann coordinate ``wave profile'' $H_{ab}$ is given by
\begin{equation}\label{eqn:HOframe}
H_{ab}  = E^{(\gamma)}{}^i{}_a  \ddot{E}^{(\gamma)}{}_{ib} \, .
\end{equation}

There is an alternative approach to the Penrose limit based on the introduction of Fermi null coordinates
\cite{Blau:2006ar}, which directly obtains the limit in Brinkmann coordinates. This Fermi null
coordinate approach relies on first constructing, in the original spacetime, a parallel propagated null
frame\footnote{We will refer to a null frame (sometimes called a pseudo-orthonormal frame) as a frame with two
	complementary null directions, say $u^{\mu}$ and $v^{\mu}$ normalized such that $u^{\mu} v_{\mu} = 1$, in addition
	to an orthonormal basis for the transverse directions. We will reserve the terminology Newman-Penrose frame for a
	four-dimensional frame where the transverse directions are also spanned by (complexified) complementary null
	vectors.} $\left\lbrace u^{\mu}, v^{\mu}, E^{\mu}_a \right \rbrace$ on $\gamma$, where $u^{\mu}$ is tangent to
$\gamma$ (affinely parameterized). Then, the wave profile $H_{ab}$ of the Penrose limit plane wave in Brinkmann coordinates \eqref{eqn:Brinkmann} can be identified as \cite{Blau:2006ar,blau2011plane}
\begin{equation}\label{eqn:penroseframe}
H_{ab}(u) = \left(R_{\mu \nu \alpha \beta} E^{\mu}_a u^{\nu} E^{\alpha}_b u^{\beta} \right)|_{\gamma} \, .
\end{equation}
From the perspective of the Penrose limit metric in Brinkmann coordinates \eqref{eqn:Brinkmann}, \eqref{eqn:penroseframe} is consistent with the relation between the Riemann tensor $R^{(\gamma)}_{\mu \nu \alpha \beta}$ of the Penrose limit plane wave and its wave profile, given by
\begin{equation}\label{eqn:HRplanewave}
H_{ab}(u) = R^{(\gamma)}_{\mu \nu \alpha \beta} \left(\partial_a\right)^{\mu} (\partial_u)^{\nu}  \left(\partial_b\right)^{\alpha} (\partial_u)^{\beta}   \, .
\end{equation}

Moreover, $H_{ab}$ entirely specifies the plane wave metric. In Brinkmann coordinates, $\partial_u$ is only null at the origin of the transverse space $x^a = 0$. We therefore introduce
\begin{equation}\label{eqn:udefplane}
u^{(\gamma)} = \partial_u + \frac{1}{2} H_{ab}x^a x^b \partial_v \, ,
\end{equation}
which together with the coordinate vector fields
\begin{equation}\label{eqn:vdefplanewave}
v^{(\gamma)} = \partial_v \, , \quad E^{(\gamma)}_a = \partial_a \, ,
\end{equation}
forms a null frame $\left\lbrace u^{(\gamma)}{}^{\mu}, v^{(\gamma)}{}^{\mu}, E^{(\gamma)}{}^{\mu}_a \right \rbrace$
for the plane wave. From $R^{(\gamma)}_{\mu \nu \alpha \beta}v^{(\gamma)}{}^{\mu} = 0$, it follows that \eqref{eqn:HRplanewave} is unchanged by replacing $\partial_u$ with $u^{(\gamma)}$. In addition, note the difference between $E^{(\gamma)}{}^{\mu}_a$ and the previously introduced transverse frame $E^{(\gamma)}{}^{i}_a$ relating Rosen and Brinkmann coordinates in \eqref{eqn:RtoB}. These transverse frames are related by
\begin{equation}
E^{(\gamma)}{}_a = E^{(\gamma)}{}^i_{a}\partial_i + \dot{E}^{(\gamma)}{}^i_a E^{(\gamma)}{}_{ib} x^b 	\partial_V \, .
\end{equation}

In four dimensions, an equivalent two-spinor approach to \eqref{eqn:penroseframe} was suggested in \cite{Tod:2019urw}. This two-spinor approach will be very useful in conjunction with the Weyl double copy so we will review it in some detail. First, introduce the complex coordinate $z$ for the two transverse directions of the four-dimensional plane wave in Brinkmann coordinates
\begin{equation}\label{eqn:complexz}
z = \frac{1}{\sqrt{2}}\left(x^1 + i x^2\right) \, , \quad \bar{z} = \frac{1}{\sqrt{2}}\left(x^1 - i x^2\right)  \, ,
\end{equation}
such that \eqref{eqn:Brinkmann} with $a,b \in \left\lbrace  1,2\right \rbrace$ becomes
\begin{equation}\label{eqn:Brinkmannc}
ds_{\gamma}^2 = 2 du dv + \left(\Psi(u) z^2 + 2 \Phi(u) z \bar{z} + \bar{\Psi}(u) \bar{z}^2\right) du^2 - 2 dz d\bar{z}  \, .
\end{equation}
$\Psi(u)$ and $\Phi(u)$ in \eqref{eqn:Brinkmannc} are related to $H_{ab}(u)$ in \eqref{eqn:Brinkmann} through
\begin{equation}\label{eqn:HBrinkmannincomplex}
H_{ab} = - \begin{pmatrix}
\frac{1}{2}\left(\Psi(u)+ 2 \Phi(u)  + \bar{\Psi}(u)\right) & 	\frac{i}{2}\left(\Psi(u) - \bar{\Psi}(u)\right)\\
\frac{i}{2}\left(\Psi(u) - \bar{\Psi}(u)\right) 	&  \frac{1}{2}\left(-\Psi(u)+ 2 \Phi(u)  - \bar{\Psi}(u)\right) 
\end{pmatrix}  \, .
\end{equation}
In terms of the real functions $\Psi_{+},\Psi_{\times},\Phi$, where
\begin{equation}
\Psi(u) = \Psi_{+}(u) + i \Psi_{\times}(u) \, ,
\end{equation}
the wave profile suggestively takes the form of the ``+'' ($\Psi_{+}$) and ``$\times$'' ($\Psi_{\times}$) gravitational wave polarizations
\begin{equation}
H_{ab} =  \begin{pmatrix}
- \Psi_{+}(u) & 	\Psi_{\times}(u) \\
\Psi_{\times}(u) 	&   \Psi_{+}(u) 
\end{pmatrix} - \begin{pmatrix}
\Phi(u)  & 	0 \\
0 	&   \Phi(u) 
\end{pmatrix} \, .
\end{equation}
The trace, $2\Phi$, will vanish by the vacuum Einstein equations and therefore will not play a significant role in much of the main text\footnote{Although we discuss only the classical double copy for vacuum spacetimes, see \cite{Easson:2021asd,Easson:2022zoh} for extensions to include matter.}. However, we do not need to impose vacuum Einstein equations to take the Penrose limit so we will not impose $\Phi(u)$ to vanish here.

To take the Penrose limit along a null geodesic $\gamma$ in four dimensions, instead of using the tangent vector $u^{\mu}$ and a null frame $\left\lbrace u^{\mu}, v^{\mu}, E^{\mu}_a \right \rbrace$ as in \eqref{eqn:penroseframe}, we can use a tangent, parallel propagated two-spinor $\alpha^{A}$. Explicitly, $\alpha^{A}$ and its complex conjugate $\bar{\alpha}^{A'}$ are chosen to satisfy
\begin{equation}\label{eqn:tangentspinor}
u^{\alpha} = \sigma^{\alpha}_{AA'} \alpha^{A}\bar{\alpha}^{A'} \, , \quad	u^{\mu}\nabla_{\mu} \alpha^A = 0 \, , \quad u^{\mu}\nabla_{\mu} \bar{\alpha}^{A'} = 0 \, ,
\end{equation}
where $\sigma^{\alpha}_{AA'}$ are the Infeld-Van der Waerden symbols. See Appendix \ref{app:conventions} for more details and our spinor conventions. Using $\alpha^{A}$, we can obtain the Penrose limit in terms of the curvature spinors by computing \cite{Tod:2019urw}
\begin{equation}\label{eqn:penrosespinor}
\Psi(u) = \left(\Psi_{ABCD}\alpha^A \alpha^B \alpha^C \alpha^D\right)|_{\gamma} \, , \quad \Phi(u) = \left(\Phi_{ABA'B'}\alpha^A \alpha^B \bar{\alpha}^{A'} \bar{\alpha}^{B'}\right)|_{\gamma} \, .
\end{equation}
Here, $\Psi_{ABCD}$ and $\Phi_{ABA'B'}$ are respectively the Weyl spinor and Ricci curvature spinor of the original spacetime while $\Psi$ and $\Phi$ are the functions determining the plane wave spacetime in (complex) Brinkmann coordinates by \eqref{eqn:Brinkmannc}. 

In order to derive the expression for the wave profile \eqref{eqn:penroseframe} from \eqref{eqn:penrosespinor}, introduce a complementary, parallel propagated spinor $\beta^A$ to $\alpha^A$ to have a two-spinor basis (or dyad) $\left\lbrace \alpha^A, \beta^A\right\rbrace$ along the null geodesic. Then, after constructing the Newman-Penrose frame $\left\lbrace u^{\mu}, v^{\mu}, m^{\mu},\bar{m}^{\mu}  \right \rbrace$ from the dyad $\left\lbrace \alpha^A, \beta^A\right\rbrace$, \eqref{eqn:penrosespinor} can be derived directly from \eqref{eqn:penroseframe} with
\begin{equation}\label{eqn:complextransversebasis}
m^{\mu} = -\frac{1}{\sqrt{2}}\left(E^{\mu}_1 - i E^{\mu}_2 \right) \, , \quad 	\bar{m}^{\mu} = -\frac{1}{\sqrt{2}}\left(E^{\mu}_1 + i E^{\mu}_2 \right) \, .
\end{equation} 
See Appendix \ref{app:NP} for the explicit calculation.

From the perspective of the Penrose limit metric in (complex) Brinkmann coordinates \eqref{eqn:Brinkmannc}, we similarly first go from the null frame $\left\lbrace u^{(\gamma)}{}^{\mu}, v^{(\gamma)}{}^{\mu}, E^{(\gamma)}{}^{\mu}_1 , E^{(\gamma)}{}^{\mu}_2 \right \rbrace$ to the Newman-Penrose frame $\left\lbrace u^{(\gamma)}{}^{\mu}, v^{(\gamma)}{}^{\mu}, m^{(\gamma)}{}^{\mu}_1 , \bar{m}^{(\gamma)}{}^{\mu}_2 \right \rbrace$ with
\begin{equation}\label{eqn:mdefplanewave}
m^{(\gamma)}{}^{\mu} = -\frac{1}{\sqrt{2}}\left(E^{(\gamma)}{}^{\mu}_1 - i E^{(\gamma)}{}^{\mu}_2 \right) \, , \quad 	\bar{m}^{(\gamma)}{}^{\mu} = -\frac{1}{\sqrt{2}}\left(E^{(\gamma)}{}^{\mu}_1 + i E^{(\gamma)}{}^{\mu}_2 \right) \, ,
\end{equation} 
and we introduce an associated dyad $\left\lbrace \alpha^{(\gamma)}{}^A , \beta^{(\gamma)}{}^A\right\rbrace$, meaning that the Newman-Penrose frame is given by
\begin{equation}\label{eqn:NPindyad}
u^{(\gamma)}{}^{AA'} = \alpha^{(\gamma)}{}^A \bar{\alpha}^{(\gamma)}{}^{A'}\, ,  \quad v^{(\gamma)}{}^{AA'} = \beta^{(\gamma)}{}^A \bar{\beta}^{(\gamma)}{}^{A'}\, , \quad m^{(\gamma)}{}^{AA'} =\alpha^{(\gamma)}{}^A \bar{\beta}^{(\gamma)}{}^{A'} \, .
\end{equation}
In particular, $\beta^{(\gamma)}{}^{A}$  can be chosen to be a Killing spinor \cite{Freedman:2012zz}\footnote{See in particular Appendix 22A.}
\begin{equation}
\nabla_{\mu} \beta^{(\gamma)}{}^{A} = \nabla_{\mu} \bar{\beta}^{(\gamma)}{}^{A'}  = 0 \, . 
\end{equation}
Now $\left\lbrace \alpha^{(\gamma)}{}^A , \beta^{(\gamma)}{}^A\right\rbrace$ is a principal dyad for the plane wave. Therefore, as the plane wave is of Petrov type N, the only independent Weyl scalar is $\Psi^{(\gamma)}_0$, which can be computed to be
\begin{equation}\label{eqn:NPsi}
\Psi^{(\gamma)}_0 = \Psi^{(\gamma)}_{ABCD}\alpha^{(\gamma)}{}^A \alpha^{(\gamma)}{}^B \alpha^{(\gamma)}{}^C \alpha^{(\gamma)}{}^D = - C^{(\gamma)}_{\alpha \beta \gamma \delta} u^{(\gamma)}{}^{\alpha}  m^{(\gamma)}{}^{\beta} u^{(\gamma)}{}^{\gamma} m^{(\gamma)}{}^{\delta} = \Psi(u) \, .
\end{equation}
On the other hand, the only independent Ricci spinor is
\begin{equation}\label{eqn:Phi00penrose}
\Phi^{(\gamma)}_{00} = \Phi^{(\gamma)}_{ABC'D'}\alpha^{(\gamma)}{}^A \alpha^{(\gamma)}{}^B \bar{\alpha}^{(\gamma)}{}^{C'} \bar{\alpha}^{(\gamma)}{}^{D'} = \frac{1}{2}R^{(\gamma)}_{\alpha \beta} u^{(\gamma)}{}^{\alpha} u^{(\gamma)}{}^{\beta} = \Phi(u) \, .
\end{equation}
When evaluated at the origin of the transverse plane $z = \bar{z} = 0$, the expressions \eqref{eqn:NPsi} and
\eqref{eqn:Phi00penrose} are the plane wave version of the two-spinor prescription to take the Penrose limit \eqref{eqn:penrosespinor}. Moreover, the resulting $\Psi(u) $ and $\Phi(u)$ fully specify the four-dimensional plane wave.

The Penrose limit was originally proposed in the context of pure pseudo-Riemannian geometry, that is including only a metric. However the Penrose limit has since been generalized to include various types of matter, specifically the massless matter fields of the supergravity theories arising from the low-energy limit of the perturbative string theories  \cite{Gueven:2000ru}.  The generalized Penrose limit including extra matter is sometimes referred to as the Penrose-G\"uven limit. Importantly for our purposes, there is a Penrose-G\"uven limit for massless scalars $S(x)$ and gauge fields $A_{\mu}(x)$. 

For a massless scalar field, after going to adapted coordinates in general dimension $d$ \eqref{eqn:adapted}, the Penrose-G\"uven limit is trivially 
\begin{equation}
S(x) \to S^{(\gamma)}\left(U\right) = S|_{V=0, X^i=0} \, .
\end{equation}
On the other hand, for the gauge fields, in addition to the adapted coordinates, we need to impose an analogous ``adapted'' gauge choice \cite{Gueven:2000ru}
\begin{equation}\label{eqn:adaptedgauge}
A_{U}(x) = 0 \, .
\end{equation}
With this choice of gauge in the adapted coordinates, the gauge field is $O(\epsilon)$ to leading order with the Penrose limit scaling $\left\lbrace U, V, X^i \right\rbrace \sim \left\lbrace \epsilon^0, \epsilon^2, \epsilon^1 \right\rbrace$. An argument for \eqref{eqn:adaptedgauge}, given by G\"uven \cite{Gueven:2000ru}, is that the $O(\epsilon)$ scaling of the gauge field and an $O(\epsilon^0)$ scaling of the scalar together with the $O(\epsilon^2)$ scaling of the metric both imply a $O(\epsilon^{d-2})$ scaling similarity of the action. Such scaling similarities are not true symmetries of the theory but they can nevertheless be very useful, see for instance \cite{Biggs:2023sqw} and references therein.

Taking the Penrose limit with the gauge condition \eqref{eqn:adaptedgauge}, the ``Rosen-gauge'' Penrose limit gauge field is
\begin{equation}\label{eqn:ARosen}
A^{(\gamma)}_{\mu} dx^{\mu} = A_i|_{V=0, X^i=0}  dX^i  \, .
\end{equation}
The gauge transformation
\begin{equation}
A^{(\gamma)}_{\mu} dx^{\mu} \to A^{(\gamma)}_{\mu} dx^{\mu} - d( X^i A_i) \, , \quad \dot{A}_a = E^{(\gamma)}{}^i_a \dot{A}_i \, ,
\end{equation}
results in the ``Brinkmann-gauge'' field
\begin{equation}\label{eqn:Abrinkmann}
A^{(\gamma)}_{\mu} dx^{\mu} = -x^a \dot{A}_a du \, , \quad  F^{(\gamma)}_{\nu \mu} dx^{\nu} \wedge dx^{\mu} =  \dot{A}_a du \wedge dx^a\, .
\end{equation}
$E^{(\gamma)}{}^i_a$ here is the frame that was introduced to go from Rosen to Brinkmann coordinates in \eqref{eqn:RtoB}. 

As for the metric, there is a more direct way to obtain the Penrose-G\"uven limit of a gauge field in Brinkmann coordinates from a Fermi null coordinate expansion. Analogous to \eqref{eqn:penroseframe}, using again a parallel propagated null frame $\left\lbrace u^{\mu}, v^{\mu}, E^{\mu}_a \right \rbrace$ on the reference geodesic $\gamma$
\begin{equation}\label{eqn:penroseframeF}
\dot{A}_a(u) = \left(F_{\mu \nu} u^{\nu}  E^{\mu}_a \right)|_{\gamma} \, .
\end{equation}

In four dimensions, \eqref{eqn:penroseframeF} can also be expressed in terms of the parallel propagated dyad  $\left\lbrace \alpha^A, \beta^A\right\rbrace$ analogously to \eqref{eqn:penrosespinor} for the metric. First, in terms of the complex Brinkmann coordinates \eqref{eqn:complexz},  \eqref{eqn:Abrinkmann} is given by 
\begin{equation}
F^{(\gamma)}_{\nu \mu} dx^{\nu} \wedge dx^{\mu} =  \bar{\cA} \,  d\bar{z} \wedge du   + \cA  \, dz   \wedge du \, ,  \quad  \cA = -\frac{1}{\sqrt{2}}\left(\dot{A}_1 - i \dot{A}_2\right) \, .
\end{equation}
Now, we use the Newman-Penrose frame \eqref{eqn:complextransversebasis} in \eqref{eqn:penroseframeF}, instead of the real orthonormal transverse basis, and write this Newman-Penrose frame in terms of the associated dyad \eqref{eqn:NPindyad}. Finally, we write the field strength in terms of spinors
\begin{equation}
F_{AA' BB'} = f_{AB}\epsilon_{A'B'} + \bar{f}_{A'B'}\epsilon_{AB} \, ,
\end{equation}
and find that the expression for Penrose-G\"uven limit  \eqref{eqn:penroseframeF} in terms of two-spinors becomes
\begin{equation}\label{eqn:penrosespinF}
\begin{aligned}
\cA =  \left(f_{CD} \alpha^C \alpha^D \right)|_{\gamma}\, .
\end{aligned}
\end{equation}
Similar to the Penrose limit metric, this field is ``type N'' in the sense that, in the expansion of the field strength in terms of the principal dyad $\left\lbrace \alpha^{(\gamma)}{}^A , \beta^{(\gamma)}{}^A\right\rbrace$
\begin{equation}
f^{(\gamma)}_{AB} = f_2  \alpha^{(\gamma)}{}_A   \alpha^{(\gamma)}{}_B - 2 f_1    \alpha^{(\gamma)}{}_{(A}  \beta^{(\gamma)}{}_{B)} + f_0 \beta^{(\gamma)}{}_A   \beta^{(\gamma)}{}_B = \cA \beta^{(\gamma)}{}_A   \beta^{(\gamma)}{}_B\, ,
\end{equation}
only $f_0$ is non-vanishing and it is given by \eqref{eqn:penrosespinF}.

To conclude this brief review of the Penrose limit, let us emphasize that the Penrose limit is only the leading order of an expansion around a reference null geodesic. Higher orders of such a ``Penrose expansion'', interpreted in terms of a Fermi null coordinate system, moreover exhibit a ``peeling'' behavior of the curvature analogous to expansions around asymptotically flat space \cite{Blau:2006ar}. Nevertheless, it is interesting that even the leading order of the Penrose expansion, the Penrose limit itself is not just flat space. The Penrose limit isometry group is thus not the Poincar\'e group but there are still at least $2d+1$ Killing vectors in $d$-dimensions \cite{blau2011plane}.

We will not review the isometry group of plane wave spacetimes here in detail, but simply note that there are different useful perspectives. A first point of view, which emphasizes the relation of the Penrose limit to Jacobi vector fields and thus the tangent space to the phase space of null geodesics, is that the isometry group of plane waves are generated (at least) by a Heisenberg algebra \cite{blau2011plane}. Alternatively, as done in  \cite{Duval:2017els}, one can understand the isometries of a $d$-dimensional plane wave in terms of a $(d-1)$-dimensional Carroll group (generally without rotations). This Carrollian perspective of the plane wave emphasizes the way plane waves are naturally foliated in null hypersurfaces, which in turn are closely related to Carrollian geometry \cite{Ciambelli:2023mir}.

\section{The classical double copy of the Penrose limit}\label{sec:dcandPenrose}

Plane wave spacetimes are an excellent testing ground for going beyond flat space both on account of their simplicity and because, as just reviewed in Section \ref{sec:Penrosereview}, any curved spacetime is approximated by a plane wave around null geodesics. In the introduction, we have mentioned for instance prominent applications in string theory \cite{Horowitz:1989bv,Horowitz:1990sr,Kiritsis:1993jk,Russo:2002rq,Papadopoulos:2002bg,Berenstein:2002jq,Eberhardt:2018exh}, which make use of plane waves as a bridge to more complicated background geometries, but the same logic has recently been fruitful in the context of the double copy \cite{Adamo:2017nia,Adamo:2018mpq,Adamo:2020qru}, where plane waves have been one of the first examples where one can double copy both the non-perturbative backgrounds and amplitudes on top of them. 

\begin{figure}[h]
	\begin{tikzcd}
		\makebox[\lenf em][c]{Double copy spacetime}  \arrow[r,leftrightarrow,shorten=-4mm, "\text{Weyl DC}"] \arrow[dd,"\text{Penrose limit}{}_{\gamma}"]& \makebox[\lenf em][c]{single copy gauge field} \arrow[dd,dashed,"\text{?}"] \\
		& \\
		\makebox[\lenf em][c]{plane wave spacetime}  \arrow[r,leftrightarrow,shorten=-4mm, "\text{Weyl DC}"] &  \makebox[\lenf em][c]{plane wave single copy} 
	\end{tikzcd}
	\caption{To understand how double copy results on plane wave backgrounds can inform more general
		backgrounds, from which the plane waves arise as Penrose limits, we complete the above diagram.}\label{fig:schematicquestion}
\end{figure}
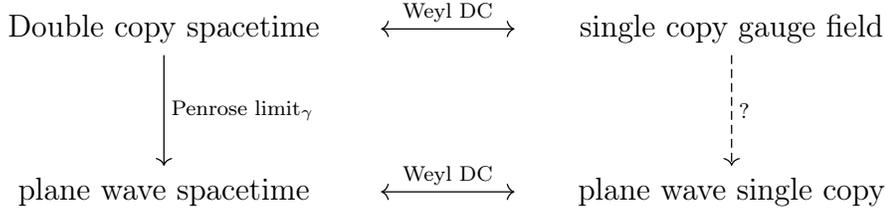

On the other hand, to understand how such results for plane waves inform more general spacetimes, one must know the interplay between the double copy and the Penrose limit. It is for this reason that we now investigate the interplay between the classical Weyl double copy and the Penrose limit. Specifically, this means we first need to complete the diagram in Figure \ref{fig:schematicquestion}.

Combining the Weyl double copy prescription in two-spinor language \eqref{WeylDCRelation} with the two-spinor Penrose limit curvatures \eqref{eqn:penrosespinor}, a natural proposal for the Penrose limit on the single and zeroth copy is
\begin{equation}\label{eqn:dcPenroselimit}
f^{(\gamma)}_{AB} = \left(f_{CD}\alpha^C \alpha^D \right)|_{\gamma} \beta^{(\gamma)}_A  \beta^{(\gamma)}_B   \, , \quad S^{(\gamma)} = S|_{\gamma} \, .
\end{equation}
Then, if by the Weyl double copy \eqref{WeylDCRelation}
\begin{equation}
\psi_{ABCD} = \frac{1}{S} f_{(AB} f_{CD)} \, ,
\end{equation}
it follows immediately from the two-spinor Penrose limit \eqref{eqn:penrosespinor} that
\begin{equation}
\psi^{(\gamma)}_{ABCD} = \frac{1}{S^{(\gamma)}} f^{(\gamma)}_{(AB} f^{(\gamma)}_{CD)}  \, .
\end{equation}
When $f_{AB}$ and $S$ are interpreted as living on the curved background, \eqref{eqn:dcPenroselimit} is simply the spinorial formulation of the Penrose-G\"uven limit. The single and zeroth Weyl copy satisfy the Maxwell and massless free scalar wave equation on the curved background with curvature set by $\Psi_{ABCD}$, and so do their Penrose-G\"uven limits with respect to the Penrose limit of the background with curvature $\Psi^{(\gamma)}_{ABCD}$. The single copy Penrose limit \eqref{eqn:dcPenroselimit} is thus perfectly consistent and well-known as the Penrose-G\"uven limit. However, to interpret $f_{AB}$ and $S$ as living on a Minkowski background in line with the original double copy, the meaning of the reference null geodesic $\gamma$ and its associated spinor $\alpha^A$ in \eqref{eqn:dcPenroselimit} requires further clarification. 

We use superscripts, e.g. $\gamma^{(g)}$ and $\gamma^{(\eta)}$, to denote respectively structures on the curved and flat backgrounds. As pointed out in the original construction of the Weyl double copy \cite{Luna:2018dpt}, to say more about the single copy on the flat background we need some input from the Kerr-Schild double copy. Specifically, the coordinate map between the curved and the flat spacetime comes from Kerr-Schild coordinates. Thus, $\gamma^{(\eta)}$ can be identified in the Minkowski background in terms of its expression in Kerr-Schild coordinates. That is, if there exists a Kerr-Schild double copy prescription, we can write the metric in the form \eqref{eqn:KSmetric}
\begin{equation}
g_{\mu \nu} = \eta_{\mu \nu} + \phi k_{\mu} k_{\nu} \, , \quad  g^{\mu \nu} = \eta^{\mu \nu} - \phi k^{\mu} k^{\nu}  \, ,
\end{equation}
and identify $\gamma^{(\eta)}$ and  $\gamma^{(g)}$ in these Kerr-Schild coordinates. For the tangent vectors, we then find $u^{(\eta)}{}^{\mu} = u^{(g)}{}^{\mu}$.\footnote{Equalities written between quantities on the curved and on the flat background are, with some abuse of notation, always to be understood as implying that they map to each other under the double copy.}

On the other hand, a null geodesic  $\gamma^{(g)}$ in the curved background will no longer be null when mapped to the Minkowski space curve $\gamma^{(\eta)}$. Instead, the norm of the tangent $u^{(\eta)}{}^{\mu}$ to $\gamma^{(\eta)}$ becomes
\begin{equation}\label{eqn:flatmass}
\eta_{\mu \nu} u^{(\eta)}{}^{\mu} u^{(\eta)}{}^{\nu} = - \phi \left(k_{\mu} u^{(\eta)}{}^{\mu}\right)^2 \, .
\end{equation}
As a result, no flat space spinor $\alpha^{(\eta)}{}^A$ can square to the tangent vector as in curved spacetime because\footnote{Recall that $\sigma^{\alpha}_{AA'}$ are the Infeld-Van der Waerden symbols and further details on the two-spinor conventions can be found in Appendix \ref{app:conventions}.} $\alpha^{(\eta)}{}^A \bar{\alpha}^{(\eta)}{}^{A'} \neq \sigma^{(\eta)}{}{}^{AA'}_{\mu} u^{(\eta)}{}^{\mu}$ for any non-null $u^{(\eta)}{}^{\mu}$. We will thus need to find a natural null vector that can be used to identify, from the single copy perspective in flat space, the two-spinor in the single copy Penrose limit \eqref{eqn:dcPenroselimit}. To do so, we will first establish a relation between spinors on the curved and flat spacetimes under the Kerr-Schild map.

One way to understand the Kerr-Schild map is in terms of a shift of an appropriate null frame; there exist frames $\left\lbrace  k^{(g)},  n^{(g)}, m^{(g)}, \bar{m}^{(g)}\right \rbrace$ and $\left\lbrace  k^{(\eta)},  n^{(\eta)}, m^{(\eta)}, \bar{m}^{(\eta)}\right \rbrace$ with
\begin{equation}\label{eqn:NPframemetric}
g_{\mu \nu} = 2k^{(g)}_{(\mu} n^{(g)}_{\nu)} - 2m^{(g)}_{(\mu} \bar{m}^{(g)}_{\nu)} \, , \quad  	\eta_{\mu \nu} =2k^{(\eta)}_{(\mu} n^{(\eta)}_{\nu)} - 2m^{(\eta)}_{(\mu} \bar{m}^{(\eta)}_{\nu)} \, ,
\end{equation}
such that, under the Kerr-Schild identification map
\begin{equation}\label{eqn:KSmap}
n^{(g)}_{\mu}= n^{(\eta)}_{\mu} + \frac{\phi}{2} k^{(\eta)}_{\mu} \, , \quad  k^{(g)}_{\mu} = k^{(\eta)}_{\mu} = k_{\mu} \, , \quad m^{(g)}_{\mu} = m^{(\eta)}_{\mu} \, , \quad \bar{m}^{(g)}_{\mu} = \bar{m}^{(\eta)}_{\mu}   \, .
\end{equation}
Some care is needed with raising and lowering, which can differ between the curved and the flat space. For the basis vectors one has instead
\begin{equation}\label{eqn:frameshift}
n^{(g)}{}^{\mu}= n^{(\eta)}{}^{\mu} - \frac{\phi}{2} k^{(\eta)}{}^{\mu} \, , \quad  k^{(g)}{}^{\mu} = k^{(\eta)}{}^{\mu} = k^{\mu} \, , \quad m^{(g)}{}^{\mu} = m^{(\eta)}{}^{\mu} \, , \quad \bar{m}^{(g)}{}^{\mu} = \bar{m}^{(\eta)}{}^{\mu}   \, .
\end{equation}

To express \eqref{eqn:NPframemetric} and \eqref{eqn:frameshift} on the level of two-spinors,   define the Infeld-Van der Waerden symbols $\sigma{}{}^{AA'}_{\mu}$ in both the flat and curved spacetime implicitly by the choice of a compatible dyad basis $\left\lbrace o^A,\iota^A \right \rbrace$ such that
\begin{equation}\label{eqn:associateddyad}
o^A\bar{o}^{A'} = \sigma{}^{AA'}_{\mu}k^{\mu} \, , \quad o^A\bar{\iota}^{A'} = \sigma{}^{AA'}_{\mu}m^{\mu} \, , \quad \iota^A\bar{o}^{A'} = \sigma{}^{AA'}_{\mu}\bar{m}^{\mu} \, , \quad \iota^A\bar{\iota}^{A'} = \sigma{}^{AA'}_{\mu}n^{\mu} \, .
\end{equation}
Then, we can identify the spinors $\left\lbrace o^{(\eta)}{}^A,\iota^{(\eta)}{}^A \right \rbrace$ and $\left\lbrace
o^{(g)}{}^A,\iota^{(g)}{}^A \right \rbrace$ under the double copy, despite \eqref{eqn:frameshift}, on the condition
that the flat and curved spinors connect differently to their respective spacetime tangent spaces. Specifically,
just like the Newman-Penrose frames, the Infeld-van der Waerden symbols shift in the map from the curved to the
flat spacetime  \cite{Lobo:2017bfh} 
\begin{equation}\label{eqn:IvW}
\sigma^{(g)}{}{}^{AA'}_{\mu}   = \sigma^{(\eta)}{}{}^{AA'}_{\mu} + \frac{\phi}{2} k_{\mu} \sigma^{(\eta)}{}{}^{AA'}_{\nu} k^{\nu} \, .
\end{equation}
This relation between Infeld-van der Waerden symbols exactly ensures \eqref{eqn:associateddyad} is consistent with \eqref{eqn:KSmap} when we identify the flat and curved spinors. For instance
\begin{equation}
\sigma^{(g)}{}{}^{AA'}_{\mu} n^{(g)}{}^{\mu} = \left(\sigma^{(\eta)}{}{}^{AA'}_{\mu} + \frac{\phi}{2} k_{\mu} \sigma^{(\eta)}{}{}^{AA'}_{\nu} k^{\nu}\right)\left(n^{(\eta)}{}^{\mu} - \frac{\phi}{2} k^{(\eta)}{}^{\mu}\right) = \sigma^{(\eta)}{}{}^{AA'}_{\mu} n^{(\eta)}{}^{\mu}  \, .
\end{equation}

Now we have established that the flat space, single copy tangent $u^{(\eta)}{}^{\mu}$ to $\gamma^{(\eta)}$ cannot directly be used to define the relevant two-spinor appearing in \eqref{eqn:dcPenroselimit}. On the other hand, we have shown that we can identify spinors between the flat and curved spacetimes in the dyads \eqref{eqn:associateddyad} on the condition that we shift the Infeld-Van der Waerden symbols. Therefore, we can identify $\alpha^{(\eta)}{}^A$ by $\alpha^{(\eta)}{}^A=\alpha^{(g)}{}^A$ and find that the associated flat space null vector $p^{(\eta)}{}^{\mu}$ is given by
\begin{equation}\label{eqn:pflat}
p^{(\eta)}{}^{\mu}  =  u^{(\eta)}{}^{\mu}+ \frac{\phi}{2} \left(k_{\nu} u^{(\eta)}{}^{\nu}\right) k^{\mu} \, ,
\end{equation}
such that
\begin{equation}
p^{(\eta)}{}^{\mu} \sigma^{(\eta)}{}^{AA'}_{\mu}  = \alpha^{(\eta)}{}^A \alpha^{(\eta)}{}^{A'} = \alpha^{(g)}{}^A \alpha^{(g)}{}^{A'} = u^{(g)}{}^{\mu} \sigma^{(g)}{}^{AA'}_{\mu} \, .
\end{equation}

Moreover, as a consistency check, from \eqref{eqn:pflat} and \eqref{eqn:flatmass} it follows immediately that $\eta_{\mu \nu} p^{(\eta)}{}^{\mu}p^{(\eta)}{}^{\nu} = 0$. In addition, as the notation suggests, $p^{(\eta)}{}^{\mu}$ has a natural physical interpretation as a momentum. On the gravity side, $u^{(g)}{}^{\mu}$ plays the roles of both the kinematical velocity as well as the dynamical momentum. That is no longer true for $u^{(\eta)}{}^{\mu}$ on the gauge theory side. It is thus rather unexpected that a momentum-type object such as $p^{(\eta)}{}^{\mu}$ appears to play an important role in addition to the velocity $u^{(\eta)}{}^{\mu}$. Specifically, it is in terms of this momentum that $\alpha^{(\eta)}{}^A$ can be independently understood for the single copy.

In the above, following \cite{Luna:2018dpt}, our starting point to understand the proposed single copy Penrose limit \eqref{eqn:dcPenroselimit} on a flat background was the Kerr-Schild map of \cite{Monteiro:2014cda}. Specifically, \eqref{eqn:dcPenroselimit} required the single copy of the null geodesic $\gamma^{(g)}$ and its tangent spinor $\alpha^{(g)}{}^A$. On the other hand, a ``geodesic'' double copy was already proposed in \cite{Gonzo:2021drq} by relating conserved charges between the double copy geodesic equations and the single copy Wong equations, which govern a test charge in a Yang-Mills field  (generalizing the Lorenz force law). This ``geodesic'' double copy was further developed in \cite{Ball:2023xnr} with an emphasis on the hidden symmetry.

The mapping of conserved charges proposed in \cite{Gonzo:2021drq} implies in particular that the mass of test-particles is preserved and null trajectories on the gravity side are mapped to null trajectories on the gauge theory side. This contrasts with our ``local'' application of the Kerr-Schild map where \eqref{eqn:flatmass} implied that massless geodesics on the gravity side would not (generically) map to massless worldlines for gauge theories on flat space. The local nature of the Kerr-Schild double copy is a strange feature compared to the traditional double copy on the level of amplitudes, which may not hold beyond special background field configurations \cite{Luna:2022dxo}.  The identification  of $\gamma^{(\eta)}$ with $\gamma^{(g)}$ in Kerr-Schild coordinates may thus be questioned in favor of an identification of conserved charges as in \cite{Gonzo:2021drq,Luna:2022dxo}. However, for the zeroth copy scalar in particular, it is hard to see an alternative for $S^{(\gamma)} = S|_{\gamma}$. 

As we will discuss next, in plane wave spacetimes, this zeroth copy scalar is highly ambiguous within the classical
double copy but, as we will discuss in the examples, up to conserved quantities, the entire functional dependence
of the Penrose limit for type D spacetimes, not just the single copy scalar, is governed by a scalar quantity
$\zeta$ which behaves as $\zeta^{(\gamma)} = \zeta|_{\gamma}$ in the Penrose limit. Therefore, we are led to
believe that within a philosophy of mapping between conserved charges as in \cite{Gonzo:2021drq,Ball:2023xnr},
without a relation $\gamma^{(\eta)} = \gamma^{(g)}$ under a suitable coordinate identification, no single copy
Penrose limit exists compatible with the original background and plane wave classical double copy. We will thus
proceed with the ``local'' identifications of the Kerr-Schild map \cite{Monteiro:2014cda}, where we have argued
\eqref{eqn:dcPenroselimit} is consistent. This identification, using the map between the null frames
\eqref{eqn:frameshift}, is also the approach taken by \cite{Chawla:2023bsu,He:2023iew} to argue
for a double copy interpretation of apparent horizons.  

\subsection{On the type N ambiguity}\label{sec:ambiguity}

The ambiguity in the Weyl double copy for type N spacetimes identified in \cite{Luna:2018dpt}, which amounts to changing the zeroth copy $S^{(\gamma)}$ by an arbitrary function of $u$, obviously applies to the special case of plane waves. On the other hand, if the plane wave is obtained from a Penrose limit of a spacetime where no such ambiguity is present, \eqref{eqn:dcPenroselimit} implies a natural choice. However, this is \emph{not} the ``natural'' choice suggested by \cite{Luna:2018dpt}. If we denote for the single copy of a type N spacetime
\begin{equation}
	f_{AB} = \cA^{(\gamma)}\beta^{(\gamma)}_A \beta^{(\gamma)}_B \, ,
\end{equation}
analogous to \eqref{eqn:NPsi} for the plane wave, whose Weyl spinor we write here as
\begin{equation}
\Psi_{ABCD} = \Psi^{(\gamma)}  \beta^{(\gamma)}_A \beta^{(\gamma)}_B \beta^{(\gamma)}_C \beta^{(\gamma)}_D \, ,
\end{equation}
then  it was argued in \cite{Luna:2018dpt} that we should fix the type N Weyl double copy by
\begin{equation}\label{eqn:oldtypeNscaling}
\Psi^{(\gamma)} = \left(\cA^{(\gamma)}\right)^{3/2} = (S^{(\gamma)})^3 \, .
\end{equation}
We find instead
\begin{equation}\label{eqn:typeNscaling}
\Psi^{(\gamma)} = \left(\cA^{(\gamma)}\right)^{5/3} = (S^{(\gamma)})^5 \, .
\end{equation}
Motivated by the type D Weyl double copy, \cite{Luna:2018dpt} argues \eqref{eqn:oldtypeNscaling} is natural by first rewriting the type D Weyl spinor and the single copy fields in terms of the valence-2 Killing spinor $\chi_{CD}$ 
\begin{equation}\label{eqn:Cvalence2decomposition}
C_{ABCD} = [\chi]^{-5} \chi_{(AB} \chi_{CD)} \, , \quad  f_{AB} = [\chi]^{-3}  \chi_{AB} \, ,\quad [\chi] = S^{-1} = (\chi_{AB}\chi^{AB})^{1/2} \, .
\end{equation}
The logic is then that
\begin{equation}
C_{ABCD} \sim [\chi]^{-3} \sim  (f_{AB})^{3/2} \sim S^3 \, ,
\end{equation}
which are the scalings of \eqref{eqn:oldtypeNscaling}.

On the other hand, from a Penrose limit perspective $\chi_{AB}\alpha^A \alpha^B$ is a conserved quantity, so just a constant evaluated on a null geodesic, while $\chi_{AB}\alpha^A \alpha^B/[\chi]$ is not. Therefore, we believe it is more natural to consider the unnormalized valence-2 Killing spinors as the fundamental invariant spinorial building blocks and letting the $u$-dependent scalings of $\Psi^{(\gamma)}$, $\cA^{(\gamma)}$, and $S^{(\gamma)}$ be determined simply by the prefactors $[\chi]^{-5}$, $[\chi]^{-3}$, $[\chi]^{-1}$ of these Killing spinors in \eqref{eqn:Cvalence2decomposition}, implying the scalings \eqref{eqn:typeNscaling}. \\


Nevertheless, we should concede here that perhaps both \eqref{eqn:oldtypeNscaling} and \eqref{eqn:typeNscaling} are reasonable choices. Indeed, in some sense it is the freedom in parameterization of the null geodesic that seems to give rise to the ambiguity in the plane wave double copy from the Penrose limit perspective. Consider a variant of the adapted coordinates 
\begin{equation}\label{eqn:adaptedmod}
ds^2 = 2 Z(x^{\mu}) dU dV + A(x^{\mu}) dV^2 + B_i(x^{\mu}) dV dX^i - C_{ij}(x^{\mu})dX^i dX^j \, ,
\end{equation}
such that the vector field $\partial_U$ satisfies
\begin{equation}
\nabla_{\partial_U}\partial_U = \frac{d\left(\log Z\right)}{dU} \partial_U \, .
\end{equation}
In words, \eqref{eqn:adaptedmod} represents an adapted coordinate system for a non-affinely parameterized geodesic congruence. Now, going through the same steps to take the Penrose limit as in \cite{penrose1976any} (reviewed below \eqref{eqn:adapted}), we obtain the Penrose limit plane wave in Rosen coordinates with a ``rescaled'' $U$
\begin{equation}\label{eqn:Rosenmod}
ds_{\gamma}^2 = 2 Z(U)dU dV - C_{ij}(U)dX^i dX^j  \, ,
\end{equation}
with $Z(U) = Z(x^{\mu})|_{V=0,X^i=0}$. Going to affine coordinates at this point is a rather trivial operation, which also readily translates between Rosen and Brinkmann coordinates. The Kerr-Schild form of the latter is clearly independent of such changes, hence the ambiguity. 

It is of course natural to choose an affine parameterization, in which case $\chi_{AB} \alpha^A \alpha^B$ is conserved and \eqref{eqn:typeNscaling} is appropriate. However, other parametrizations also appear naturally in the algebraically special types of spacetimes for which the Weyl double copy works. Indeed, the choice where $\chi_{AB} \alpha^A \alpha^B \sim [\chi]$ is closely related to Mino time, which is arguably a better choice than affine time. Specifically, to obtain the metric in the form \eqref{eqn:adaptedmod} with $U$ in Mino time is drastically easier than to try to obtain a metric of the form of the adapted coordinates \eqref{eqn:adapted}.

In light of the previous, it is unsurprising that the above mentioned choices are not the only ones discussed in
the literature. For instance, one option for plane waves suggested in \cite{Godazgar:2020zbv}, who discuss in
detail the freedom in different type N solutions, is to make a choice such that the zeroth copy scalar field is a
constant. From a different point of view, from the twistor constructions of \cite{White:2020sfn, Chacon:2021wbr}
one also seems to land more generically on the choice \eqref{eqn:typeNscaling}, which is what we will use in the remainder. 

However, importantly, in \cite{Godazgar:2020zbv} it has been argued that one can't actually impose choices such as \eqref{eqn:typeNscaling} and \eqref{eqn:oldtypeNscaling} for more general type N spacetimes with a double copy than pp-waves. Nevertheless, as we illustrate in more detail in Section \ref{sec:typeN}, the discussion in \cite{Godazgar:2020zbv} does not account for the full functional freedom we have to rescale $\beta_A^{(\gamma)}$. Indeed, while the Killing spinor in pp-waves implies an obvious choice to (partially) fix that frame ambiguity, for more general type N spacetimes there is an additional freedom to be accounted for. In that case, $\cA^{(\gamma)}$ and $\Psi^{(\gamma)}$ by themselves carry little invariant content.

\section{Examples} \label{sec:examples}

We will consider two types of examples. First we will discuss vacuum (with or without cosmological constant) Petrov type D spacetimes, whose Weyl double copy was discussed in \cite{Luna:2018dpt}. Specifically, we present the construction of Section \ref{sec:dcandPenrose} for Schwarzschild as a simple illustrative example, then we move on to Kerr as the main example of astrophysical interest, and finally we present the general result for the full class of Petrov type D vacuum metrics. Second we turn to some examples of type N spacetimes whose Weyl double copy is discussed in \cite{Godazgar:2020zbv}. Further technical details related to the examples are relegated to Appendix \ref{app:examples}.

\subsection{The Schwarzschild black hole}\label{sec:Schwarzschild}

As a first example, consider the Schwarzschild black hole with line element
\begin{equation}\label{eqn:Smetric}
ds^2 = f(r)dt^2-\frac{dr^2}{f(r)} -r^2 \left(d\theta^2 + \sin^2\theta d \phi^2\right) \, , \quad f(r) = 1-\frac{2M}{r} \, .
\end{equation}
Let
\begin{equation}
d\hat{t} =  dt + \frac{2M}{r}\frac{dr}{f(r)} \, ,
\end{equation}
then the line element takes the Kerr-Schild form \eqref{eqn:KSmetric}
\begin{equation}\label{eqn:SmetricKS}
ds^2 = d\hat{t}^2- dr^2-r^2 \left(d\theta^2 + \sin^2\theta d \phi^2\right) -\frac{2M}{r}(d\hat{t} + dr)^2  \, .
\end{equation}
As in the previous section, we use the principal null frame $\left\lbrace  k^{(g)},  n^{(g)}, m^{(g)}, \bar{m}^{(g)}\right \rbrace$ together with its flat, single copy counterpart $\left\lbrace  k^{(\eta)},  n^{(\eta)}, m^{(\eta)}, \bar{m}^{(\eta)}\right \rbrace$, which for a Schwarzschild black hole can be given explicitly in Kerr-Schild coordinates by
\begin{alignat}{3}
k^{(g)}_{\mu}dx^{\mu} &= k^{(\eta)}_{\mu}dx^{\mu} &=& \quad \frac{1}{\sqrt{2}}\left(d\hat{t} +dr\right)  \, , \nonumber \\
n^{(g)}_{\mu}dx^{\mu} &= n^{(\eta)}_{\mu}dx^{\mu} - \frac{2M}{r}   k^{(\eta)}_{\mu}dx^{\mu}  \quad &=& \quad \frac{1}{\sqrt{2}}\left( d\hat{t} - dr\right)- \frac{\sqrt{2}M}{r}  \left(d\hat{t} +dr\right) \, , \nonumber \\
m^{(g)}_{\mu}dx^{\mu} &= m^{(\eta)}_{\mu}dx^{\mu} &=& \quad \frac{r}{\sqrt{2}} \left(d\theta + i \sin{\theta}d\phi\right) \, , \nonumber \\
\bar{m}^{(g)}_{\mu}dx^{\mu} &= \bar{m}^{(\eta)}_{\mu}dx^{\mu} &=& \quad \frac{r}{\sqrt{2}} \left(d\theta -i \sin{\theta}d\phi\right) \, . \label{eqn:SPNframe}
\end{alignat}

As first shown in \cite{Monteiro:2014cda}, the general discussion of the Kerr-Schild double copy, reviewed in Section \ref{sec:dc}, will thus go through for the Schwarzschild black hole with\footnote{We use here $\phi^{(\rm KS)}$ instead of the $\phi$ from the previous subsection to avoid confusion with the conventional use of $\phi$ as the angular coordinate.}
\begin{equation}\label{eqn:SKSsinglecopy}
k_{\mu} = k^{(g)}_{\mu} = k^{(\eta)}_{\mu} = \frac{1}{\sqrt{2}}\left(d\hat{t} +dr\right) \, , \quad \phi^{(\rm KS)} =  -\frac{4M}{r}  \, .
\end{equation}
In the two-spinor discussion, we will use the dyad $\left\lbrace o^A, \iota^A \right\rbrace$, with complex conjugate $\left\lbrace \bar{o}^{A'}, \bar{\iota}^{A'} \right\rbrace$, associated to the principle null frames defined in \eqref{eqn:SPNframe} as described for the general case in \eqref{eqn:associateddyad}.

A key geometrical observation is that the only (independent\footnote{Naturally, such statements are always up to symmetry and complex conjugation.}) non-trivial curvature projection on the frame \eqref{eqn:SPNframe} is given by
\begin{equation}\label{eqn:SPsi2}
\Psi_2 = -C_{\mu \nu \alpha \beta} k^{(g)}{}^{\mu} m^{(g)}{}^{\nu}\bar{m}^{(g)}{}^{\alpha} n^{(g)}{}^{\beta}= -\frac{M}{r^3} \, .
\end{equation}
The other curvature scalars vanish because the Schwarzschild black hole is of Petrov type D and
\eqref{eqn:SPNframe} is a principal null frame. The general form of a Petrov type D Weyl spinor
\eqref{eqn:spinortypeD} together with \eqref{eqn:SPsi2} implies that this Weyl spinor for a Schwarzschild black
hole is given in terms of the principal dyad $\left\lbrace o^A, \iota^A \right\rbrace$ by\footnote{In the spinor version of \eqref{eqn:SPsi2} there are $4$ contributing contractions out of the $24$ permutations, which results in the overall factor $6$ in \eqref{eqn:SPsiABCDpre}.}
\begin{equation}\label{eqn:SPsiABCDpre}
\Psi_{ABCD} = -6 \frac{M}{r^3} o_{(A} o_B \iota_C \iota_{D)} \, .
\end{equation}

Penrose and Walker showed that vacuum Petrov type D spacetimes admit a solution to the valence-2 Killing spinor equation \cite{Walker:1970un} (Lemma 1)
\begin{equation}\label{eqn:valence2}
	\nabla_{(A}{}^{A'} \chi_{BC)} = 0 \, .
\end{equation}
For the Schwarzschild black hole in particular, such a Killing spinor is given by
\begin{equation}\label{eqn:Schi}
	 \chi_{AB} =  -\frac{\sqrt{2} r}{M^2} o_{(A} \iota_{B)} \, .
\end{equation}
We will find it useful to write the Weyl spinor \eqref{eqn:SPsiABCDpre} in terms of the Killing spinor \eqref{eqn:Schi} as
\begin{equation}\label{eqn:SPsiABCD}
\Psi_{ABCD} = -\frac{3 M^5}{r^5} \chi_{(AB}\chi_{CD)} \, . 
\end{equation}

For the convenience of the reader, we will next summarize some facts about the null geodesics and Penrose limits of this spacetime, with a focus on the perhaps less familiar spinor formulation. On the other hand, the results are well-known \cite{blau2011plane} so we omit several intermediate derivations and the reader familiar with these results can skip to \ref{sec:SdcPenrose}.

\subsubsection{Null geodesics and Penrose limits}

A solution to the null geodesic equation on a Schwarzschild background of fixed energy $\hat{E}$, total angular momentum $L$, and angular momentum $L_{\phi}$ associated to the azimuthal angle $\phi$ is given by
\begin{equation}\label{eqn:Su}
u_{\mu}dx^{\mu} = \hat{E}d\hat{t} - R'(r)dr - \Theta'(\theta)d\theta -L_{\phi} d\phi \, ,
\end{equation}
where $R(r)$ and $\Theta(\theta)$ solve the separated null Hamilton-Jacobi equations
\begin{equation}\label{eqn:uschwarzschild}
\begin{aligned}
-L^2 &= -r^2 \hat{E}^2+r^2 \left(R'(r)\right)^2-2 M r \left(R'(r)+\hat{E}\right)^2 \, , \\
L^2 &= \left(\csc\theta L_{\phi}\right)^2+\left(\Theta'(\theta)\right)^2  \, .
\end{aligned}
\end{equation}
We will take both $\hat{E}$ and $L_{\phi}$ to be positive. Consistency of course requires that $L^2 \geq L_{\phi}^2$. 

The fundamental equations that the tangent spinors $\alpha^A$ and $\bar{\alpha}^{A'}$ to $u^{\mu}$  should satisfy are given in \eqref{eqn:tangentspinor}, which we repeat here for convenience
\begin{equation}\label{eqn:Stangentspinor}
u^{\mu} \sigma_{\mu}^{AA'} = \alpha^A \bar{\alpha}^{A'} \, , \quad u^{\mu} \nabla_{\mu} \alpha^A = 0 \, .
\end{equation}
A solution of \eqref{eqn:Stangentspinor} in terms of the dyad $\left\lbrace o^A , \iota^A \right\rbrace$ compatible with the null frame $\left\lbrace  k^{(g)},  n^{(g)}, m^{(g)}, \bar{m}^{(g)}\right \rbrace$ (as in \eqref{eqn:associateddyad}) is given by
\begin{equation}\label{eqn:alpha}
\begin{aligned}
\alpha^A = & \left(\frac{\Theta'-i \frac{L_{\phi}}{\sin{\theta}}}{\Theta'+i \frac{L_{\phi}}{\sin{\theta}}}\right)^{1/4}\left(\sqrt{\left|\frac{1}{\sqrt{2}}\left(\hat{E} - R'\right)+ \frac{1}{2}\frac{4M}{r}\frac{1}{\sqrt{2}}\left(\hat{E} + R'\right)\right|}o^A  \right. \\  &- \left.\sqrt{\frac{1}{\sqrt{2}}\left|\hat{E} + R'\right|} \left(\Theta'+i \frac{L_{\phi}}{\sin{\theta}}\right)\frac{\iota^A}{L}  \right)\, .
\end{aligned}
\end{equation}
The relation $u^{\mu} \sigma_{\mu}^{AA'} = \alpha^A \bar{\alpha}^{A'}$ is just algebraic and is readily checked (see \eqref{eqn:alphadecomp} and \eqref{eqn:alphainu} for a general solution). On the other hand, this algebraic relation leaves the freedom to rescale with a general spacetime dependent phase $\alpha^A \to e^{i S(x^{\mu})} \alpha^A$, which however is constrained by the differential equation $u^{\mu} \nabla_{\mu} \alpha^A = 0$. We verify in the Appendix \ref{app:examples} explicitly that this differential equation is solved by \eqref{eqn:alpha}. Here  instead, we will simply point out a trick that will also be very useful in more complicated cases. 

The trick to verify that $\alpha^A $ given by \eqref{eqn:alpha} is parallel propagated exploits the existence of the valence-2 Killing spinor and can thus be used for general vacuum type D spacetimes \cite{Walker:1970un}. We simply observe that \eqref{eqn:valence2} and \eqref{eqn:Stangentspinor} imply that $\chi_{AB} \alpha^A \alpha^B$ should be a constant of motion of the null geodesic with tangent $u^{\mu}$. On the other hand, this contraction of the tangent spinor with the Killing spinor is obviously no longer constant when we shift $\alpha^A \to e^{i S(x^{\mu})} \alpha^A$ for a non-constant function $S(x)$. Using \eqref{eqn:alpha} and \eqref{eqn:Schi}, we find
\begin{equation}\label{eqn:Skillingspinorconserved}
\chi_{AB} \alpha^A \alpha^B =  -\frac{r}{\sqrt{2} M^2} 2 o_{(A} \iota_{B)}\alpha^A \alpha^B = - \frac{L}{M^2} \, ,
\end{equation}
which is a constant, as desired for $\alpha^A$ to be parallel propagated. 

Now, using the form of the Weyl spinor \eqref{eqn:SPsiABCD} and the spinor prescription to take the Penrose limit \eqref{eqn:penrosespinor}, we immediately find that the Penrose limit of the Schwarzschild black hole with respect to the null geodesic $\gamma$ generated by \eqref{eqn:Su} is given by
\begin{equation}\label{eqnS:Brinkmannc}
ds_{\gamma}^2 = 2 du dv - \frac{3 M^5}{r(u)^5} \frac{L^2}{M^4}\left(z^2 + \bar{z}^2\right) du^2 - 2 dz d\bar{z}  \, ,
\end{equation}
where $r(u) = r|_{\gamma}$ satisfies the radial geodesic equation. Explicitly, using a solution $R'(r)$ to the (separated) Hamilton-Jacobi equations \eqref{eqn:uschwarzschild} and with $u^{\mu}$ given by \eqref{eqn:Su}, the function $r(u)$ in \eqref{eqnS:Brinkmannc} satisfies
\begin{equation}\label{eqn:Srgeod}
u^r = \frac{dr}{du} = \left(1-\frac{2M}{r(u)}\right)R'(r(u)) - \frac{2M \hat{E}}{r(u)} \, .
\end{equation}
More generally, when the coordinates of the original spacetime (here Schwarzschild \eqref{eqn:SmetricKS}) show up in the Penrose limit metric, we interpret these coordinates as being evaluated along the null geodesic $\gamma$, with respect to which the Penrose limit was taken.

In terms of the real transverse variables \eqref{eqn:complexz}, the metric \eqref{eqnS:Brinkmannc} becomes the familiar Schwarzschild Penrose limit in Brinkmann coordinates \cite{blau2011plane}
\begin{equation}\label{eqnS:Brinkmann}
ds_{\gamma}^2 = 2 du dv - \frac{3 M^5}{r(u)^5} \frac{L^2}{M^4}\left(x_1^2-x_2^2\right) du^2 - dx_1^2 - dx_2^2 \, .
\end{equation}
Before moving on, let us consider other approaches that lead to this same result for the Penrose limit, and discuss their relative merit.

First, let us construct a two-spinor $\beta^A$ to complete $\left\lbrace \alpha^A, \beta^A \right\rbrace$ into a dyad of its own such that 
\begin{equation}\label{eqn:betadef}
\epsilon_{AB} \alpha^A \beta^B = 1 \, , \quad \alpha_B u^{\mu}\nabla_{\mu}\beta^B = 0 \, . 
\end{equation}
In Appendix \ref{app:examples}, we discuss how to construct such a $\beta^A$ from $\alpha^A$, see specifically \eqref{eqn:betadecomp}. Here we simply quote the result:
\begin{equation}\label{eqn:beta}
\begin{aligned}
\beta^A = & \frac{r}{\sqrt{2} L}\left(\frac{\Theta'+i \frac{L_{\phi}}{\sin{\theta}}}{\Theta'-i \frac{L_{\phi}}{\sin{\theta}}}\right)^{1/4} \left(\sqrt{\left|\frac{1}{2}\left(\hat{E} - R'\right)+ \frac{1}{2}\frac{2M}{r} \left(\hat{E} + R'\right)\right|}o^A \right. \\ &+ \left. \sqrt{\left|\hat{E} + R'\right|} \left(\Theta'+i \frac{L_{\phi}}{\sin{\theta}}\right)\frac{\iota^A}{L} \right)\, .
\end{aligned}
\end{equation}
The second equation in \eqref{eqn:betadef} , $\alpha_B u^{\mu}\nabla_{\mu}\beta^B = 0$, is simply a consequence of $\epsilon_{AB} \alpha^A \beta^B = 1$ because $\alpha^A$ is parallel transported along the geodesic, as imposed in \eqref{eqn:Stangentspinor}, and $\epsilon_{AB}$ is compatible with the covariant derivative. 

We can complete the tangent vector field $u^{\mu}$ to a full Newman-Penrose frame $\left\lbrace u^{\mu}, v^{\mu} , m^{\mu} , \bar{m}^{\mu}  \right\rbrace$ associated to the dyad $\left\lbrace \alpha^A, \beta^A \right\rbrace$  (again this means imposing a relation like \eqref{eqn:NPindyad}):
\begin{equation}\label{eqn:Sframec}
\begin{aligned}
u_{\mu}dx^{\mu} &= \hat{E}d\hat{t} - R' dr - \Theta' d\theta -L_{\phi} d\phi \, , \\
v_{\mu}dx^{\mu} &= \frac{r^2}{2 L^2}\left(\hat{E}d\hat{t} - R' dr + \Theta' d\theta +L_{\phi} d\phi\right)\, , \\
m_{\mu}dx^{\mu} &= \frac{r}{\sqrt{2} L}\left\lbrack -\left( R' -\frac{2M}{r}\left(\hat{E} + R'\right)\right) d\hat{t} + \left( \hat{E} +\frac{2M}{r}\left(\hat{E} + R'\right)\right) dr \right. \\ &\quad \quad + \left. i\left(-\frac{L_{\phi}}{\sin \theta}d\theta+ \Theta' \sin{\theta} d\phi \right) \right\rbrack \, , \\
\bar{m}_{\mu}dx^{\mu} &= \frac{r}{\sqrt{2} L}\left\lbrack -\left( R' -\frac{2M}{r}\left(\hat{E} + R'\right)\right) d\hat{t} + \left( \hat{E} +\frac{2M}{r}\left(\hat{E} + R'\right)\right) dr \right. \\ &\quad \quad - \left. i\left(-\frac{L_{\phi}}{\sin \theta}d\theta + \Theta' \sin{\theta} d\phi \right) \right\rbrack \, .
\end{aligned}
\end{equation}
After making the transverse directions into a real orthonormal frame, as in \eqref{eqn:complextransversebasis}, we recover \eqref{eqnS:Brinkmann} from \eqref{eqn:penroseframe}, that is
\begin{equation}\label{eqn:SHframe}
H_{ab}(u) = \left(R_{\mu \nu \alpha \beta} E^{\mu}_a u^{\nu} E^{\alpha}_b u^{\beta} \right)_{|\gamma} = \begin{pmatrix}
\frac{3L^2 M}{r^5} & 0 \\
0 &  -\frac{3L^2 M}{r^5}
\end{pmatrix}\, .
\end{equation}
Now, the transverse directions arise explicitly from the transverse frame
\begin{equation}\label{eqn:Sframe}
\begin{aligned}
E_1{}_{\mu}dx^{\mu} &= \frac{r}{L} \left\lbrack \left( R'(r) -\frac{2M}{r}\left(\hat{E} + R'(r)\right)\right) d\hat{t} - \left( \hat{E} +\frac{2M}{r}\left(\hat{E} + R'(r)\right)\right) dr \right\rbrack \, ,\\
E_2{}_{\mu}dx^{\mu} &= \frac{r}{L}\left(-\frac{L_{\phi}}{\sin{\theta}} d\theta +  \sin\theta \Theta'(\theta)d\phi \right)\, .   \\
\end{aligned}
\end{equation}
As a check of the signs, note the ``unstable'' transverse direction (in the harmonic oscillator equation \eqref{eqn:HOframe}) is associated to the $\hat{t}$, $r$ motion while the ``stable'' transverse dimension is in the angular $\theta$, $\phi$ directions.

Note that the frame $\left\lbrace u^{\mu}, v^{\mu} , m^{\mu} , \bar{m}^{\mu}  \right\rbrace$, given in \eqref{eqn:Sframec}, is in fact not parallel propagated, while \eqref{eqn:penroseframe} as described in Section \ref{sec:Penrosereview} required a parallel propagated null frame. However, it turns out that this did not affect the result \eqref{eqn:SHframe}. To see why, first observe that the frame \eqref{eqn:Sframec} constructed from the dyad $\left\lbrace \alpha^A, \beta^A \right\rbrace$ was not automatically parallel propagated because we did not impose that $\beta^A$ was parallel propagated, instead $\beta^A$ satisfies the weaker condition \eqref{eqn:betadef}. On the other hand, given $\beta^A$  as in \eqref{eqn:beta}, and a solution $I_{\beta}(u)$ to
\begin{equation}\label{eqn:Idef}
u^{\mu}\nabla_{\mu} I_{\beta}(u) = \beta_B u^{\mu}\nabla_{\mu}\beta^B  \, ,
\end{equation}
along the null geodesic $\gamma$, we can readily construct $\tilde{\beta}^B$ satisfying \begin{equation}\label{eqn:betadef2}
\epsilon_{AB} \alpha^A \tilde{\beta}^B = 1 \, , \quad  u^{\mu}\nabla_{\mu}\tilde{\beta}^B = 0 \, . 
\end{equation}
 by
\begin{equation}\label{eqn:ppdyad}
\tilde{\beta}^B =  \beta^B +  I_{\beta}(u) \alpha^B \, .
\end{equation}
Then $\lbrace \alpha^A,\tilde{\beta}^B\rbrace$, as well as the associate Newman-Penrose frame, are parallel transported along $\gamma$. On the other hand, the shift that $\beta^A \to \tilde{\beta}^A$ induces into the frame \eqref{eqn:Sframec}, specifically a shift of the transverse basis proportional to $u^{\mu}$, does not contribute to \eqref{eqn:penroseframe} by the antisymmetry of the Riemann tensor. Therefore, while it is formally important to have a parallel propagated dyad to take the Penrose limit using the associated parallel propagated frame with \eqref{eqn:penroseframe}, in practice the result is the same here. As is clear from the spinorial approach \eqref{eqn:penrosespinor}, the Penrose limit in the end does not depend on the choice of $\beta^B$. 

Finally, in order to connect to Penrose's original construction, consider the following change of coordinates  from the Kerr-Schild  $\lbrace \hat{t},r,\phi,\theta \rbrace$  to $\lbrace U,V,\hat{t}_0,\hat{\phi}_0 \rbrace$: \footnote{See Appendix \ref{app:examples} for a derivation.}
\begin{equation}\label{eqn:Schwarzschildadapted}
\begin{aligned}
U &= \int^{r(U)}_{r_{\rm ref}} \frac{d\rho}{R'(\rho) -\frac{2M}{\rho} \left(\hat{E}+R'(\rho)\right)}   \, , \\ 
\hat{t}(U,\hat{t}_0) &= \hat{t}_0 + \int^{r(U)}_{r_{\rm ref}} d\rho \, \frac{\hat{E}+\frac{2M}{\rho} \left(\hat{E}+ R'(\rho)\right)}{R'(\rho) -\frac{2M}{\rho} \left(\hat{E}+R'(\rho)\right)} \, , \\
\theta_0(V,\hat{t}_0) &= \frac{1}{L}\left(\hat{E}\hat{t}_0  - V\right) \, , \\
\theta(U,V, \hat{t}_0) &= \theta_0(V,\hat{t}_0)+ L\int^{r(U)}_{r_{\rm ref}} d\rho \, \frac{1}{\rho^2}\frac{1}{R'(\rho) -\frac{2M}{\rho} \left(\hat{E}+R'(\rho)\right)} \, ,\\
\phi(U,V,\phi_0,\hat{t}_0) &= \phi_0  \, .
\end{aligned}
\end{equation}
Here, $r_{\rm ref}$ is an arbitrarily chosen constant while $r(U) = r(U;\hat{E},L^2)$ is a radial solution to the geodesic equations on a Schwarzschild spacetime, which in particular satisfies \eqref{eqn:Srgeod} in terms of an affine parameter $U$ with initial condition $r(0) = r_{\rm ref}$. In deriving \eqref{eqn:Schwarzschildadapted}, we have used the spherical symmetry to take $L_{\phi} = 0$ without loss of generality. 

The Schwarzschild metric in the coordinates $\lbrace U,V,\hat{t}_0,\hat{\phi}_0 \rbrace$, defined in \eqref{eqn:Schwarzschildadapted}, takes the form
\begin{equation}\label{eqn:Sadapted}
\begin{aligned}
ds^2 =  
2 dU dV&- \left(r(U)^2 \frac{\hat{E}^2}{L^2}-\left(1-\frac{2M}{r(U)}\right)\right)d\hat{t}_0^2 - r(U)^2 \sin^2\left\lbrack\theta(U,V,\hat{t}_0)\right\rbrack d \phi_0^2  \\ &-\frac{r(U)^2}{L^2} dV^2+2\hat{E} \frac{r(U)^2}{L^2} dV d\hat{t}_0 \, .\\
\end{aligned}
\end{equation}
We should stress that this metric is still the exact Schwarzschild black hole metric; we have only made a change of coordinates. No approximations have been made so far, although the new adapted coordinates may not chart as much of the spacetime as the Kerr-Schild coordinates\footnote{As a consistency check, we have verified for instance that \eqref{eqn:Sadapted} is still a solution to the vacuum Einstein equations. In addition, we have computed several scalars such as $R_{\mu \nu \alpha \beta}R^{\mu \nu \alpha \beta} = \frac{48 M^2}{r(U)^6}$ etc.}.

Our motivation to introduce the coordinates $\lbrace U,V,\hat{t}_0,\hat{\phi}_0\rbrace$ was to put the Schwarzschild metric in the adapted coordinate form \eqref{eqn:adapted} and the line element \eqref{eqn:Sadapted} therefore justifies the choice \eqref{eqn:Schwarzschildadapted}. Nevertheless, let us briefly interpret the definitions in \eqref{eqn:Schwarzschildadapted} to clarify their origin. First, the coordinates $\lbrace U,\hat{t}_0,\theta_0,\hat{\phi}_0 \rbrace$ were constructed such that at $U = 0$, $\lbrace\hat{t}_0,\theta_0,\hat{\phi}_0\rbrace$ simply parameterizes the $r = r_{\rm ref}$ hypersurface in the Kerr-Schild coordinates $\lbrace\hat{t},\theta,\hat{\phi} \rbrace$. Now, at non-zero $U$, the point $p =( U,\hat{t}_0,\theta_0,\hat{\phi}_0)$ is such that a null geodesic with energy $\hat{E}$ and total angular momentum $L$ ($L_{\phi} = 0$) reaches $p$ from the point $p_0 = ( \hat{t}_0 , r_{\rm ref},\theta_0,\hat{\phi}_0 )$ in Kerr-Schild coordinates in affine time $U$. $\lbrace\hat{t}_0,\theta_0,\hat{\phi}_0 \rbrace$ thus parameterize a null geodesic congruence. The only remaining step to get to $\lbrace U,V,\hat{t}_0,\hat{\phi}_0 \rbrace$ is to appreciate that we can use the action of the null geodesics as one of the coordinates in order to obtain the form \eqref{eqn:adapted} \cite{Patricot:2003dh, blau2011plane}. For the chosen null geodesic congruence, the action is simply $V = \hat{E}\hat{t}_0 - L \theta_0$ and, as an example, we have thus chosen to replace $\theta_0$ by $L \theta_0= \hat{E}\hat{t}_0  - V$.  

From the Schwarzschild metric  in adapted coordinates \eqref{eqn:Sadapted}, we find the Penrose limit in Rosen coordinates
\begin{equation}\label{eqn:Sadaptedlimit}
\begin{aligned}
ds^2_{\gamma} =  
2 dU dV&- \left(r(U)^2 \frac{\hat{E}^2}{L^2}-\left(1-\frac{2M}{r(U)}\right)\right)d\hat{t}_0^2 - r(U)^2 \sin^2\theta(U) d \phi_0^2  \, ,
\end{aligned}
\end{equation}
where $\theta(U) = \theta(U,V,\hat{t}_0)|_{V=\hat{t}_0=0}$ is a solution to the geodesic equation which, analogously to the radial equation \eqref{eqn:Srgeod} and using \eqref{eqn:Su} and \eqref{eqn:uschwarzschild} (with $L_{\phi} = 0$), can be expressed as
\begin{equation}
u^{\theta} = \frac{d\theta}{dU} = \frac{L}{r(U)^2} \, .
\end{equation}

For more general black holes, the approach to the Penrose limit through adapted coordinates is a lot more complicated. Therefore, in this simple setting, we would like to illustrate why it is nevertheless useful. The key point is that we now have a version of \eqref{eqnS:Brinkmann} in Rosen coordinates. Therefore, by constructing a frame $E^{(\gamma)}{}^i_{a}$, as in \eqref{eqn:frame} and \eqref{eqn:Esymm}, such that
\begin{equation}
C_{ij} = E^{(\gamma)}{}_i{}^{a} \delta_{ab} E^{(\gamma)}{}_i{}^{b} =\begin{pmatrix}
\sqrt{r(U)^2 \frac{\hat{E}^2}{L^2}-\left(1-\frac{2M}{r(U)}\right)} &  0 \\
0 & r(U) \sin \theta(U)  
\end{pmatrix}^2 \, ,
\end{equation} 
by \eqref{eqn:HOframe}, we have that $E^{(\gamma)}{}^1_{1}$ and $E^{(\gamma)}{}^2_{2}$ respectively solve 
\begin{equation}\label{eqn:HOS}
\frac{d^2f_1}{dU^2} = \frac{3 M L^2}{\left(r(U)\right)^5}f_1 \, ,  \quad  	\frac{d^2f_2}{dU^2} = -\frac{3 M L^2}{\left(r(U)\right)^5}f_2\, .
\end{equation}
Such solutions are an essential input to any physical process on the plane wave background, determining the Killing vectors and solutions to the wave equations. They would have been hard to find in such an elegant form from directly solving \eqref{eqn:HOS}.

\subsubsection{The double copy}\label{sec:SdcPenrose}

We first recall explicitly the double copy constructions for the Schwarzschild black hole \eqref{eqn:Smetric} and the plane wave \eqref{eqnS:Brinkmann} independently of each other to then show that the spinor Penrose limit proposal \eqref{eqn:dcPenroselimit} holds.

First, for the black hole, from the Weyl spinor \eqref{eqn:SPsiABCD}
\begin{equation}\label{eqn:Spsidc}
\Psi_{ABCD} = -\frac{3 M^5}{r^5} \chi_{(AB}\chi_{CD)} \, ,
\end{equation}
it follows that the Weyl single copy and zeroth copy of the Schwarzschild black hole are 
\begin{equation}\label{eqn:Ssingle}
f_{AB} =   -\frac{2 M^3}{r^3} \chi_{AB} \, , \quad S = -\frac{2M}{r} \, . 
\end{equation}
We have slightly deviated from \eqref{WeylDCRelation} here in favor of keeping $S$, which is only defined up to a constant, free of combinatorial factors. As shown in \cite{Luna:2018dpt}, the above Weyl double copy is consistent with the Kerr-Schild approach. Explicitly, the Kerr-Schild single copy can be derived from \eqref{eqn:SKSsinglecopy} to be
\begin{equation}\label{eqn:SKSsingle}
A =  -\frac{2 \sqrt{2} M}{r}\left(d\hat{t} +dr\right) \, , \quad F =   \frac{2 \sqrt{2} M}{r^2}    dr \wedge d\hat{t} \, .
\end{equation} 
The Maxwell spinor \eqref{eqn:Ssingle} can then be found from \eqref{eqn:SPsiABCD} and
\begin{equation}
\frac{F_{\mu \nu}}{2} \left(k^{(\eta)}{}^{\mu} n^{(\eta)}{}^{\nu} - m^{(\eta)}{}^{\mu}\bar{m}^{(\eta)}{}^{\nu}\right) =  -\frac{\sqrt{2} M}{r^2} \, , \quad	0 = F_{\mu \nu} k^{(\eta)}{}^{\mu} m^{(\eta)}{}^{\nu} =  F_{\mu \nu} n^{(\eta)}{}^{\mu} \bar{m}^{(\eta)}{}^{\nu} \, ,
\end{equation}
while the relation between the Kerr-Schild and Weyl zeroth copy is given by
\begin{equation}
S + S^* = \phi^{(\rm KS)} \, ,
\end{equation}
as suggested in \cite{Luna:2018dpt}.

For the plane wave \eqref{eqnS:Brinkmannc}, it follows from \eqref{eqn:NPsi} that the Weyl spinor is given by
\begin{equation}\label{eqn:Sgammapsidc}
\Psi^{(\gamma)}_{ABCD} = -\frac{3 M^5}{r(u)^5}  \frac{L^2}{M^4}  \beta^{(\gamma)}_A \beta^{(\gamma)}_B  \beta^{(\gamma)}_C  \beta^{(\gamma)}_D   \, .
\end{equation}
As emphasized in the discussion around \eqref{eqnS:Brinkmannc}, for the plane wave, the function $r(u) = r|_\gamma$ here satisfies the geodesic equation \eqref{eqn:Srgeod} of the reference null geodesic. The Weyl double copy with our choice  \eqref{eqn:typeNscaling} to fix the (functional) type N ambiguity then implies\footnote{In \eqref{eqn:Sppscweyl}, we are treating the numerical prefactors in the same way as in the black hole case.}
\begin{equation}\label{eqn:Sppscweyl}
f^{(\gamma)}_{AB} =   \frac{M^3}{r(u)^3}  \frac{L}{M^2} 2 \beta^{(\gamma)}_A \beta^{(\gamma)}_B  \, , \quad S^{(\gamma)} = -\frac{2M}{r(u)} \, . 
\end{equation}

The connection to the Kerr-Schild construction is more subtle than in the black hole case. Following the original Kerr-Schild double copy proposal of \cite{Monteiro:2014cda} yields
\begin{equation}\label{eqn:KSplanewaveoriginal}
\phi_{(\gamma)}^{(\rm KS)} = -\frac{3 M^5}{ r(u)^5}  \frac{L^2}{M^4} (x_1^2 - x_2^2) \, , \quad A_{(\gamma)}^{(\rm KS)} = 	\phi_{(\gamma)}^{(\rm KS)}  du \, .
\end{equation}
However, this construction of the single copy for the plane wave is not consistent with \eqref{eqn:Sppscweyl}. As discussed in \cite{Ilderton:2018lsf, Luna:2018dpt}, the identification \eqref{eqn:KSplanewaveoriginal} will not relate a gravitational plane wave with a gauge theory plane wave\footnote{Rather, \eqref{eqn:KSplanewaveoriginal} relates a gravitational plane wave to the gauge theory ``vortex'' solutions of \cite{Bialynicki-Birula:2004bvr}.}. Following \cite{Luna:2018dpt} and \cite{Adamo:2017nia}, we will take another approach which does impose that a gravitational plane wave is related by the double copy to a gauge theory plane wave. If we additionally fix the remaining type N ambiguity with \eqref{eqn:typeNscaling}, we find
\begin{equation}\label{eqn:SKS0matchingWeyl}
\phi_{(\gamma)}^{(\rm KS)} = -\frac{4M}{r(u)} \, ,
\end{equation}
and the (Kerr-Schild) single copy becomes
\begin{equation}\label{eqn:SKS1matchingWeyl}
A^{(\gamma)} = -\frac{2 M^3}{r(u)^3}  \frac{L}{M^2} (z + \bar{z}) du \, , \quad  F^{(\gamma)} = -  \frac{2 M^3}{r(u)^3}  \frac{L}{M^2}  ( dz + d\bar{z}) \wedge du \, .
\end{equation} 
From the projections on the null frame given by\footnote{Observing that the Brinkmann coordinates are in Kerr-Schild form, the frame in \eqref{eqn:SKS1matchingWeyl} is the flat, single copy map of the gravitational plane wave frame defined in \eqref{eqn:udefplane}, \eqref{eqn:vdefplanewave}, and \eqref{eqn:mdefplanewave}. We avoid adding an additional ${}_{(\eta)}$ for obvious reasons.}
\begin{equation}
k^{(\gamma)}_{\mu} dx^{\mu} =  du  \, , \quad n^{(\gamma)}_{\mu} dx^{\mu} = dv  \, , \quad m^{(\gamma)}_{\mu} dx^{\mu} = dz \, ,
\end{equation}
we can again compute
\begin{equation}
F_{\mu \nu} n^{\mu}_{(\gamma)} \bar{m}^{\nu}_{(\gamma)} =   -\frac{2  M^3}{r(u)^3}  \frac{L}{M^2}  \, , \quad	0 = F_{\mu \nu} k^{\mu}_{(\gamma)} m^{\nu}_{(\gamma)} = \frac{F_{\mu \nu}}{2} \left(k_{(\gamma)}^{\mu} n_{(\gamma)}^{\nu} - m_{(\gamma)}^{\mu}\bar{m}_{(\gamma)}^{\nu}\right)   \, ,
\end{equation}
which, when translated into spinor components, matches \eqref{eqn:Sppscweyl}. Therefore, \eqref{eqn:SKS0matchingWeyl} and \eqref{eqn:SKS1matchingWeyl} indeed provide the Kerr-Schild single and zeroth copy of the plane wave consistent with the Weyl double copy, as opposed to \eqref{eqn:KSplanewaveoriginal} (the original Kerr-Schild double copy proposal \cite{Monteiro:2014cda}).

Note that one of the main reasons for us to connect back to the Kerr-Schild double copy of \cite{Monteiro:2014cda}, was to identify the flat space map such that the single copy equations of motion are satisfied. For the plane waves, we don't follow  \cite{Monteiro:2014cda} but these equations of motion are still satisfied with respect to the flat space metric in Brinkmann coordinates. 

If we now apply the proposal \eqref{eqn:dcPenroselimit} for the Penrose limit of the single and zeroth copy to the Schwarzschild fields \eqref{eqn:Ssingle} (using also \eqref{eqn:Skillingspinorconserved}), we find exactly the plane wave result \eqref{eqn:Sppscweyl} as desired. The spinorial approach to the Penrose limit in conjunction with the Weyl double copy makes this procedure, summarized in Fig. \ref{fig:schematic}, look straightforward. On the other hand, from the perspective of the ``original'' Kerr-Schild double copy for the black hole and the ``original'' Penrose limit through adapted coordinates, the relation between the double copy of the black hole and that of the plane wave would not have been obvious at all.

\begin{figure}[h]
	\begin{tikzcd}
		\makebox[\lenff em][c]{$\Psi_{ABCD} = S^5 \chi_{(AB} \chi_{CD)}$} \arrow[r,leftrightarrow,shorten=-3mm, "\text{Weyl DC}"] \arrow[dd,"\text{Penrose limit}{}_{\gamma}"]& \makebox[\lenf em][c]{$f_{AB} = S^3 \chi_{AB}$} \arrow[dd,"\text{Penrose limit}_{\gamma}"] \\
		& \\
			\makebox[\lenff em][c]{$\Psi^{(\gamma)}_{ABCD} = \left(S^5 |_{\gamma}\right) \beta^{(\gamma)}_A \beta^{(\gamma)}_B \beta^{(\gamma)}_C \beta^{(\gamma)}_D$} \arrow[r,leftrightarrow,shorten=-3mm, "\text{Weyl DC}"] & \makebox[\lenf em][c]{$f^{(\gamma)}_{AB} = \left(S^3 |_{\gamma}\right) \beta^{(\gamma)}_A \beta^{(\gamma)}_B$}
	\end{tikzcd}
	\caption{Schematic summary of the interplay between the Weyl double copy and the Penrose limit. The functional spacetime dependence is in some sense entirely controlled by the zeroth copy $S = -\frac{2M}{r}$ (for a Schwarzschild black hole) and its evaluation $S|_{\gamma}$ on the null geodesic $\gamma$, with respect to which the Penrose limit is taken. The spinorial structure on the other hand is fixed by the valence-2 Killing spinor $\chi_{AB}$ for the Petrov type D black hole while on the Petrov type N plane wave it is fixed by the Killing spinor $\beta^{(\gamma)}_A$.} \label{fig:schematic}
\end{figure}

\subsection{The Kerr black hole}\label{sec:Kerr}

We now add angular momentum to the black hole from the previous example and study the Kerr black hole. The
technical details are more algebraically involved for rotating black holes but these complications also highlight
the added value of the two-spinor approach. As for the Schwarzschild black hole, we will want to work with
Kerr-Schild coordinates for our discussion of the double copy. We nevertheless use the line element in
Boyer-Lindquist coordinates as our starting point, because these coordinates are ubiquitous in applications, but it is safe to skip to  \eqref{eqn:KKSmetric} (skimming the definitions in \eqref{eqn:Sigma}).

In Boyer-Lindquist coordinates $\left\lbrace t, r, \theta, \phi \right\rbrace$, the Kerr metric for a black hole of mass $M$ and angular momentum $J = a M$ is given by \cite{boyer1967maximal}
\begin{equation}
\begin{aligned}
ds^2 &=  \frac{\Delta_r}{\Sigma} \left(dt -  a \sin^2{\theta} d \phi \right)^2 - \frac{\Sigma}{\Delta_r} dr^2 - \Sigma d \theta^2 - \frac{\sin^2\theta}{\Sigma} \left\lbrack \left(r^2 + a^2\right) d \phi - a dt \right\rbrack^2 \, , 
\end{aligned}
\end{equation}
where
\begin{equation}
\\
\Delta_r = r^2 - 2 M r + a^2 \, , \qquad  \qquad \Sigma = r^2 + a^2 \cos^2{\theta} \, .
\end{equation}
As an intermediate step to connect the Boyer-Lindquist coordinates to the Kerr-Schild coordinates, we use the canonical coordinates $\left\lbrace \tau, r, y, \psi \right\rbrace$ reviewed in detail in \cite{Frolov:2017kze}. The relation between the canonical coordinates and Boyer-Lindquist coordinates is given by
\begin{equation}\label{eqn:canonicalcoord}
\tau = t - a \phi  \, , \quad y = a \cos{\theta} \, , \quad \psi = \phi/a \, .
\end{equation}
In canonical coordinates, the metric takes the form
\begin{equation}
ds^2 = \frac{\Delta_r}{\Sigma}\left(d\tau + y^2 d\psi \right)^2 - \frac{\Delta_y}{\Sigma}\left(d\tau - r^2 d\psi \right)^2 - \frac{\Sigma}{\Delta_r} dr^2 - \frac{\Sigma}{\Delta_y} dy^2 \, ,  
\end{equation}
with
\begin{equation}\label{eqn:Sigma}
\begin{aligned}
\Sigma = r^2 + y^2  \, , \quad \Delta_y = -y^2+a^2 \, . 
\end{aligned}
\end{equation}
Finally, the Kerr-Schild coordinates $\lbrace \hat{\tau}, r, y, \hat{\psi} \rbrace$ and the canonical coordinates are related by \cite{Frolov:2017kze} 
\begin{equation}\label{eqn:KScoord}
d\hat{\tau} = d\tau + \frac{r^2}{\Delta_r} dr - \frac{y^2}{\Delta_y} dy \, , \quad d\hat{\psi} = d \psi + \frac{dr}{\Delta_r} + \frac{dy}{\Delta_y} \, .
\end{equation}
The metric in Kerr-Schild coordinates is given by
\begin{equation}\label{eqn:KKSmetric}
\begin{aligned}
ds^2 =  ds^2_{0} - \frac{2 M r}{\Sigma} \left(d\hat{\tau} + y^2 d\hat{\psi} \right)^2 \, ,
\end{aligned}
\end{equation}
where the flat space line element $ds_{0}$ is given by
\begin{equation}
\frac{\Sigma ds^2_{0}}{r^2+a^2}	= \left(d\hat{\tau} + y^2 d\hat{\psi} \right)^2 - \frac{2 \Sigma dr}{r^2+a^2}  \left(d\hat{\tau} + y^2 d\hat{\psi} \right) - \left(d\hat{\tau}-r^2 d\hat{\psi} + \frac{\Sigma dy}{a^2-y^2}  \right)^2 -  \frac{\Sigma^2 dy^2}{\left(a^2-y^2\right)^2} \, .
\end{equation}
The basic objects to construct the Kerr-Schild double copy for the Kerr black hole are thus
\begin{equation}
k_{\mu} = d\hat{\tau} + y^2 d\hat{\psi} \, , \quad \phi^{(\rm KS)} =  - \frac{2 M r}{\Sigma}  \, .
\end{equation}
A full Newman-Penrose principal null frame, supplementing $k_{\mu} = k^{(g)}_{\mu} = k^{(\eta)}_{\mu}$ as in Section \ref{sec:dcandPenrose} and the Schwarzschild example,  is given by
\begin{alignat}{3}
k^{(g)}_{\mu}dx^{\mu} &= k^{(\eta)}_{\mu}dx^{\mu} &=& \quad  d\hat{\tau} + y^2 d\hat{\psi}\, , \nonumber \\
n^{(g)}_{\mu}dx^{\mu} &= n^{(\eta)}_{\mu}dx^{\mu} -\frac{M r}{\Sigma} k^{(\eta)}_{\mu}dx^{\mu} \quad  &=& \quad  -dr+\frac{r^2+a^2}{2 \Sigma}\left(d\hat{\tau} + y^2 d\hat{\psi}\right)  - \frac{M r}{\Sigma} \left(d\hat{\tau} + y^2 d\hat{\psi}\right) \, , \nonumber \\
m^{(g)}_{\mu}dx^{\mu} &= m^{(\eta)}_{\mu}dx^{\mu} &=& \quad  \sqrt{\frac{\Sigma}{2\Delta_y}}\left(dy+i \frac{\Delta_y}{\Sigma} \left(r^2 d\hat{\psi}-d\hat{\tau}-\frac{\Sigma}{\Delta_y}dy\right)\right)\, ,  \nonumber \\
\bar{m}^{(g)}_{\mu}dx^{\mu} &=  \bar{m}^{(\eta)}_{\mu}dx^{\mu} &=&  \quad  \sqrt{\frac{\Sigma}{2\Delta_y}}\left(dy-i \frac{\Delta_y}{\Sigma} \left(r^2 d\hat{\psi}-d\hat{\tau}-\frac{\Sigma}{\Delta_y}dy\right)\right) \, . \label{eqn:Kerrframe}
\end{alignat}
We will also again introduce the dyad $\left\lbrace o^A, \iota^A \right\rbrace$, with complex conjugate $\left\lbrace \bar{o}^{A'}, \bar{\iota}^{A'} \right\rbrace$, associated to Newman-Penrose frame as in \eqref{eqn:associateddyad}.

The Kerr black hole is a vacuum Petrov type D spacetime. Thus the only non-trivial curvature projection withe respect to the principal Newman-Penrose frame \eqref{eqn:Kerrframe} is
\begin{equation}\label{eqn:KPsi2}
\Psi_2 = -C_{\mu \nu \alpha \beta} k^{(g)}{}^{\mu} m^{(g)}{}^{\nu}\bar{m}^{(g)}{}^{\alpha} n^{(g)}{}^{\beta}= -\frac{M}{\zeta^3} \, \, , \quad \zeta = r - i y \, .
\end{equation}
The Weyl spinor is thus given by
\begin{equation}
\Psi_{ABCD} = -\frac{6 M}{\zeta^3} o_{(A} o_B \iota_C \iota_{D)} \, ,
\end{equation}
which can be expressed in terms of a valence-2 Killing spinor $\chi_{AB}$, satisfying \eqref{eqn:valence2}, as
\begin{equation}\label{eqn:KPsiABCD}
\Psi_{ABCD} = -\frac{3 M^5}{\zeta^5}\chi_{(AB}\chi_{CD)} \, , \quad \chi_{AB} =  -\frac{\sqrt{2}\zeta}{M^2} o_{(A} \iota_{B)}  \, .
\end{equation}
These expressions for the Weyl spinors are identical to those of the Schwarzschild black hole with $r \to \zeta$
and this relation underlies the so-called Newman-Janis trick \cite{Newman:1965tw}\footnote{For another, but
related, point of view on the origin of the Newman-Janis trick see \cite{Arkani-Hamed:2019ymq,Guevara:2020xjx}.}. 

We now continue to discuss relevant aspects of null geodesics on Kerr black holes and their Penrose limits. While most of the discussion is a review of well-known results, to the best of our knowledge, the two-spinor approach to the Penrose limit for a Kerr black hole has not been worked out explicitly and, while implicitly or formally known, the simple final expressions \eqref{eqn:KBrinkmannc} and \eqref{eqn:KBrinkmann} for Penrose limits of \emph{generic} null geodesics seem to not be readily available in the literature.

\subsubsection{Null geodesics and Penrose limits}

As for the Schwarzschild black hole, there are many excellent references on the (null) geodesics of  Kerr spacetimes \cite{Carter:1968rr,Bardeen:1973tla,Chandrasekhar:1985kt,Gralla:2019ceu,Compere:2021bkk,Cieslik:2023qdc} and we shall only review what we need. References on Penrose limits with respect to arbitrary null geodesics are somewhat more scarce but see \cite{Igata:2019pgb,Fransen:2023eqj,Hollowood:2009qz} for photon sphere and equatorial orbits. See also \cite{Papadopoulos:2020qik} for the construction of adapted coordinates and \cite{Kubiznak:2008zs} for the construction of a parallel propagated frame. \cite{Papadopoulos:2020qik} formally presents the Penrose limit of Kerr black holes for arbitrary geodesics in Rosen coordinates, while the form in Brinkmann coordinates can be more directly constructed from \cite{Kubiznak:2008zs} and \eqref{eqn:penroseframe}. 

As illustrated for the Schwarzschild example, the combination of the Brinkmann and Rosen form of the Penrose limit plane wave is equivalent to finding and solving the geodesic deviation equations. This problem of the geodesic deviation equations in Kerr has also been directly discussed in \cite{Cariglia:2018erv}. As a result, implicit in \cite{Cariglia:2018erv} are again both the Brinkmann and Rosen forms of the Penrose limit plane waves. Nevertheless, using the two-spinor approach we find a particularly simple form of the Penrose limit. Moreover, we can then directly build a parallel propagated Newman-Penrose frame from a dyad, as a complementary approach to \cite{Kubiznak:2008zs}.

The tangent to the null geodesic of fixed energy $\hat{E}$, separation constant $K$ and angular momentum $L_{\psi}$ along $\hat{\psi}$ is given by
\begin{equation}\label{eqn:ukerr}
u_{\mu}dx^{\mu} = \hat{E}d\hat{\tau} - R'(r)dr - Y'(y)dy -L_{\psi} d\hat{\psi} \, ,
\end{equation}
where $R(r)$ and $Y(y)$ solve the separated null Hamilton-Jacobi equations
\begin{equation}\label{eqn:uKerr}
\begin{aligned}
-K^2 &= 2 \left(L_{\psi}-\hat{E} r^2 \right) \left(R'\left(r\right)\right)+\Delta_r \left(R'\left(r\right)\right)^2 \, , \\
K^2 &= 2(L_{\psi}+\hat{E}y^2)Y'\left(y\right)+\Delta_y \left(Y'\left(y\right)\right)^2 + 2\frac{\left(L_{\psi}+\hat{E}y^2 \right)^2}{a^2-y^2} \, .
\end{aligned}
\end{equation}
For later reference, in terms of the affine coordinate $u$, the $r$ and $y$ components of the motion satisfy
\begin{equation}\label{eqn:Kgeod}
\begin{aligned}
\left(r^2+y^2\right)\frac{dr}{du} &=  \left(L_{\psi}-\hat{E} r^2 \right) + \Delta_r R'(r) \, , \\
\left(r^2+y^2\right)\frac{dy}{du} &=  \left(L_{\psi}+\hat{E} y^2 \right) + \Delta_y Y'(y) \, .
\end{aligned}
\end{equation}

A two-spinor formulation of $u^{\mu}$ in terms of the dyad $\left\lbrace o^A , \iota^A \right\rbrace$ compatible with the null frame $\lbrace k^{(g)},n^{(g)},m^{(g)},\bar{m}^{(g)} \rbrace$ is given by
\begin{equation}\label{eqn:Kalpha}
\begin{aligned}
\alpha^A = & \left(\frac{\bar{\zeta}}{\zeta} \, \frac{\sqrt{2}(L_{\psi}+\hat{E}y^2)e^{-i\pi/4}+\Delta_y Y'}{\sqrt{2}(L_{\psi}+\hat{E}y^2)e^{i\pi/4}+\Delta_y Y'}\right)^{1/4}\left(\sqrt{\left|\frac{1}{2\Sigma }\left(2L_{\psi}-2\hat{E}r^2+\Delta_r R'\right)\right|}o^A  \right. \\  &- \left.\sqrt{\left|R'\right|} \left(\sqrt{2}(L_{\psi}+\hat{E}y^2)e^{i\pi/4}+\Delta_y Y'\right)\frac{\iota^A}{K \sqrt{\Delta_y}}  \right)\, .
\end{aligned}
\end{equation}
In addition to squaring to the tangent $u^{\mu} \sigma_{\mu}^{AA'} = \alpha^A \bar{\alpha}^{A'}$, this spinor and its conjugate satisfy the parallel transport equations, as before \eqref{eqn:tangentspinor}
\begin{equation}
u^{\mu} \nabla_{\mu} \alpha^A = 0 \, \quad  u^{\mu} \nabla_{\mu} \bar{\alpha}^A = 0 .
\end{equation}
As a consequence of this parallel transport, and the valence-2 Killing spinor equation, we again have the constant of motion
\begin{equation}\label{eqn:Kkillingspinorconserved}
\chi_{AB} \alpha^A \alpha^B =  -\frac{\zeta}{\sqrt{2} M^2} 2 o_{(A} \iota_{B)}\alpha^A \alpha^B = - \frac{K}{M^2} \, .
\end{equation}
It is in fact easiest to derive \eqref{eqn:Kalpha} by requiring that \eqref{eqn:Kkillingspinorconserved} should be constant, to avoid solving \eqref{eqn:tangentspinor}. We have nevertheless checked explicitly that \eqref{eqn:tangentspinor} is satisfied but leave the details to Appendix \ref{app:Kerr}.

From the form of the Weyl spinor \eqref{eqn:KPsiABCD} and the spinor prescription to take the Penrose limit \eqref{eqn:penrosespinor}, we find
\begin{equation}\label{eqn:KBrinkmannc}
ds_{\gamma}^2 = 2 du dv - \frac{3 K^2}{M^4}\left(M^5\zeta^{-5} z^2 + M^5 \bar{\zeta}^{-5}\bar{z}^2\right) du^2 - 2 dz d\bar{z}  \, ,
\end{equation}
with $\zeta(u) = \zeta|_{\gamma} = r(u)-i y(u)$, and where $r(u)$, $y(u)$ satisfy the geodesic equations \eqref{eqn:Kgeod}. In terms of real transverse coordinates, as in \eqref{eqn:complexz}, the metric \eqref{eqn:KBrinkmannc} becomes 
\begin{equation}\label{eqn:KBrinkmann}
\begin{aligned}
ds_{\gamma}^2 &= 2 du dv  - dx_1^2 - dx_2^2  \\
&- \frac{3 K^2 M}{\Sigma^5}\left\lbrack \left(r^5-10r^3 y^2+5ry^4\right)\left(x_1^2- x_2^2\right) -2  \left(5r^4 y-10r^2 y^3 + y^5 \right)x_1 x_2 \right\rbrack du^2 \, .
\end{aligned}
\end{equation} 
This result is consistent (up to rescaling of $u$ and $v$) with equatorial and photon sphere results presented in
\cite{Igata:2019pgb,Fransen:2023eqj}, and with the eigenvalues of the ``wave profile'' ($H_{ab}$) presented in
\cite{Hollowood:2009qz} for general Petrov type D spacetimes as well as for Kerr in particular.\footnote{In fact, \cite{Hollowood:2009qz} may be considered to give a full alternative derivation for \eqref{eqn:KBrinkmannc}  when, in the right places, expressions are kept complex and the transverse directions are interpreted in terms of $z$ and $\bar{z}$; see specifically their (7.46). However, this does not seem to have been what \cite{Hollowood:2009qz} had in mind.} 

We will not describe here a parallel propagated null frame, derived from a dyad $\lbrace\alpha,\beta \rbrace$, nor the adapted coordinate approach as we did for the Schwarzschild black hole. We do wish to note, that after we computed \eqref{eqn:KBrinkmannc} and \eqref{eqn:KBrinkmann}, we became aware of the reference \cite{penrose19952}, where Penrose performs essentially the same calculation of \eqref{eqn:penrosespinor} for vacuum Petrov type D spacetimes. However, Penrose doesn't discuss it in the context of the Penrose limit or the geodesic deviation equations but rather as representing the ``null datum'' for the characteristic initial value problem \cite{penrose1980golden}. Along the same lines, it was known (e.g. \cite{penrose1984spinors}, Section 5.11) that \eqref{eqn:penrosespinor} is a way to compute the ``astigmatic'', $\Psi(u)$, and ``anastigmatic'', $\Phi(u)$, part of the geodesic deviation but we are unaware of other sources than \cite{Tod:2019urw} emphasizing the simplicity of this approach, or actually using it to take a Penrose limit. 


\subsubsection{The double copy}\label{sec:KdcPenrose}

The Weyl double copy structure of the Kerr black hole, as described in \cite{Luna:2018dpt}, is a straightforward generalization of the Schwarzschild example presented in \eqref{eqn:Spsidc} and \eqref{eqn:Ssingle}
\begin{equation}\label{eqn:Kpsidc}
\Psi_{ABCD} = -\left(\frac{M}{\zeta}\right)^5 3\chi_{(AB}\chi_{CD)} \, , \quad f_{AB} =   -\left(\frac{M}{\zeta}\right)^3 2\chi_{AB} \, , \quad S = - 2\left(\frac{M}{\zeta}\right) \, .
\end{equation}
Similarly, the double copy of the plane wave \eqref{eqn:KBrinkmannc} is a straightforward generalization of
\eqref{eqn:Sgammapsidc} and \eqref{eqn:Sppscweyl}. If we fix the type N ambiguity in the Weyl double copy by \eqref{eqn:Sppscweyl}, then the Weyl single copy of \eqref{eqn:KBrinkmann}, with Weyl scalar
\begin{equation}\label{eqn:Kgammapsidc}
	\Psi^{(\gamma)}_{ABCD} = -\left(\frac{M}{\zeta(u)}\right)^5 \frac{3 K^2}{M^4}  \beta^{(\gamma)}_A \beta^{(\gamma)}_B  \beta^{(\gamma)}_C  \beta^{(\gamma)}_D   \, ,
\end{equation}
 is given by
\begin{equation}\label{eqn:Kgammapsisc}
 f^{(\gamma)}_{AB} =\left(\frac{M}{\zeta(u)}\right)^3 \frac{2K}{M^2} \beta^{(\gamma)}_A \beta^{(\gamma)}_B \, , \quad S^{(\gamma)} = -\frac{2M}{\zeta(u)}  \, .
\end{equation}
Together with the conservation equation \eqref{eqn:Kkillingspinorconserved}, \eqref{eqn:Kpsidc} and \eqref{eqn:Kgammapsisc} immediately imply the validity of the proposed prescription \eqref{eqn:dcPenroselimit} for the single copy Penrose limit. Said differently, the Kerr black hole, just like the Schwarzschild black hole, behaves perfectly according to the schematic summary in Figure \ref{fig:schematic}.

We will not repeat the full analysis of the relation to the Kerr-Schild double copy, which can be found in \cite{Monteiro:2014cda} for the black hole, and in \cite{Luna:2018dpt} for the plane wave.  The difference between \cite{Monteiro:2014cda} and \cite{Luna:2018dpt} for the plane wave, Kerr-Schild double copy was described in detail for the Schwarzschild example. We use the ``plane wave-to-plane wave'' rule of \cite{Luna:2018dpt}, and believe that the consistency with the Penrose limit of this rule adds some support for its validity.

The result is that the Kerr-Schild single copy of the Penrose limits of an arbitrary null geodesic on a Kerr black hole is given by 
\begin{equation}\label{eqn:Ksingle}
\begin{aligned}
	A^{(\gamma)} &= -\frac{K}{M^2}\left( M^3 \zeta(u)^{-3} z + M^3 \bar{\zeta}(u)^{-3} \bar{z}\right)du \\ &=  -\frac{\sqrt{2} M K}{\left(r(u)^2+y(u)^2\right)^3} \left\lbrack \left(r(u)^3 -3 r(u) y(u)^2\right)x_1 + \left(-y(u)^3 +3 r(u)^2 y(u)\right)x_2\right\rbrack du \, .
\end{aligned}
\end{equation}
The zeroth copy is given by
\begin{equation}\label{eqn:Kzero}
\phi^{(\rm KS)}_{(\gamma)} = - 2\left( M \zeta(u)^{-1} + M \bar{\zeta}(u)^{-1}\right) =- \frac{4 M r(u)}{r(u)^2+y(u)^2} \, .
\end{equation}
Recall that the functions $r|_{\gamma} = r(u)$, and  $y|_{\gamma} = y(u)$ satisfy the geodesic equations \eqref{eqn:Kgeod}. 

The canonical coordinates and the closely related Kerr-Schild coordinates for the Kerr black hole, used throughout this example and introduced in \eqref{eqn:canonicalcoord} and \eqref{eqn:KScoord}, are not well-suited to take the Schwarzschild limit $a \to 0$. However, taking $y \sim a \to 0$, the Penrose limit plane wave \eqref{eqn:KBrinkmann}, as well as the single copy \eqref{eqn:Ksingle} and zeroth copy \eqref{eqn:Kzero} readily reduce the their equivalents for the Schwarzschild black holes, respectively \eqref{eqnS:Brinkmann}, \eqref{eqn:SKS1matchingWeyl}, and \eqref{eqn:SKS0matchingWeyl}.

\subsection{Vacuum Petrov type D spacetimes}\label{sec:typeD}

The Schwarzschild and Kerr examples above readily generalize to the full class of vacuum Petrov type D spacetimes. These spacetimes all have Weyl spinors of the form \cite{penrose19952}
\begin{equation}\label{eqn:dcD}
\Psi_{ABCD} = -\left(\frac{\tilde{M}}{\zeta}\right)^5 3 \chi_{(AB}\chi_{CD)} \, ,
\end{equation}
where $\chi_{AB}$ is a valence-2 Killing spinor, $\zeta$ is a non-trivial function on the spacetime generalizing (up to constants) $r$ for Schwarzschild and $r+i a \sin{\theta}$ for Kerr, and $\tilde{M}$ similarly is a suitable generalization of the mass.

 In \cite{penrose19952} for instance, it is discussed how it is appropriate to think of $\zeta$ as a ``complex luminosity'' parameter along the principal null directions, while $\tilde{M}$ can differ from the mass by a NUT-charge contribution. To be more explicit, for the Plebanski-Demianski family of metrics 
\begin{equation}\label{eqn:generaltypeDexamples}
\tilde{M} = M-iN \, , \quad \zeta = -\frac{q-ip}{1-pq} \, .
\end{equation}
We will  not actually need these expressions for $\tilde{M}$ and $\zeta$ here. Instead, we refer to Appendix \ref{app:PlebanskiDemianski} for some additional details, such as the definition\footnote{$\zeta$ and $\bar{\zeta}$ are arguably the more fundamental quantities. Therefore, (the inversion of) \eqref{eqn:generaltypeDexamples} can actually be viewed as giving a more invariant meaning to the coordinates $p$ and $q$.} of the (real) coordinates $p$ and $q$. 

The Weyl single and zeroth copy of vacuum Petrov type D spacetimes \eqref{eqn:dcD}, in our conventions, are minimally different from the Kerr example \eqref{eqn:Kpsidc} 
\begin{equation}\label{eqn:zcD}
f_{AB} = -\left(\frac{\tilde{M}}{\zeta}\right)^3 2 \chi_{AB} \, , \quad S = - \frac{2 \tilde{M}}{\zeta} \, .
\end{equation}
Note that the inverse of $\zeta$ always represents the spacetime dependent piece of the zeroth copy and $\zeta^2 \propto \chi_{AB} \chi^{AB}$, as illustrated by the Schwarzschild \eqref{eqn:Schi} and Kerr \eqref{eqn:KPsiABCD} examples. The entire structure and consistency of the Weyl double copy for this class of vacuum Petrov type D spacetimes is thus based on the Killing spinor $\chi_{AB}$.

The existence of the valence-2 Killing spinors is also enough for the null geodesic equations to separate \cite{Walker:1970un}. We will not go through this explicitly again, but rather assume a null geodesic $\gamma$ and a tangent spinor $\alpha^A$ were found as in the previous examples. Then we have the conserved quantity\footnote{For discussions on the relation between the conserved quantity in \eqref{eqn:chialphalpha} to the Carter constant, see for instance the original work \cite{Walker:1970un}, the review \cite{Frolov:2017kze} (Appendix F, focused on the relation to the Killing-Yano tensor), or \cite{Aksteiner:2014zyp} (beginning of Chapter 5, with an eye towards a field theory generalization).} (defined by analogy with \eqref{eqn:Kkillingspinorconserved})
\begin{equation}\label{eqn:chialphalpha}
\chi_{AB} \alpha^{A}\alpha^{B} = - \frac{\tilde{K}}{\tilde{M}^2} \, ,
\end{equation}
and the Penrose limit follows immediately from
\begin{equation}
\left(\Psi_{ABCD}  \alpha^{A}\alpha^{B} \alpha^{A}\alpha^{B}\right)_{|\gamma}  = -\frac{3\tilde{K}^2}{\tilde{M}^4} \left(\frac{\tilde{M}}{\zeta(u)}\right)^5 \, .
\end{equation}
Explicitly, the Penrose limit plane wave for an arbitrary null geodesic in a vacuum Petrov type D spacetime is given by
\begin{equation}\label{eqn:DBrinkmannc}
ds_{\gamma}^2 = 2 du dv - \left\lbrack \left(\frac{3\tilde{K}^2}{\tilde{M}^4} \left(\frac{\tilde{M}}{\zeta(u)}\right)^5\right) z^2 + \left(\frac{3\tilde{K}^2}{\tilde{M}^4} \left(\frac{\tilde{M}}{\zeta(u)}\right)^5\right)^* \bar{z}^2\right\rbrack du^2 - 2 dz d\bar{z}  \, .
\end{equation}
Here, we have left implicit that $\zeta(u) = \zeta|_{\gamma}$ is to be evaluated along the geodesic, which will generically mean that the $u$-dependence of \eqref{eqn:DBrinkmannc} will be rather complicated in practice.

The single copy of the plane wave spacetime \eqref{eqn:DBrinkmannc} is given by
\begin{equation}
f_{AB}^{(\gamma)} = \left(\frac{\tilde{M}}{\zeta(u)}\right)^3 \frac{\tilde{K}}{\tilde{M}^2} 2 \beta^{(\gamma)}_{A}\beta_B^{(\gamma)} \, , \quad S^{(\gamma)} = - \frac{2 \tilde{M}}{\zeta(u)} \, .
\end{equation}
consistent with the proposed Weyl double copy of the Penrose limit. Let us conclude our discussion of the vacuum Petrov type D cases by reiterating, in Figure \ref{fig:schematicD}, the general schematic structure of the results Figure \ref{fig:schematic} explicitly in terms of the key valence-2 Killing spinor. 

Note for the normalizations that, in addition to the overall, constant ambiguity in the Weyl double copy $(S, f_{AB}) \to (C_{W}^2 S, C_{W} f_{AB})$ for a constant $C_{W}$, the normalization of the valence-2 Killing spinor is not fixed a priori, so we can also rescale $\chi_{AB} \to C_{\chi}\chi_{AB}$ for constant $C_{\chi}$. Finally, a different choice of affine time of the null geodesic will rescale $\alpha$ and therefore $\beta^{(\gamma)}$ (to preserve the dyad basis normalization \eqref{eqn:dyadnorm}). On the other hand, both the normalization of tangent spinor (and related $\beta^{(\gamma)}$) , as well as that of $\chi_{AB}$ enter into $\tilde{K}$,  say $(\beta^{(\gamma)}, \tilde{K}) \to (C_u \beta^{(\gamma)}, \frac{C_{\chi}}{C_u^2}\tilde{K})$. All in all, we have the following rescaling freedom
\begin{equation}
\Psi_{ABCD} \to C^{-3}_{\chi} \Psi_{ABCD} \, , \quad f_{AB} \to C_{W} C^{-2}_{\chi} f_{AB} \, , \quad  C_{W}^2 C^{-1}_{\chi} S \, ,
\end{equation}
Therefore, we are indeed free to impose, by our choice of $\chi_{AB}$ and the Weyl double copy scaling 
\begin{equation}\label{eqn:typeDclean}
\Psi_{ABCD} =  \frac{\chi_{(AB} \chi_{CD)}}{\left(\chi_{EF}\chi^{EF}\right)^{5/2}} \, , \quad f_{AB} = \frac{\chi_{AB}}{\left(\chi_{EF}\chi^{EF}\right)^{3/2}} , \quad S = \left(\chi_{EF}\chi^{EF}\right)^{-1/2}  \, .
\end{equation}
Then, from the definition of $\tilde{K}$ in \eqref{eqn:chialphalpha}, it follows that
\begin{equation}
\begin{aligned}
\Psi^{(\gamma)}_{ABCD} &=  \left(\frac{\tilde{K}}{\tilde{M}^2}\right)^2 \frac{\beta^{(\gamma)}_A \beta^{(\gamma)}_B \beta^{(\gamma)}_C \beta^{(\gamma)}_D}{\left. \left\lbrack\left(\chi_{EF}\chi^{EF}\right)^{5/2}\right\rbrack \right |_{\gamma}}  \, , \quad f^{(\gamma)}_{AB} = -\left(\frac{\tilde{K}}{\tilde{M}^2}\right) \frac{\beta^{(\gamma)}_A \beta^{(\gamma)}_B}{\left. \left\lbrack\left(\chi_{EF}\chi^{EF}\right)^{3/2}\right\rbrack \right|_{\gamma}} , \\ S^{(\gamma)} &= \left.\left\lbrack\left(\chi_{EF}\chi^{EF}\right)^{-1/2}\right\rbrack \right|_{\gamma}  \, .
\end{aligned}
\end{equation}
On the other hand, rescaling $\chi_{AB}$ while keeping $\tilde{M}$ fixed, will implicitly change the definition of $\zeta$. In particular examples, like the previously discussed Schwarzschild and Kerr black holes, we preferred not to include extra combinatorial factors into this quantity such the reality is slightly less clean than \eqref{eqn:typeDclean}.

From the plane wave perspective, one can absorb $\tilde{K}/\tilde{M}^2$ into the plane wave coordinate $u$. However, from our understanding of the plane wave as arising from a Penrose limit, we know that $\tilde{K}/\tilde{M}^2$ is not independent of this choice of affine time. Therefore, we prefer to keep it.

\begin{figure}[t!]
	\begin{tikzcd}
		 \makebox[\lenfff em][c]{$\begin{aligned}
	\Psi_{ABCD} =  \frac{\chi_{(AB} \chi_{CD)}}{\left(\chi_{EF}\chi^{EF}\right)^{5/2}} 
	 		\end{aligned}$}
 		  \arrow[r,leftrightarrow, "\text{Weyl DC}"] \arrow[ddd,shorten=6mm,"\text{Penrose limit}{}_{\gamma}"]&  \makebox[\lenfff em][c]{$\begin{aligned}
	  S &=& \left(\chi_{EF}\chi^{EF}\right)^{-1/2} \\[0.5em] 
	  f_{AB} &=& \frac{\chi_{AB}}{\left(\chi_{EF}\chi^{EF}\right)^{3/2}} \quad
	 \end{aligned}$} 	\arrow[ddd,shorten=5mm,"\text{Penrose limit}_{\gamma}"]	  \,  \\
&  \\ & \\
		\makebox[\lenfff em][c]{$\begin{aligned}
		 \Psi^{(\gamma)}_{ABCD} = \left(\frac{\tilde{K}}{\tilde{M}^2}\right)^2 \frac{\beta^{(\gamma)}_A \beta^{(\gamma)}_B \beta^{(\gamma)}_C \beta^{(\gamma)}_D}{\left. \left\lbrack\left(\chi_{EF}\chi^{EF}\right)^{5/2}\right\rbrack \right|_{\gamma}} 
		 	\end{aligned}$}
	 	 \arrow[r,leftrightarrow, "\text{Weyl DC}"] & \makebox[\lenfff em][c]{$\begin{aligned}
	 	 			 	S^{(\gamma)} &=&  \left. \left\lbrack\left(\chi_{EF}\chi^{EF}\right)^{-1/2}\right\rbrack\right|_{\gamma} \qquad \qquad \\[0.5em] 
		 	f^{(\gamma)}_{AB} &=& - \left(\frac{\tilde{K}}{\tilde{M}^2}\right) \frac{\beta^{(\gamma)}_A \beta^{(\gamma)}_B}{ \left. \left\lbrack\left(\chi_{EF}\chi^{EF}\right)^{3/2}\right\rbrack \right|_{\gamma}}  
		 \end{aligned}$}  
	\end{tikzcd}
	\caption{The Weyl double copy and Penrose limit of general vacuum Petrov type D spacetimes, in  terms of the valence-2 Killing spinor $\chi_{AB}$ and the associated constant of motion $\tilde{K}/\tilde{M}^2$ of the null geodesic $\gamma$, as defined in \eqref{eqn:chialphalpha}.} \label{fig:schematicD} 
\end{figure}

\subsection{Type N examples} \label{sec:typeN}

In the previous Petrov type D examples, we saw the outsized role played by the Killing spinor $\chi_{AB}$, in the
Weyl double copy, the Penrose limit, and in their interplay. Let us therefore explore what happens in a
qualitatively different class of spacetimes. We focus on radiative type N solutions of the vacuum Einstein
equations. In the context of the classical double copy, type N solutions beyond the Kundt class were studied in \cite{Godazgar:2020zbv}. 

The Weyl spinors for type N spacetimes are of the form
\begin{equation}\label{eqn:WeyltypeN}
\Psi_{ABCD} = \Psi(x^{\mu}) \, \iota_A \iota_B \iota_C \iota_D \, ,
\end{equation}
for some function $\Psi(x^{\mu})$ on the spacetime and with $\iota_A$ a principal spinor. In keeping with our convention for the plane wave (see \eqref{eqn:NPsi} and also \eqref{eqn:psidefs}), the form of \eqref{eqn:WeyltypeN} is such that only the Weyl scalar $\Psi_0 = \Psi(x^{\mu})$ is non-vanishing with respect to a dyad $\left\lbrace o^A, \iota^A \right\rbrace$. The Weyl double copy implies the existence of a scalar $S$ and Maxwell spinor $f_{AB}$ of the form
\begin{equation}\label{eqn:ftypeN}
f_{AB} = f(x^{\mu}) \, \iota_A \iota_B \, . 
\end{equation}
Therefore, the Weyl double copy for type N spacetimes can be stated as
\begin{equation}\label{eqn:typeNdc}
\Psi = \frac{f^2}{S} \, .
\end{equation}
A condition on $S$ for the compatibility of the Bianchi identity for $\Psi$ with the Maxwell equation for $f$, on the curved background, for general type N spacetimes was established in \cite{Godazgar:2020zbv}. The equations of motion for the field strength associated to $f$ and the scalar $S$ on the flat spacetime instead were established on a case by case basis. For the ``twisted'' type N metrics, where the principal null geodesic congruence has a non-vanishing twist, the flat space equations of motion were found not to be satisfied \cite{Godazgar:2020zbv}.

Focusing then on the non-twisting spacetimes with single copies on flat space, these type N metrics can naturally
be given in terms of an adapted coordinate system\footnote{In the example of the plane wave, the Brinkmann
coordinates \eqref{eqn:Brinkmann} are an adapted coordinate system for the principal, null geodesic congruence
$\partial_v$. Hence, we avoid using the notation of  \eqref{eqn:adapted} for adapted coordinate systems in which we would have $(U,V,X_1,X_2) = (v,u,x_1,x_2)$.} (of the form \eqref{eqn:adapted}) to the principal null geodesic congruence. A null affine time $v$ along the principal null geodesic congruence is then used as a coordinate time and the principal null spinor can be chosen such that $(\partial_v)^{AA'} = \iota^A \bar{\iota}^{A'}$. Fixing the principal null spinor, it was shown in \cite{Godazgar:2020zbv} that the functional ambiguity of the Weyl double copy can not in general be fixed by choices like \eqref{eqn:typeNscaling} or \eqref{eqn:oldtypeNscaling}. 

On the other hand, the choice of principal null spinor is not unique. If we only impose that $o^A$ is such that
\eqref{eqn:WeyltypeN} holds, $\iota_A$ can be rescaled by an arbitrary scalar function. We will call this rescaling
a Geroch-Held-Penrose (GHP) transformation \cite{geroch1973space}, after the GHP, or compacted spin-coefficient,
formalism in which one works covariantly with respect to such rescalings \cite{penrose1984spinors}. We will say a
quantity $X$ has boost-weight $b$ and spin-weight $s$ if under $o^A \to \lambda e^{i \varphi}o^A$ and $\iota^A \to
\lambda^{-1} e^{-i \varphi} \iota^A$, for real $\lambda$ and $\varphi$, it transforms as $X \to \lambda^{2b} e^{2 i
s \varphi}  X$. The zeroth copy scalar $S$, the field strength spinor $f_{AB}$, and the Weyl spinor $\Psi_{ABCD}$
are invariant under GHP transformations but the (spacetime) scalars $f$ and $\Psi$ have boost-and spin weight $1$
and $2$ respectively. Moreover, the governing Bianchi and Maxwell equations for $\Psi$ and $f$ in \cite{Godazgar:2020zbv} are GHP-covariant, although they were not written in a manifestly covariant way\footnote{The GHP-covariance is one way to understand the ``clear structure'' in these massless free field equations for $\Psi$ an $f$ remarked upon in \cite{Godazgar:2020zbv}.}. 

Therefore, while in Section \ref{sec:ambiguity} we discussed the functional ambiguity in the double copy for type N spacetimes with a focus on plane waves, the choices \eqref{eqn:oldtypeNscaling} and \eqref{eqn:typeNscaling} of the form
\begin{equation}\label{eqn:typeNscalingN}
\Psi = S^{2a+1} \, , \quad f = S^{a+1} \, , 
\end{equation}
can be made for general type N spacetimes. Simply let $\iota^A \to \left(\frac{\Psi}{S^{2a+1}}\right)^{1/4} \iota^A$.
Then, $\Psi \to S^{2a+1}$ and $f \to S^{a+1}$. The GHP-covariance ensures the equations of motion for $\Psi$ and
$f$ are still satisfied while the spacetime tensors and spinors associated to $\Psi$ and $f$ are GHP-invariant.
Thus, the flat space wave equations are also still satisfied. Of course, in this way the choices
\eqref{eqn:oldtypeNscaling} and \eqref{eqn:typeNscaling} don't actually fix the ambiguity in the double copy, which
exists on the level of a functional freedom for zeroth copy $S$. In explicit examples, we will see that there seems
to be natural choices in between an a priori fixed $\iota^A$, such as can be imposed for plane waves, and the
arbitrariness that the full GHP-covariance allows.\footnote{With regards to the type D examples, note that
$\Psi_2$, contrary to $\Psi_0$ or $\Psi_4$, is GHP-invariant.}

As before, given a geodesic $\gamma$ and a parallel transported tangent spinor $\alpha^A$,  the spinor approach to the Penrose limit \eqref{eqn:penrosespinor} yields the plane wave equivalents of $\Psi$, $f$ and $S$
\begin{equation}\label{eqn:gammatypeN}
\Psi^{(\gamma)} =\left. \left\lbrack \Psi \left(\iota_A \alpha^A\right)^4\right\rbrack  \right|_{\gamma}  \, , \quad f^{(\gamma)} = \left. \left\lbrack  f  \left(\iota_A \alpha^A\right)^2 \right\rbrack \right|_{\gamma}\, , \quad S^{(\gamma)} = S|_{\gamma} \, .
\end{equation}
Therefore, \eqref{eqn:typeNdc} implies that the Weyl double copy relationship is also satisfied for the Penrose limit plane waves
\begin{equation}\label{eqn:typeNdcpenrose}
\Psi^{(\gamma)} = \frac{\left(f^{(\gamma)}\right)^2}{S^{(\gamma)}} \, .
\end{equation}
However, contrary to the type D examples, the choice \eqref{eqn:typeNscalingN} isn't necessarily preserved, up to a constant, by the Penrose limit. 

One way to understand that, say, the choice \eqref{eqn:typeNscaling} to fix the functional ambiguity in type N spacetimes can not generally be imposed for both the original type N spacetime and the Penrose limit plane wave is an obstruction to choosing an affine ``light-cone gauge'' coordinate $\tau$ along the null geodesic for which
\begin{equation}\label{eqn:lightconegauge}
|\iota_A \alpha^A|^2 = \frac{d u}{d \tau} = \text{constant} \, ,
\end{equation}
with $u$ playing the role of $V$ in the adapted coordinate system \eqref{eqn:adapted} with respect to the principal
null geodesic congruence $\partial_v$ (instead of $\partial_U$ in \eqref{eqn:adapted}). One can try to nevertheless
impose \eqref{eqn:typeNscaling}, by letting go of the affine coordinates either on the level of general type N
background, generalizing the adapted coordinates to \eqref{eqn:adaptedmod}, or on the level of the Penrose limit,
by a GHP-transformation of the plane wave principal spinor. Nevertheless, for both of these options, the interplay
between the Penrose limit and the Weyl double copy comes out a little less naturally than in the type D examples and their associated Penrose limit plane waves.

The only class of type N spacetimes for which \eqref{eqn:lightconegauge} can be chosen to be unity in general, are the so-called pp-waves. We shall discuss this class of metrics first and subsequently comment on other explicit type N spacetimes. 

\subsubsection{pp-waves} \label{sec:ppwaves}

Plane-fronted waves with parallel rays, or pp-waves, are a class of type N spacetimes whose principal null direction is covariantly constant. Moreover, this principal null direction can be written as the square of a Killing spinor. As a result, the double copy and the Penrose limit for pp-waves will play out in the same way as for the Petrov type D examples but with an ordinary Killing spinor taking over the role of the valence-2 Killing spinor. 

Plane wave spacetimes themselves are a special class of pp-waves. We will discuss these first and show that the ``iterated'' Penrose limit, or a Penrose limit of a plane wave, will either be the same plane wave or flat space. Therefore, we will not additionally discuss the double copy, as it either remains identical or becomes trivial. Subsequently we will discuss more general pp-waves.

To discuss the Penrose limits of plane waves, there is no need to impose field equations so we keep both curvature spinors $\Psi_{ABCD}$ and $\Phi_{ABA'B'}$ in their general form
\begin{equation}\label{eqn:psiNplane}
\Psi_{ABCD} = \Psi(u) \iota_{A}\iota_{B}\iota_{C}\iota_{D} \, , \quad \Phi_{ABC'D'} = \Phi(u) \iota_{A}\iota_{B}\bar{\iota}_{C'}\bar{\iota}_{D'} \, .
\end{equation}
As mentioned, $\iota^A$ can be chosen to be a Killing spinor which squares to the covariantly constant null vector $\partial_v = \iota \bar{\iota}$. As a result, if $\alpha^{A}$ is parallel propagated and tangent to a null geodesic $\gamma$ in the plane wave, then $\alpha_{A} \iota^{A} = c$ is a constant of motion. Therefore, the Penrose limit along $\gamma$ is given by
\begin{equation}\label{eqn:penroselimitplane}
    \Psi^{(\gamma)}  = c^4 \Psi(u)|_{\gamma}  \, , \quad \Phi^{(\gamma)}  = (c \bar{c})^{2} \Phi(u)|_{\gamma} \, .
\end{equation}
Let $\tau$ be the affine parameter along the chosen null geodesic $\gamma$, from $\partial_v = \iota \bar{\iota}$ and $\left(\partial_v\right)_{\mu}dx^{\mu} = du$, it follows that
\begin{equation}\label{eqn:dtaudu}
\frac{du}{d \tau} = (\alpha \bar{\alpha})^{\mu}\left(\partial_v\right)_{\mu} = |c|^2  \, .
\end{equation}
Without loss of generality, we can rotate $\iota^A \to e^{i \lambda} \iota^A$ and $\bar{\iota}^A \to e^{-i \lambda} \bar{\iota}^A$, for a constant $\lambda$, such that $c$ is real. Then, the resulting Penrose limit is indeed the plane wave we started with, up to a rescaling of the coordinates $(\tau,\tau',z,\bar{z}) \to (c^{-2} u, c^{2} v,z,\bar{z})$. The only exception occurs for $c = 0$, when $\alpha^{A}$ becomes proportional to $\iota^A$. In this case of a null geodesic along the principal null direction, the plane wave degenerates to flat space in the Penrose limit. 

For general pp-waves, the main difference compared to plane waves is that, in the equivalent of
\eqref{eqn:psiNplane}, the functions $\Psi$ and $\Phi$ can still depend on the transverse coordinates. That is,
coordinates $(u,v,z,\bar{z})$ exist for pp-waves such that the metric generalizes \eqref{eqn:Brinkmannc}
to\footnote{\eqref{eqn:Brinkmanncpp} is not the most general pp-wave; we can also introduce a non-trivial metric on
the transverse space if we do not impose the vacuum Einstein equation.}
\begin{equation}\label{eqn:Brinkmanncpp}
ds_{\gamma}^2 = 2 du dv -H(z,\bar{z},u) du^2  - 2 dz d\bar{z}  \, ,
\end{equation}
where $H(z,\bar{z},u)$ is no longer restricted to be quadratic in $z$ and $\bar{z}$. The curvature spinors of pp-waves are given by
\begin{equation}\label{eqn:psiNpp}
\Psi_{ABCD} = \Psi(u,z,\bar{z}) \iota_{A}\iota_{B}\iota_{C}\iota_{D} \, , \quad \Phi_{ABC'D'} = \Phi(u,z,\bar{z}) \iota_{A}\iota_{B}\bar{\iota}_{C'}\bar{\iota}_{D'} \, ,
\end{equation}
where
\begin{equation}\label{eqn:ppcurvaturespinors}
\Psi(u,z,\bar{z}) = -\frac{1}{2}\partial_z^2 H \, , \quad  	\Phi(u,z,\bar{z}) = -\frac{1}{2}\partial_z \partial_{\bar{z}} H \, ,
\end{equation}
are still dependent on $z$, $\bar{z}$ in general. Therefore, the Penrose limit now also depends on $z(\tau)$, which, from the geodesic equation, satisfies
\begin{equation}
\frac{d^2 z(\tau)}{d \tau^2} = \frac{c^2}{2}\left(\partial_z H\right)|_{\gamma}\, .
\end{equation}
On the other hand, for a null geodesic $\gamma$ and parallel transported tangent spinor $\alpha^A$, the conservation of $\alpha_{A} \iota^{A}$ along the null geodesic as well as  \eqref{eqn:dtaudu} still hold.

Restricting ourselves to the vacuum case, \eqref{eqn:ppcurvaturespinors} implies that the wave profile is of the form
\begin{equation}
H(u,z,\bar{z}) = -2h(u,z)-2\bar{h}(u,\bar{z})\, ,
\end{equation}
for arbitrary ``holomorphic'' and ``anti-holomorphic'' functions $h(u,z)$ and $\bar{h}(u,\bar{z})$. As discussed in \cite{Godazgar:2020zbv}, the single and zeroth copy of the pp-wave satisfy the flat space wave equations if they are of the form
\begin{equation}
S = S(u,z)  \, , \quad f = \sqrt{S(u,z)\partial_z^2 h} \, .
\end{equation}
In particular, as in \eqref{eqn:typeNscaling}, we are free to choose  $S = \left(\partial_z^2 h\right)^{1/5}$ such that 
\begin{equation}\label{eqn:ppchoice}
S = \left(\partial_z^2 h\right)^{1/5}  \, , \quad f =  \left(\partial_z^2 h\right)^{3/5} \, , \quad \Psi =  \partial_z^2 h \, .
\end{equation}
With the choice \eqref{eqn:ppchoice} to fix the remaining functional ambiguity of the Weyl double copy, we summarize the results for the interplay between the Weyl double and the Penrose limit for pp-waves in Figure \ref{fig:schematicpp}. 

\begin{figure}[t!]
	\begin{tikzcd}
		\begin{aligned}
			\Psi_{ABCD} =  S^5 \iota_{(A} \iota_B \iota_C \iota_{D)}
		\end{aligned}
		\arrow[r,leftrightarrow, "\text{Weyl DC}"] \arrow[dd,"\text{Penrose limit}{}_{\gamma}"]&  \quad \begin{aligned}	
			f_{AB} = S^3 \iota_{(A} \iota_{B)}
		\end{aligned} \arrow[dd,"\text{Penrose limit}_{\gamma}"] \,  \\
		&   \\
		\begin{aligned}
			\Psi^{(\gamma)}_{ABCD} =  c^4 (S|_{\gamma})^5 \beta^{(\gamma)}_{(A} \beta^{(\gamma)}_B \beta^{(\gamma)}_C \beta^{(\gamma)}_{D)}
		\end{aligned}
		\arrow[r,leftrightarrow, "\text{Weyl DC}"] & \quad \begin{aligned}
			f^{(\gamma)}_{AB} = c^2 (S|_{\gamma})^3  \beta^{(\gamma)}_{(A} \beta^{(\gamma)}_{B)} 
		\end{aligned}  
	\end{tikzcd}
	\caption{The Weyl double copy and Penrose limit of general vacuum, pp-wave spacetime, in  terms of the Killing spinor $\iota_{A}$ and the associated constant of motion $c = \alpha^A \iota_A$ of the null geodesic $\gamma$ with tangent, parallel propagated spinor $\alpha^A$.} \label{fig:schematicpp} 
\end{figure}

\subsubsection{Other type N solutions}


The pp-waves fall into the broader Kundt class of type N spacetimes, which are characterized by a vanishing expansion and twist of the principal null geodesic
congruence. Compared to pp-waves, even in this slightly broader Kundt class, a subtlety already arises in the
application of \eqref{eqn:typeNscaling} to fix the functional ambiguity of the double copy. As was shown in \cite{Godazgar:2020zbv}, for the Kundt class metrics of the form (adapted to the null geodesic congruence $\partial_v$)
\begin{equation}\label{eqn:otherkundt}
ds^2 = 2du dv - (z+\bar{z})\left(2h(u,z)+2\bar{h}(u,\bar{z}) +\frac{v^2}{(z+\bar{z})^3}\right)du^2 -\frac{4v  \, du dz}{z+\bar{z}} -\frac{4v \, du d\bar{z}}{z+\bar{z}}  -dz d\bar{z} \, ,
\end{equation}
the single and zeroth copy that satisfy the flat space wave equation are given by
\begin{equation}\label{eqn:kundtsc}
S = \frac{S_0(u,z)}{z+\bar{z}}  \, , \quad f = \sqrt{S_0(u,z)\partial_z^2 h} \, ,
\end{equation}
for arbitrary $S_0(u,z)$. On the other hand, $\Psi =  (z+\bar{z})\partial_z^2 h$ is such that we can not impose
\eqref{eqn:typeNscaling} in the formulation of \cite{Godazgar:2020zbv}. However, this result depends on a choice of principal null frame. 

Using $\left(z+\bar{z}\right)^3\partial_v$ instead of $\partial_v$ as part of the principal null frame, which moreover is still geodesic,
we find $\Psi =  (z+\bar{z})^{-5} \partial_z^2 h$. Therefore, we can set
\begin{equation}
S = (\Psi)^{1/5} = \frac{(\partial_z^2 h)^{1/5}}{(z+\bar{z})} \, , \quad f = (\Psi)^{3/5} = \frac{(\partial_z^2 h)^{3/5}}{(z+\bar{z})^3} \, .
\end{equation}
The form of the  $S$ is unchanged compared to \eqref{eqn:kundtsc}, such that it satisfies the flat space wave equation. Similarly, while $f$ of course has been rescaled, the associated field strength tensor is unchanged
\begin{equation}
F_{\mu \nu}  = 2 f n^{(g)}_{[\mu} \bar{m}^{(g)}_{\nu]} \, ,
\end{equation} 
and thus the single copy field equations are also still satisfied on the flat background.

For the general type N case, \eqref{eqn:oldtypeNscaling} typically seems a better choice in relation to the Penrose limit. This choice can be achieved for \eqref{eqn:otherkundt} using $\left(z+\bar{z}\right)^2\partial_v$, such that $\Psi =  (z+\bar{z})^{-3} \partial_z^2 h$ and 
\begin{equation}
S = (\Psi)^{1/3} = \frac{(\partial_z^2 h)^{1/3}}{(z+\bar{z})} \, , \quad f = (\Psi)^{2/3} = \frac{(\partial_z^2 h)^{2/3}}{(z+\bar{z})^2} \, ,
\end{equation}
Now let $u^{AA'} = \alpha^A \bar{\alpha}^{A'}$ be the tangent to a null geodesic $\gamma$, then
\begin{equation}\label{eqn:oalphakundtother}
\left(z+\bar{z}\right)^2\left(\partial_v\right)_{\mu} u^{\mu}= |\iota_A \alpha^A|^2 = \frac{du}{d\tau} (z+\bar{z})^2 \, ,
\end{equation}
and, from the geodesic equations
\begin{equation}
\frac{du}{d\tau} = \frac{K(\tau)}{(z(\tau)-\bar{z}(\tau))^2} \, , \quad K(\tau) = \left(K^{-1}_0 - 2 \int_0^{\tau} dt \, v(t) (z(t)+\bar{z}(t))^4\right)^{-1} \, ,
\end{equation}
for an integration constant $K_0$. Therefore
\begin{equation}\label{eqn:penrosekundtother}
S^{(\gamma)}  = \left. \left(\frac{\partial_z^2 h}{(z+\bar{z})^3}\right)^{1/3} \right|_{\gamma} \, , \quad f^{(\gamma)}  =  \left. K(\tau)\left(\frac{\partial_z^2 h}{(z+\bar{z})^3}\right)^{2/3} \right|_{\gamma}\, , \quad \Psi^{(\gamma)}  = \left. \left(K(\tau)\right)^2 \left(\frac{\partial_z^2 h}{(z+\bar{z})^3}\right)\right|_{\gamma} \, .
\end{equation}
As expected from the general type N discussion, \eqref{eqn:penrosekundtother} does not exactly preserve the scaling \eqref{eqn:typeNscaling} as a result of the contribution from \eqref{eqn:oalphakundtother}.

In the Robinson-Trautman class of type N spacetimes, where the principal null congruence has vanishing twist but
non-zero expansion, one can similarly change the GHP-frame compared to \cite{Godazgar:2020zbv}, in order to impose
\eqref{eqn:typeNscaling} or \eqref{eqn:oldtypeNscaling}. Moreover, there is again a natural split between the
functional dependence of the GHP-transformations and the freedom allowed by the wave equations for the single -and
zeroth copy, comparable to the powers $(z+\bar{z})^a$ as opposed to the arbitrary ``holomorphic'' $S_0(u,z)$ in the Kundt example.

\section{Discussion and outlook}\label{sec:discussion}

To summarize, we have given an account of the interplay between Penrose limits of a metric and its double copy
structure. In cases where the full spacetime has an exact double copy structure (such as type D spacetimes), we
have shown that the natural-seeming Penrose-Güven limit of the single and zeroth copies is indeed consistent with
the Penrose limit of the double copy. That is, when the geodesic $\gamma$ is identified across the double copy as
in section \ref{sec:dcandPenrose}, diagrams such as Figures \ref{fig:schematic}, \ref{fig:schematicD}, and \ref{fig:schematicpp} commute.

Along the way, we observed in section \ref{sec:ambiguity} how the Penrose limit provides a new perspective on the
scaling ambiguity of the type N double copy. Specifically, starting from a spacetime with no scaling ambiguity in
its double copy structure leads to a ``natural'' choice for its Penrose limit single copy that is distinct from the
one suggested by \cite{Luna:2018dpt}. Starting from a type N spacetime, which itself has an ambiguity in the double
copy, the Penrose limit does not single out a clear preferred choice. Nevertheless, it points towards the
importance of the choice of GHP-frame, which is closely related to the parameterization of the principal null
congruence, as the origin of the ambiguity.

We hope that this paper can serve as the foundation for a double copy proposal in arbitrary backgrounds that are
expanded about a null geodesic. We have already seen that the leading order metric around a null geodesic --- the
Penrose limit --- is a simple type N spacetime. Similar to the peeling theorem \cite{newman1968new} for
asymptotically flat spacetimes, which was leveraged in \cite{Godazgar:2021iae,Adamo:2021dfg} to propose an
asymptotic Weyl double copy for generic spacetimes, there is also a peeling theorem around null geodesics
\cite{Kunze:2004qd,Blau:2006ar}. We have examined the leading order behavior in this paper, and leave the exploration of higher
orders to future work.

While our focus in this paper was on four dimensional spacetimes, the Penrose limit in adapted coordinates can be
taken for spacetimes in any number of higher dimensions. We are unaware of any literature performing the
Penrose limit in higher dimensions using spinors as in \eqref{eqn:penrosespinor}, but an analogous computation in
higher dimensions should be possible. In five dimensions, the spinorial approach taken in
\cite{Chawla:2022ogv,Monteiro:2018xev} might be useful. We leave details of such a computation to future work.

Our results could also be generalized by considering background spacetimes with sources,
whose double copy structure was proposed in \cite{Easson:2021asd,Easson:2022zoh}. In particular, for the pp-waves
considered in section \ref{sec:ppwaves}, allowing for a non-zero source would loosen the assumption that
$\Phi_{ABC'D'} = 0$. Considering non-vacuum spacetimes would also entail an examination of how the Penrose limit
acts on the fields that source the double copy metric.

It would be interesting to see if the double copy discussed here, with the extensions outlined above, can shed a new light on various applications of the Penrose limit. By the nature of the Penrose limit, most applications strictly deal with leading corrections to the geometrical optics limit in a non-trivial background. Nevertheless, such applications could include corrections to the index of refraction \cite{Hollowood:2007ku,Hollowood:2009qz}, gravitational lensing \cite{Harte:2012jg, Harte:2015ila}, tail and memory effects \cite{harte2013tails,Zhang:2017rno,Zhang:2017geq,Flanagan:2019ezo}, asymptotics of quasinormal modes \cite{Fransen:2023eqj}, and caustics \cite{Harte:2012uw}. Moreover, the importance of small effects in long-term propagation as well as the existence of particularly significant null geodesics, such as those of the photon ring around black holes, makes some such applications potentially relevant to observations \cite{Perlman:2014cwa,Cardoso:2016rao,Baker:2017hug,Cunha:2018acu,Johnson:2019ljv,Hadar:2022xag,Paugnat:2022qzy,Cardenas-Avendano:2023dzo,Lee:2023kry,Jia:2024mlb}.

Another rich source of applications of the Penrose limit has been the discovery that plane waves are exact, solvable perturbative string theory backgrounds \cite{Gueven:1987ad,Amati:1988sa,Horowitz:1989bv,Horowitz:1990sr,Kiritsis:1993jk,Russo:2002rq,Papadopoulos:2002bg,Berenstein:2002jq,Eberhardt:2018exh}. This discovery has stimulated extensive use of the Penrose limit to study spacetime singularities and the behavior (or break-down) of effective field theories under significant tidal effects \cite{Horowitz:1990sr,Blau:2003dz,Blau:2004yi,Giddings:2007bw,Hollowood:2007ku,Craps:2008bv,Hollowood:2009qz,Hollowood:2011yh,Martinec:2020cml,Bena:2020iyw,Dodelson:2020lal,Nishii:2021ylb,Balivada:2023akk,Horowitz:2024dch,Guo:2024pvv}. In addition, in the context of the AdS/CFT correspondence \cite{Maldacena:1997re,Witten:1998qj,Aharony:1999ti}, it has been found that perturbative string theory on a Penrose limit in the gravitational ``bulk'' theory is dual to a large R-charge sector of the ``boundary'' supersymmetric Yang-Mills theory \cite{Berenstein:2002jq,Plefka:2003nb,Sadri:2003pr}. Especially given the origin of the double copy in the KLT relations between open and closed strings \cite{Kawai:1985xq}, it would be worthwhile to reconnect to these string theory applications of the Penrose limit in future work. 

\acknowledgments
S.C. and C.K.  are supported by the U.S. Department of Energy under grant number DE-SC0019470 and by the Heising-Simons Foundation ``Observational Signatures of Quantum Gravity'' collaboration grant 2021-2818. C.K. would additionally like to thank the Aspen Center for Physics, which is supported by National Science Foundation grant PHY-1607611, for support during the completion of this work. K.F. is supported by the Heising-Simons Foundation grant \#2021-2819.

\begin{appendices}

\section{Spinor identities and conventions}\label{app:conventions}

As much as possible, we will follow the conventions of Penrose and Rindler \cite{penrose1984spinors}. On matters of curvature however, we will deviate from \cite{penrose1984spinors} by a sign. The reason is that the definition of \cite{penrose1984spinors} for the curvature is equivalent to 
\begin{equation}
[\nabla_{\alpha}, \nabla_{\beta}]V^{\delta} = R^{(PR)}{}_{\alpha \beta \gamma}{}^{\delta}V^{\gamma} \, ,
\end{equation}
while we use
\begin{equation}
[\nabla_{\alpha}, \nabla_{\beta}]V^{\delta} = R^{\delta}{}_{\gamma \alpha \beta}{}V^{\gamma} \, .
\end{equation}
Otherwise, the conventions of \cite{penrose1984spinors} imply that we work in a mostly minus signature $(+,\,-,\,-,\,-)$ and we take the anti-symmetric $\epsilon_{AB} = - \epsilon_{BA}$ to lower spinor indices ``left'' and raise ``right''
\begin{equation}
\kappa^A \epsilon_{AB} = \kappa_B \, , \quad \kappa^A = \epsilon^{AB}\kappa_B \, ,
\end{equation}
which implies
\begin{equation}
\epsilon_{AB}\epsilon^{CB} = \delta^C_A \, ,
\end{equation}
or
\begin{equation}
\epsilon_{01} = \epsilon^{01} \, , \quad \epsilon_{10} = \epsilon^{10} \, .
\end{equation}
We shall refer to a spin frame or dyad as a pair $\left\lbrace o^A, \iota^A \right \rbrace$ such that
\begin{equation}\label{eqn:dyadnorm}
\epsilon_{AB} o^A \iota^B = 1\, ,
\end{equation}
or
\begin{equation}
o_A \iota^A = 1 =-\iota_A o^A \, . 
\end{equation}
We can write $\epsilon_{AB}$ in such a dyad as
\begin{equation}\label{eqn:epsdown}
\epsilon_{AB} = o_A \iota_B - o_B \iota_A \, .
\end{equation}
A general spinor $\kappa^A$ will decompose into components $\left( \kappa^o,  \kappa^{\iota}\right)$ along the dyad as
\begin{equation}
\kappa^A = \kappa^o o^A + \kappa^{\iota} \iota^A \, .
\end{equation}
We will use the same expressions for the conjugates $\left\lbrace \bar{o}^{A'}, \bar{\iota}^{A'} \right \rbrace$ and, as is conventional, on combining these together into tangent vectors we will often omit writing the Infeld-Van der Waerden symbols $\sigma^{\mu}_{AA'}$; instead interchangeably
\begin{equation}
V^{\alpha} = \sigma^{\alpha}_{AA'} \kappa^A \bar{\kappa}^{A'} = V^{AA'} = \kappa^A \bar{\kappa}^{A'} \, . 
\end{equation}
Most importantly, to a pair of conjugates dyads, $\left\lbrace o^A, \iota^A \right \rbrace$ and $\left\lbrace \bar{o}^{A'}, \bar{\iota}^{A'} \right \rbrace$ , we will associate a Newman-Penrose frame $\left\lbrace k^{AA'}, n^{AA'}, m^{AA'}, \bar{m}^{AA'} \right \rbrace$ as
\begin{equation}\label{eqn:assocaiteNPframe}
k^{AA'} = o^A \bar{o}^{A'} \, , \quad   n^{AA'} = \iota^A \bar{\iota}^{A'} \, , \quad  m^{AA'} = o^A \bar{\iota}^{A'} \, , \quad  \bar{m}^{AA'} = \iota^A \bar{o}^{A'} \, .
\end{equation}
With the choice of basis \eqref{eqn:assocaiteNPframe}, meaning $\left\lbrace k^{AA'}, n^{AA'}, m^{AA'}, \bar{m}^{AA'} \right \rbrace$ for vectors and $\left\lbrace o^A, \iota^A \right \rbrace$  for two-spinors, or
\begin{equation}
o^A = \begin{pmatrix}
1 \\ 0
\end{pmatrix} \, ,\quad 	\iota^A = \begin{pmatrix}
0 \\ 1
\end{pmatrix} \, , 
\end{equation}
and
\begin{equation}
k^{\mu} = \begin{pmatrix}
1 \\  0 \\ 0 \\0
\end{pmatrix} \, , \quad 	n^{\mu} = \begin{pmatrix}
0 \\  1 \\ 0 \\0
\end{pmatrix} \, , \quad \text{etc.}
\end{equation}
the Infeld-Van der Waerden symbols are represented by
\begin{equation}\label{eqn:IvWexplicitnull}
\sigma^{0}_{AA'} = \begin{pmatrix} 1& 0 \\ 0 & 0
\end{pmatrix} \, , \quad 	\sigma^{1}_{AA'} = \begin{pmatrix} 0& 0 \\ 0 & 1
\end{pmatrix} \, , \quad 	\sigma^{2}_{AA'} = \begin{pmatrix} 0& 1 \\ 0 & 0
\end{pmatrix} \, , \quad 	\sigma^{3}_{AA'} = \begin{pmatrix} 0& 0 \\ 1 & 0
\end{pmatrix} \, .
\end{equation}
Of course, a conventional way to translate from the Newman-Penrose frame $\left\lbrace k^{\mu}, n^{\mu}, m^{\mu}, \bar{m}^{\mu} \right \rbrace$ to an orthonormal frame $\left\lbrace E_0^{\mu}, E_1^{\mu}, E_2^{\mu}, E_3^{\mu} \right \rbrace$ is
\begin{equation}
k^{\mu} = \frac{1}{\sqrt{2}}\left(E^{\mu}_0 +  E^{\mu}_3\right) \, , \quad 	n^{\mu} = \frac{1}{\sqrt{2}}\left(E^{\mu}_0 -  E^{\mu}_3\right) \, ,
\end{equation}
\begin{equation}\label{eqn:mframechoice}
m^{\mu} = \frac{1}{\sqrt{2}}\left(-E^{\mu}_1 + i E^{\mu}_2\right) \, , \quad 	\bar{m}^{\mu} = \frac{1}{\sqrt{2}}\left(-E^{\mu}_1 - i E^{\mu}_2\right) \, .
\end{equation}
In the orthonormal basis $\left\lbrace E_0^{\mu}, E_1^{\mu}, E_2^{\mu}, E_3^{\mu} \right \rbrace$ (but still with $\left\lbrace o^A, \iota^A \right \rbrace$  for the two-spinors), the Infeld-Van der Waerden symbols are given by
\begin{equation}\label{eqn:IvWexplicitortho}
\sigma^{t}_{AA'} = \frac{1}{\sqrt{2}}\begin{pmatrix} 1& 0 \\ 0 & 1
\end{pmatrix} \, , \quad 	\sigma^{x}_{AA'} = -\frac{1}{\sqrt{2}}\begin{pmatrix} 0& 1 \\ 1 & 0
\end{pmatrix} \, , \quad 	\sigma^{y}_{AA'} = \frac{1}{\sqrt{2}}\begin{pmatrix} 0& -i \\ i & 0
\end{pmatrix} \, , \quad 	\sigma^{z}_{AA'} = \frac{1}{\sqrt{2}}\begin{pmatrix} 1& 0 \\ 0 & -1
\end{pmatrix} \, ,
\end{equation}
where we have used $t$,$x$,$y$,$z$ instead of $0$,$1$,$2$,$3$ in the labels to avoid confusion with \eqref{eqn:IvWexplicitnull}.

Let us explicitly point out that if we desire, as in \eqref{eqn:complexz},
\begin{equation}
z = \frac{1}{\sqrt{2}}(x^1 + i x^2) =  \frac{1}{\sqrt{2}}(x + i y) \, ,
\end{equation}
and
\begin{equation}
m = \partial_z \, ,
\end{equation}
then, as a consequence (due to the mostly minus signature)
\begin{equation}
m = \frac{1}{\sqrt{2}}(-\partial_x + i \partial_y) \, .
\end{equation}
Together with our choice to define, at the level of the vector field \eqref{eqn:vdefplanewave}
\begin{equation}
E_a = \partial_a \, ,
\end{equation}
this explains the signs in \eqref{eqn:mframechoice}.

Coming back to the Newman-Penrose frame $\left\lbrace k^{\mu}, n^{\mu}, m^{\mu}, \bar{m}^{\mu} \right \rbrace$, note that it is a frame for a metric of the desired signature as
\begin{equation}
g_{\alpha \beta} = g_{AA' BB'} = \epsilon_{AB}\bar{\epsilon}_{A'B'} \, ,
\end{equation} 
and therefore
\begin{equation}
g_{AA' BB'}  = \left(o_A \bar{o}_{A'} \iota_B \bar{\iota}_{B'}+o_B \bar{o}_{B'} \iota_A \bar{\iota}_{A'}\right) - \left(o_A \bar{\iota}_{A'} \iota_B \bar{o}_{B'}+o_B \bar{\iota}_{B'} \iota_A \bar{o}_{A'}\right)  \, ,
\end{equation}
from which follows
\begin{equation}\label{eqn:ginframe}
g_{\alpha \beta} = 2 k_{(\alpha}n_{\beta)} - 2 m_{(\alpha}\bar{m}_{\beta)} \, .
\end{equation}
Here $\left( \ldots \right)$ in indices denotes a symmetrization, which we take to be weighted by the appropriate permutation factor.

Instead of working from a dyad to a Newman-Penrose frame, we will usually work the other way around; starting from a Newman-Penrose frame to define a dyad. The latter dyad will not be uniquely specified by this frame choice but that (phase\footnote{The ambiguity in $\left\lbrace o^A, \iota^A \right \rbrace$ for a given Newman-Penrose frame is actually reduced to a sign taking into account that we impose \eqref{eqn:dyadnorm}.}) ambiguity will not be important for our purposes. Moreover, in the main text, there will generally be two preferred frames and associated dyads: one coming from a preferred frame attached to a general spacetime, and another related to a frame on a chosen null geodesic of such a spacetime. The latter will then subsequently play the role of the former for the Penrose limit plane wave spacetime. Throughout the main text, we will typically write the preferred dyad of a general spacetime as $\left\lbrace o^A, \iota^A \right \rbrace$ while reserving $\left\lbrace \alpha^A, \beta^A \right \rbrace$ or $\left\lbrace \alpha^{(\gamma)}{}^A, \beta^{(\gamma)}{}^A \right \rbrace$ for dyads attached to a null geodesic $\gamma$ of the related Penrose limit plane wave.

Aside from the metric and basis vector fields, a decomposition of the Riemann tensor and anti-symmetric field strengths into spinor components plays an important role in the Weyl double copy and thus in the main text. For an anti-symmetric tensor field $F_{\alpha \beta}$, the main observation is that it can be decomposed into
\begin{equation}\label{eqn:Finspinors}
F_{\alpha \beta} = f_{AB} \bar{\epsilon}_{A' B'} +  \bar{f}_{A'B'} \epsilon_{A B} \, .
\end{equation}
Similarly, the Weyl tensor can be decomposed into
\begin{equation}\label{eqn:Cinspinors}
C_{\alpha \beta \gamma \delta} = \Psi_{ABCD}\bar{\epsilon}_{A' B'}  \bar{\epsilon}_{C' D'} + \bar{\Psi}_{A'B'C'D'} \epsilon_{A B}  \epsilon_{C D} \, . 
\end{equation}
The relation between the decompositions \eqref{eqn:Finspinors} and \eqref{eqn:Cinspinors} is key to the Weyl double copy which, given additionally the complex ``zeroth copy'' scalar $S$ gives a Weyl tensor \eqref{eqn:Cinspinors} in terms of a 2-form \eqref{eqn:Finspinors} as \eqref{WeylDCRelation}
\begin{equation}
C^{(\rm Weyl \, DC)}_{\alpha \beta \gamma \delta} = \frac{1}{S}f_{(AB} f_{CD)}\bar{\epsilon}_{A' B'}  \bar{\epsilon}_{C' D'} + \frac{1}{S^*}\bar{f}_{(A'B'} \bar{f}_{C'D')} \epsilon_{A B}  \epsilon_{C D} \, . 
\end{equation}

Various curvature and connection components with respect to a null basis have been conventionally associated to particular notations in what is usually called the Newman-Penrose formalism. While we will not lean on this heavily, we next summarize these symbols as used in the main text.

\subsection{Newman-Penrose formalism}\label{app:NP}

We now introduce our conventions for the Newman-Penrose formalism, which is mostly an exercise in decomposing spinors into the dyad $\left\lbrace o^A, \iota^A \right \rbrace$ and tensors into the Newman-Penrose frame $\left\lbrace k^{\mu}, n^{\mu}, m^{\mu}, \bar{m}^{\mu} \right \rbrace$. We will make constant use of \eqref{eqn:dyadnorm}, \eqref{eqn:epsdown}, \eqref{eqn:assocaiteNPframe}, and \eqref{eqn:ginframe}.

Let us start with an anti-symmetric tensor field $F_{\alpha \beta}$, as decomposed into spinors in \eqref{eqn:Finspinors}. The Newman-Penrose field strength scalars $f_0$, $f_1$ and $f_2$ with respect to a dyad  $\left\lbrace o^A, \iota^A \right \rbrace$ can be defined in terms of the two-spinor formulation as
\begin{equation}
f_{AB} = f_2 o_A o_B - 2 f_1 o_{(A} \iota_{B)} + f_0 \iota_A \iota_B \, .
\end{equation}
Using \eqref{eqn:Finspinors}, this is expressed in terms of the tensor itself with respect to the associated null frame $\left\lbrace k^{\mu}, n^{\mu}, m^{\mu}, \bar{m}^{\mu} \right \rbrace$ as
\begin{equation}\label{eqn:FinNP}
F_{\alpha \beta} = 2 f_2 k_{[\alpha} m_{\beta]}- 2 f_1 \left(k_{[\alpha} n_{\beta]}-m_{[\alpha}\bar{m}_{\beta]}\right) + 2 f_0 \bar{m}_{[\alpha} n_{\beta]} +  cc  \, ,
\end{equation}
where we will use ``cc'' to abbreviate the complex conjugate piece of the expression. A consequence of \eqref{eqn:FinNP}, that we will regularly use for calculations, is
\begin{equation}\label{eqn:finFprojected}
f_2 = - F_{\alpha \beta} n^{\alpha} \bar{m}^{\beta} \, , \quad f_1 = - \frac{1}{2}F_{\alpha \beta} \left(n^{\alpha} k^{\beta}+  m^{\alpha} \bar{m}^{\beta}\right) \, , \quad f_0 =  F_{\alpha \beta} k^{\alpha} m^{\beta} \, .
\end{equation}

The analogous procedure for the Riemann tensor is somewhat more (algebraically) complicated but is reviewed in detail in Chapter 4 of \cite{penrose1984spinors}. First, introduce the curvature spinors $\Psi_{ABCD}$ and $\Phi_{ABC'D'}$ as well as their complex conjugates and the scalar $\Lambda$ by the decomposition
\begin{equation}
\begin{aligned}
-R_{\alpha \beta \gamma \delta} = \left(\Psi_{ABCD} 	+ \Lambda  \left(\epsilon_{AC}\epsilon_{BD}+\epsilon_{AD}\epsilon_{BC}\right) \right)\bar{\epsilon}_{A'B'}\bar{\epsilon}_{C'D'} + \Phi_{ABC'D'} \bar{\epsilon}_{A'B'} \epsilon_{CD}
+ cc\, .
\end{aligned}
\end{equation}
Now, the Weyl and Ricci scalars are again essentially just components of $\Psi_{ABCD}$ and $\Phi_{ABC'D'}$ with respect to the dyad $\left\lbrace o^A, \iota^A \right \rbrace$, similar to \eqref{eqn:FinNP} for the Maxwell scalars\footnote{Let us emphasize again that these are identical to the definitions of \cite{penrose1984spinors}, given that we introduce an extra minus sign to correct for the different curvature convention (see (4.11.9) and (4.11.10)). On the other hand, these differ in sign from say \cite{Pound:2021qin}, where a mostly plus signature is used.}
\begin{equation}\label{eqn:psidefs}
\begin{aligned}
\Psi_0 &= \Psi_{ABCD} o^Ao^Bo^Co^D =- C_{\alpha \beta \gamma \delta} k^{\alpha}m^{\beta}k^{\gamma} m^{\delta} \, , \\
\Psi_1 &= \Psi_{ABCD} o^Ao^Bo^C\iota^D = -C_{\alpha \beta \gamma \delta} k^{\alpha}n^{\beta}k^{\gamma} m^{\delta} \, , \\
\Psi_2 &= \Psi_{ABCD} o^Ao^B\iota^C\iota^D = -C_{\alpha \beta \gamma \delta} k^{\alpha}m^{\beta}\bar{m}^{\gamma} n^{\delta} \, , \\
\Psi_3 &= \Psi_{ABCD} o^A\iota^B\iota^C\iota^D = -C_{\alpha \beta \gamma \delta} k^{\alpha}n^{\beta}\bar{m}^{\gamma}n^{\delta} \, , \\
\Psi_4 &= \Psi_{ABCD} \iota^A\iota^B\iota^C\iota^D =- C_{\alpha \beta \gamma \delta} n^{\alpha}\bar{m}^{\beta}n^{\gamma} \bar{m}^{\delta} \, ,  
\end{aligned}
\end{equation}
and
\begin{equation}\label{eqn:phidefs}
\begin{aligned}
\Phi_{00} &= \Phi_{ABC'D'} o^Ao^B\bar{o}^{C'}\bar{o}^{D'} = \frac{1}{2}R_{\mu \nu} k^{\mu} k^{\nu} \, ,\\
\Phi_{10} &= \Phi_{ABC'D'} o^A\iota^B\bar{o}^{C'}\bar{o}^{D'} = \frac{1}{2}R_{\mu \nu} k^{\mu} \bar{m}^{\nu} \, ,\\
\Phi_{20} &= \Phi_{ABC'D'} \iota^A\iota^B\bar{o}^{C'}\bar{o}^{D'} = \frac{1}{2}R_{\mu \nu} \bar{m}^{\mu} \bar{m}^{\nu} \, , \\
\text{etc.}
\end{aligned}
\end{equation}
We do not enumerate all of the $\Phi_{ABC'D'}$ components as we do not need most of them explicitly. On the other hand, the component $\Phi_{00}$ is important as it is the only (potentially) non-vanishing component of the Ricci spinor in a plane wave spacetime, as discussed around \eqref{eqn:Phi00penrose} in the main text. Similarly, the only non-vanishing Weyl scalar, in our convention, of a type N spacetime is $\Psi_0$.

Given that the Weyl curvature $C_{\alpha \beta \gamma \delta}$ is given by
\begin{equation}
\begin{aligned}
-C_{\alpha \beta \gamma \delta} = \Psi_{ABCD} \bar{\epsilon}_{A'B'}\bar{\epsilon}_{C'D'} 
+ \bar{\Psi}_{A'B'C'D'} \epsilon_{AB}\epsilon_{CD} \, ,
\end{aligned}
\end{equation}
let us check some of the equalities in \eqref{eqn:psidefs}. In particular, for the main text, it is important that the null Fermi coordinate approach to the Penrose limit \eqref{eqn:penroseframe} \cite{Blau:2006ar,blau2011plane}
\begin{equation}\label{eqn:appHPenrose}
H_{ab}(u) = \left(R_{\mu \nu \alpha \beta} E^{\mu}_a u^{\nu} E^{\alpha}_b u^{\beta} \right)_{|\gamma} \, .
\end{equation}
is equivalent to the spinor prescription \eqref{eqn:penrosespinor} \cite{Tod:2019urw}
\begin{equation}\label{eqn:app:penrosespinor}
\Psi(u) = \left(\Psi_{ABCD}\alpha^A \alpha^B \alpha^C \alpha^D\right)_{|\gamma} \, , \quad \Phi(u) = \left(\Phi_{ABA'B'}\alpha^A \alpha^B \bar{\alpha}^{A'} \bar{\alpha}^{B'}\right)_{|\gamma} \, ,
\end{equation}
if we identify, from the plane wave, as in \eqref{eqn:HBrinkmannincomplex}
\begin{equation}\label{eqn:app:H}
H_{ab} = - \begin{pmatrix}
\frac{1}{2}\left(\Psi(u)+ 2 \Phi(u)  + \bar{\Psi}(u)\right) & 	\frac{i}{2}\left(\Psi(u) - \bar{\Psi}(u)\right)\\
\frac{i}{2}\left(\Psi(u) - \bar{\Psi}(u)\right) 	&  \frac{1}{2}\left(-\Psi(u)+ 2 \Phi(u)  - \bar{\Psi}(u)\right) 
\end{pmatrix}  \, ,
\end{equation}
with, on the plane wave (\eqref{eqn:NPsi} and \eqref{eqn:Phi00penrose})
\begin{equation}\label{eqn:app:Psiplane}
\Phi(u) = \Phi^{(\gamma)}_{00} \, , \quad \Psi(u) = \Psi^{(\gamma)}_0 \, ,
\end{equation}
and, as in \eqref{eqn:complextransversebasis}
\begin{equation}\label{eqn:app:uplane}
u^{\delta} = \alpha^D \bar{\alpha}^{D'} \, , \quad M^{\delta} = -\frac{1}{\sqrt{2}}\left(E^{\delta}_1 - i E^{\delta}_2 \right) = \alpha^D \bar{\beta}^{D'} \, .
\end{equation}

First observe that
\begin{equation}\label{eqn:psi0proj}
\begin{aligned}
\Psi^{(\gamma)}_0 = -C_{\alpha \beta \delta \gamma} M^{\alpha} u^{\beta} M^{\delta} u^{\gamma} &=  \left(\Psi_{ABCD}\bar{\epsilon}_{A'B'}\bar{\epsilon}_{C'D'}+cc\right)\alpha^A \alpha^B \alpha^C  \alpha^D \bar{\beta}^{A'} \bar{\alpha}^{B'}  \bar{\beta}^{C'} \bar{\alpha}^{D'} \\
&=  \left(\Psi_{ABCD}\alpha^A \alpha^B \alpha^C \alpha^D\right)  \, ,
\end{aligned}
\end{equation}
and
\begin{equation}\label{eqn:app:checkHtrace}
\begin{aligned}
\Phi^{(\gamma)}_{00} = \frac{1}{2}R_{\alpha \beta} u^{\alpha} u^{\beta}  &=  \frac{1}{2}\left(\Phi_{ABC'D'} \bar{\epsilon}_{A'B'} \epsilon_{CD} + cc\right) \alpha^B \bar{\alpha}^{B'}  \alpha^C \bar{\alpha}^{C'}  \epsilon^{AD}\bar{\epsilon}^{A'D'} \\
&=  \Phi_{ABC'D'} \alpha^B  \alpha^A \bar{\alpha}^{C'} \bar{\alpha}^{D'}    \, .
\end{aligned}
\end{equation}
\eqref{eqn:psi0proj} and \eqref{eqn:appHPenrose} imply in particular the traceless components of \eqref{eqn:app:H}
\begin{equation}
\begin{aligned}
C_{\alpha \beta \delta \gamma} E^{\alpha}_1 u^{\beta} E^{\delta}_1 u^{\gamma} &=   
-\frac{1}{2}\left(\Psi^{(\gamma)}_0  +\bar{\Psi}^{(\gamma)}_0\right) \\
C_{\alpha \beta \delta \gamma} E^{\alpha}_1 u^{\beta} E^{\delta}_2 u^{\gamma} &=   \frac{1}{2i} \left(\Psi^{(\gamma)}_0 - \bar{\Psi}^{(\gamma)}_0\right) \\
C_{\alpha \beta \delta \gamma} E^{\alpha}_2 u^{\beta} E^{\delta}_2 u^{\gamma} &=   \frac{1}{2}\left(\Psi^{(\gamma)}_0  +\bar{\Psi}^{(\gamma)}_0\right)  \, ,
\end{aligned}
\end{equation}
while the trace follow from  \eqref{eqn:app:checkHtrace}  and \eqref{eqn:appHPenrose}  by
\begin{equation}
\begin{aligned}
H_{11}+H_{22} = R_{\alpha \beta \delta \gamma} u^{\beta} u^{\gamma} \delta^{ab} E_a^{\alpha} E_b^{\delta} &= R_{\alpha \beta \delta \gamma} u^{\beta} u^{\gamma} (-g^{\alpha \delta}) \\
&= - 2 \Phi^{(\gamma)}_{00}  \, .
\end{aligned}
\end{equation}
Note that it was our choice to build the dyad $\left\lbrace \alpha^A, \beta^B \right\rbrace$ with
\begin{equation}
	\epsilon_{AB} \alpha^A \beta^B = 1 \, ,
\end{equation}
on the null geodesic together with \eqref{eqn:app:penrosespinor} and \eqref{eqn:psidefs} that led us to use the convention that type N spacetimes only have non-vanishing $\Psi_0$. 

To take covariant derivatives of spinors in a dyad basis, we need the spin coefficients
\begin{equation}\label{eqn:spinorspincoeffs}
\gamma_{AA' C}{}^{B} = \epsilon_{D}{}^B \nabla_{AA'}\epsilon_{C}{}^D \, ,
\end{equation}
such that in particular
\begin{equation}
\begin{aligned}
\nabla_{\mu}\kappa^A &= \nabla_{\mu}\left(\kappa^o o^A + \kappa^{\iota}\iota^A\right) \\
&= \left(\partial_{\mu}\kappa^o  + \gamma_{\mu o}{}^{o}\kappa^o + \gamma_{\mu \iota}{}^{o}\kappa^{\iota}\right) o^A + \left(\partial_{\mu} \kappa^{\iota}+ \gamma_{\mu o}{}^{\iota}\kappa^o + \gamma_{\mu \iota}{}^{\iota}\kappa^{\iota} \right) \iota^A \, .
\end{aligned}
\end{equation}
One way to compute \eqref{eqn:spinorspincoeffs} is using the associated Newman-Penrose frame. Introducing first the Newman-Penrose spin coefficient symbols of \cite{penrose1984spinors,Penrose:1986ca}, which give names to the individual components,
\begin{alignat}{5}
\kappa &= m^{\alpha} k^{\mu} \nabla_{\mu} k_{\alpha} \, , \quad \epsilon&=& \frac{1}{2}\left(n^{\alpha} k^{\mu} \nabla_{\mu} k_{\alpha} + m^{\alpha} k^{\mu} \nabla_{\mu} \bar{m}_{\alpha}\right) \, , \quad &\gamma'&= \frac{1}{2}\left(k^{\alpha} k^{\mu} \nabla_{\mu} n_{\alpha} + \bar{m}^{\alpha} k^{\mu} \nabla_{\mu} m_{\alpha}\right) \, , \nonumber \\
\rho &= m^{\alpha} \bar{m}^{\mu} \nabla_{\mu} k_{\alpha} \, , \quad \alpha &=& \frac{1}{2}\left(n^{\alpha} \bar{m}^{\mu} \nabla_{\mu} k_{\alpha} + m^{\alpha} \bar{m}^{\mu} \nabla_{\mu} \bar{m}_{\alpha}\right) \, , \quad &\beta'&  = \frac{1}{2}\left(k^{\alpha} \bar{m}^{\mu} \nabla_{\mu} n_{\alpha} + \bar{m}^{\alpha} \bar{m}^{\mu} \nabla_{\mu} m_{\alpha}\right) \, , \nonumber \\
\sigma &= m^{\alpha} m^{\mu} \nabla_{\mu} k_{\alpha} \, , \quad \beta &=& \frac{1}{2}\left(n^{\alpha} m^{\mu} \nabla_{\mu} k_{\alpha} + m^{\alpha} m^{\mu} \nabla_{\mu} \bar{m}_{\alpha}\right) \, , \quad &\alpha'&= \frac{1}{2}\left(k^{\alpha} m^{\mu} \nabla_{\mu} n_{\alpha} + \bar{m}^{\alpha} m^{\mu} \nabla_{\mu} m_{\alpha}\right) \, , \nonumber \\
\tau &= m^{\alpha} n^{\mu} \nabla_{\mu} k_{\alpha} \, , \quad \gamma&=& \frac{1}{2}\left(n^{\alpha} n^{\mu} \nabla_{\mu} k_{\alpha} + m^{\alpha} n^{\mu} \nabla_{\mu} \bar{m}_{\alpha}\right) \, , \quad &\epsilon'&= \frac{1}{2}\left(k^{\alpha} n^{\mu} \nabla_{\mu} n_{\alpha} + \bar{m}^{\alpha} n^{\mu} \nabla_{\mu} m_{\alpha}\right) \, , \nonumber \\
\tau' &= \bar{m}^{\alpha} k^{\mu} \nabla_{\mu} n_{\alpha} \, , \quad \sigma' &=& \bar{m}^{\alpha} \bar{m}^{\mu} \nabla_{\mu} n_{\alpha} \, , \quad \rho'  = \bar{m}^{\alpha} m^{\mu} \nabla_{\mu} n_{\alpha} \, , \quad &\kappa'&  =\bar{m}^{\alpha} n^{\mu} \nabla_{\mu} n_{\alpha} \, ,  \label{eqn:spincoeffssymbols}
\end{alignat}
then
\begin{alignat}{7}
\gamma_{oo' o}{}^{o} &= \epsilon \, , \quad \gamma_{oo' o}{}^{\iota} &=& -\kappa \, , \quad \gamma_{oo' \iota}{}^{o} &=& -\tau' \, , \quad \gamma_{oo' \iota}{}^{\iota} &=& \gamma' \nonumber \\
\gamma_{\iota o' o}{}^{o} &= \alpha \, , \quad \gamma_{\iota o' o}{}^{\iota} &=& -\rho \, , \quad \gamma_{\iota o' \iota}{}^{o} &=& -\sigma' \, , \quad \gamma_{\iota o' \iota}{}^{\iota} &=& \beta'  \nonumber  \\
\gamma_{o\iota' o}{}^{o} &= \beta \, , \quad \gamma_{o\iota' o}{}^{\iota} &=& -\sigma \, , \quad \gamma_{o\iota' \iota}{}^{o} &=& -\rho' \, , \quad \gamma_{o\iota' \iota}{}^{\iota} &=& \alpha'  \nonumber  \\
\gamma_{\iota \iota' o}{}^{o} &= \gamma \, , \quad \gamma_{\iota \iota' o}{}^{\iota} &=& -\tau \, , \quad \gamma_{\iota \iota' \iota}{}^{o} &=& -\kappa' \, , \quad \gamma_{\iota \iota' \iota}{}^{\iota} &=& \epsilon'  \, .  \label{eqn:spincoeffs}
\end{alignat}
Although we will not generally need the expressions \eqref{eqn:spincoeffs} and symbols \eqref{eqn:spincoeffssymbols} in the main text, they are useful performing explicit calculations for the examples. Moreover, the expressions above quoted from \cite{penrose1984spinors,Penrose:1986ca} retain the possibility to work with an unnormalized dyad, while we always impose \eqref{eqn:dyadnorm}. Therefore
\begin{equation}
\epsilon = - \gamma' \, , \quad \alpha = -\beta' \, , \quad \beta = - \alpha' \, , \quad \gamma = - \epsilon' \, .
\end{equation}

\section{Examples: further details}\label{app:examples}

In this appendix, we gather some technical details that are used in Section \ref{sec:examples}. We start with two, closely related, common problems one encounters in all the examples. These end up being essentially a combination of simple algebraic problems and first order linear, ordinary differential equations. However, the solutions will generically still become rather intractable symbolically and may not have a very satisfying analytic form.

The first of these problems is to express
\begin{equation}\label{eqn:uisalpha2}
u^{A A'} = \alpha^A \bar{\alpha}^{A'} \, .
\end{equation}
With respect to a null basis $\left\lbrace K,N,M,\bar{M} \right\rbrace$ and the associated dyad $\left\lbrace o^A, \iota^{A} \right\rbrace$, such that in particular $\alpha^A = \alpha^o o^A + \alpha^{\iota} \iota^A$
\begin{equation}\label{eqn:alphainu}
\begin{pmatrix}
u^{\mu}N_{\mu} & -u^{\mu}\bar{M}_{\mu}  \\
-u^{\mu}M_{\mu} & u^{\mu}K_{\mu}
\end{pmatrix} =  \begin{pmatrix}
\alpha^o \bar{\alpha}^o & \alpha^{o} \bar{\alpha}^{\iota} \\
\alpha^{\iota} \bar{\alpha}^{o} & \alpha^{\iota} \bar{\alpha}^{\iota} 
\end{pmatrix} = \begin{pmatrix}
\alpha^o \\ \alpha^{\iota}
\end{pmatrix} \begin{pmatrix} \bar{\alpha}^o  & \bar{\alpha}^{\iota} \end{pmatrix}\, .
\end{equation}
The absolute values of the spinor components and a relative phase are immediate. For instance
\begin{equation} \label{eqn:alphadecomp}
\alpha^A = e^{i S}\left(|\alpha^o| o^A + e^{i \delta}|\alpha^{\iota}| \iota^A\right)
\end{equation}
where\footnote{We assume $u^{\mu}$ is generic; $u^{\nu}N_{\mu}$, $u^{\mu}K_{\mu} \neq 0$. If this does not hold the problem simplifies even further.}
\begin{equation}\label{eqn:spinorsfromu} 
|\alpha^o|^2 = u^{\mu}N_{\mu} \, , \quad 	|\alpha^{\iota}|^2 = u^{\mu}K_{\mu} \, , \quad e^{i \delta} = -\frac{u^{\mu}M_{\mu}}{\sqrt{|u^{\mu}N_{\mu} ||u^{\mu}K_{\mu}|}} \, .
\end{equation}
In addition to $\alpha^A$, we often want to find a $\beta^A$ such that
\begin{equation}
\epsilon_{AB}\alpha^A \beta^B = 1 \, .
\end{equation}
For an $\alpha^A$ as in \eqref{eqn:alphadecomp}, such a $\beta^A$ can be given by
\begin{equation}\label{eqn:betadecomp}
\beta^A = \frac{e^{-i \delta-i S}}{2|\alpha^{\iota}| |\alpha^o|}\left( |\alpha^o| o^A - e^{i \delta} |\alpha^{\iota}| \iota^A\right)\, .
\end{equation}
A fundamental ambiguity will always remain in the choice of $e^{i S}$, in the sense that it is simply not fixed by only specifying $u^{\mu}$. However, we will want to impose that 
\begin{equation}\label{eqn:alphaode}
u^{\mu}\nabla_{\mu} \alpha^A = 0 \, .
\end{equation}
Therefore this ambiguity cannot take an arbitrary functional form, as would be allowed by \eqref{eqn:uisalpha2}. Let $\tilde{\alpha}^A$ and $\tilde{\beta}^A$ be two-spinors constructed as in \eqref{eqn:alphadecomp} and \eqref{eqn:betadecomp} with $S = 0$. From \eqref{eqn:alphaode}, the appropriate choice of $S$ can be constructed by solving
\begin{equation}\label{eqn:Seqn}
u^{\mu} \nabla_{\mu} S  = i\epsilon_{AB}  \left(u^{\mu}\nabla_{\mu}\tilde{\alpha}^A\right)\tilde{\beta}^B \, .
\end{equation}
Note that by \eqref{eqn:uisalpha2}, \eqref{eqn:Seqn} can be written in terms of the spin-coefficient $\epsilon$  (see \eqref{eqn:spincoeffssymbols}) with respect to the dyad $\left\lbrace \tilde{\alpha}, \tilde{\beta} \right\rbrace$ (we therefore denote it as $\tilde{\epsilon}$) as
\begin{equation}\label{eqn:Seqn2}
u^{\mu} \nabla_{\mu} S  - i \tilde{\epsilon} = 0\, .
\end{equation}

The second reoccurring algebraic problem is to find basis vectors $E_i{}^a$ which square to a given (real) symmetric positive definite matrix $C_{ij}$
\begin{equation}\label{eqn:eqrefbasis}
C_{ij} = E_i{}^a \delta_{ab} E_j{}^b = \sum_a E_i{}^a E_j{}^a = {\boldsymbol E}\cdot \left({\boldsymbol{E}}\right)^T \, .
\end{equation}
One natural approach would be to first diagonalize $C_{ij}$, solve the trivial version of the above, and subsequently transform back. Slightly more efficiently, one can simply make an ansatz for $E_i{}^a$ to be triangular and solve the resulting set of equations from \eqref{eqn:eqrefbasis} step-by-step. In other words, one can construct $E_i{}^a$ from a type of Gram-Schmidt orthogonalization procedure.

As with the previous problem however, the complication for our purposes actually arises because in addition to \eqref{eqn:eqrefbasis}, we need $E_i{}^a$ to be parallel propagated which, as discussed in the main text, results in the condition \eqref{eqn:Esymm}
\begin{equation}\label{eqn:Esymmetryconstraint}
\dot{E}^i{}_a  E_{ib} = \dot{E}^i{}_b  E_{ia} \, .
\end{equation}
Given a solution to $\tilde{E}_i{}^a$ which does not solve the above, the problem is to find a time-dependent orthogonal transformation such that
\begin{equation}
E_j{}^a  = O_j{}^i \tilde{E}_i{}^a \, .
\end{equation}
does satisfy \eqref{eqn:Esymmetryconstraint}. 

In a four dimensional spacetime, there are only two transverse dimensions and one can rather easily parameterize explicitly
\begin{equation}
{\boldsymbol O} = \begin{pmatrix}
\cos{\theta} & \sin{\theta} \\
-\sin{\theta} & \cos{\theta} 
\end{pmatrix} \, ,
\end{equation}
and solve the ordinary differential equation for $\theta$ that will ensure \eqref{eqn:Esymmetryconstraint}. We will not work this out explicitly here but in practice this ends up being identical to the first problem we discussed, as once we have found a parallel propagated dyad  $\left\lbrace \alpha^A, \beta^{A} \right\rbrace$, we can readily construct a parallel propagated frame.

\subsection{The Schwarzschild black hole}\label{app:Schwarzschild}

In this section, for completeness, we provide some additional details related to the Schwarzschild example in Section \ref{sec:Schwarzschild}. First, using the frame \eqref{eqn:SPNframe} and definitions \eqref{eqn:spincoeffssymbols}, the spin coefficients are given by 
\begin{alignat}{7}
\kappa &= 0 \, ,  \quad \sigma &=& 0 \, , \quad &\rho& = \frac{1}{\sqrt{2} r} \, , \quad &\tau& = 0  \, , \nonumber \\
\kappa' &= 0 \, ,  \quad \sigma' &=& 0 \, , \quad &\rho'& = -\frac{1}{\sqrt{2} r}(1-\frac{2M}{r}) \, , \quad &\tau'& = 0 \, , \nonumber  \\
\epsilon &= 0 \, ,  \quad \epsilon' &=& \frac{M}{\sqrt{2}r^2} \, , \quad &\beta& = -\frac{\cot{\theta}}{2\sqrt{2}} \, , \quad &\beta'& = -\frac{\cot{\theta}}{2\sqrt{2}}  \, ,  \label{eqn:Sspincoeffssymbols}
\end{alignat}
For comparisons, using instead the flat frame, given also in  \eqref{eqn:SPNframe}, on the flat background in
Kerr-Schild coordinates, we of course just find \eqref{eqn:Sspincoeffssymbols} with $M \to 0$.

The flat space spin coefficients are just telling us that the null directions in flat space generate radially propagating null congruences. Compared to  \eqref{eqn:Sspincoeffssymbols}, the main difference is that the expansion of the outgoing congruence is reduced by the black hole. In the Geroch-Held-Penrose (or compacted spin coefficient) formalism \cite{geroch1973space}, $\beta$ and $\beta'$ act as gauge fields for rotations in the transverse planes, $m^{\mu} \to e^{i \varphi(x)} m^{\mu}$ for some function $\varphi(x)$. In this spherically symmetric Schwarzchild example these are related to the induced connection on the transverse spheres. $\epsilon$ and $\epsilon'$ play a similar role for the timelike surface spanned by the time and radial coordinates, $\hat{t}$ and $r$, and the transformations $k^{\mu} \to e^{ \varphi(x)} k^{\mu}$ and $n^{\mu} \to e^{-\varphi(x)} n^{\mu}$. In the flat case, this surface is just a flat plane, so $\epsilon$ and $\epsilon'$ vanish, but the plane is curved by the black hole inducing also a non-zero value for $\epsilon'$. The fact that, contrary to $\rho'$ and $\epsilon'$, $\rho$ and $\epsilon$ are unchanged between the black hole and the flat space is a consequence of the Kerr-Schild form of the metric and the frame.

In order to check the solution \eqref{eqn:Schi} to the valence-2 Killing spinor equations \eqref{eqn:valence2}, let us write out these equations explicitly for the ansatz
\begin{equation}\label{eqn:chiansatz}
\chi_{AB} = 2 \chi_{01}(r) o_{(A} \iota_{B)} \, .
\end{equation}
When we evaluate the Killing spinor equations \eqref{eqn:valence2} on \eqref{eqn:chiansatz}, setting the vanishing spin coefficients in \eqref{eqn:Sspincoeffssymbols} to zero, we find
\begin{equation}\label{eqn:S2killingeqn}
\begin{aligned}
\left(\frac{d}{dr} - \sqrt{2}\rho\right)\chi_{01} = 0  \, , \quad \left(\left(1-\frac{2M}{r}\right)\frac{d}{dr} + \sqrt{2} \rho'\right)\chi_{01} = 0 \, .
\end{aligned}
\end{equation}
A solution to  \eqref{eqn:valence2} is thus given by $\chi_{01}(r) \propto r$. 

Next, we provide a first step to check that the two-spinor $\alpha^A$, given in \eqref{eqn:alpha}, satisfies the
parallel transport equation \eqref{eqn:Stangentspinor} along the null geodesic $\gamma$ with tangent $u^{\alpha} =
\alpha^A \bar{\alpha}^{A'} $, given in Kerr-Schild coordinates in \eqref{eqn:Su} and \eqref{eqn:uschwarzschild}.
Simply writing out the parallel transport equation while using that the components of $\alpha^A$ have no explicit
dependence on $\hat{t}$ and $\phi$, and setting the vanishing spin coefficients in \eqref{eqn:Sspincoeffssymbols}
to zero already, we find,
\begin{equation}\label{eqn:Salphaequations}
\begin{aligned}
-\iota_A	\alpha^B \bar{\alpha}^{B'} \nabla_{BB'}\alpha^A &= \left\lbrack -\frac{1}{\sqrt{2} r} |\alpha^o|^2  \frac{\partial \alpha^o}{\partial r} + |\alpha^{\iota}|^2 \left( \frac{1}{\sqrt{2}} \left(1-\frac{2M}{r}\right) \frac{\partial \alpha^{o}}{\partial r} - \epsilon' \alpha^{o}\right) \right \rbrack \\
&+  \left\lbrack \alpha^o\bar{\alpha}^{\iota'} \left(-\frac{1}{\sqrt{2} r}\frac{\partial \alpha^o}{\partial \theta} -\rho' \alpha^{\iota} + \beta \alpha^o \right)+  \alpha^{\iota}\bar{\alpha}^{o'} \left(-\frac{1}{\sqrt{2} r}\frac{\partial \alpha^o}{\partial \theta} -\beta' \alpha^o \right) \right\rbrack \, , \\
o_A	\alpha^B \bar{\alpha}^{B'} \nabla_{BB'}\alpha^A &= \left\lbrack -\frac{1}{\sqrt{2} r}|\alpha^o|^2  \frac{\partial \alpha^{\iota}}{\partial r} + |\alpha^{\iota}|^2 \left( \frac{1}{\sqrt{2}} \left(1-\frac{2M}{r}\right) \frac{\partial \alpha^{\iota}}{\partial r} + \epsilon' \alpha^{\iota}\right) \right \rbrack  \\
&+ \left\lbrack \alpha^o\bar{\alpha}^{\iota'} \left(-\frac{1}{\sqrt{2} r}\frac{\partial \alpha^{\iota}}{\partial \theta} -\beta \alpha^{\iota}\right)+  \alpha^{\iota}\bar{\alpha}^{o'} \left(-\frac{1}{\sqrt{2} r}\frac{\partial \alpha^{\iota}}{\partial \theta} -\beta' \alpha^{\iota} - \rho \alpha^o \right) \right\rbrack \, .
\end{aligned}
\end{equation}
At this point we can just insert \eqref{eqn:alpha} and grind through the algebra, which we have done. Alternatively, we could pretend for a moment that we did not have or know of the Killing spinor $\chi_{AB}$ from which we could derive \eqref{eqn:alpha} by demanding \eqref{eqn:Skillingspinorconserved} is constant. Then, we would instead use the ansatz \eqref{eqn:spinorsfromu} together with  \eqref{eqn:alphadecomp} and solve for the unkown overall phase $S$. In that case, there is still a non-trivial partial differential equation to be solved, in stark contrast with the simple algebraic problem that results from demanding $\chi_{AB}\alpha^A \alpha^B$ is constant.

Finally, let us indicate how to derive the adapted coordinates \eqref{eqn:Schwarzschildadapted}, although see also \cite{blau2011plane}. The general strategy is to first choose a hypersurface in spacetime which parameterizes the initial conditions for the null geodesics of fixed momenta $\hat{E}$, $L$, $L_{\phi} = 0$. A coordinate system on this initial condition hypersurface together with the affine time parameterizing the evolution along the geodesics is then almost in adapted form. To get a coordinate system into adapted form, we use that the action as one of the hypersurface coordinates. The reason this use of the action as a coordinate yields an adapted coordinate system is that the induced coordinate one-form of the action is the momentum of the congruence while the metrically dual vector field is the tangent vector field along the geodesic congruence, which by construction is the null coordinate vector field $\partial_U$ of the affine parameter $U$. Therefore $g_{UV}  =  1$ and $g_{UU} = 0$. Moreover, again by construction, the other transverse coordinate vector fields keep $V$ fixed such that $g_{Ui} = 0$. As a result, ``the gauge conditions'' for the adapted coordinates \eqref{eqn:adapted} are satisfied.

There are many ways in which we could choose an initial condition hypersurface. One natural choice would be to take a slice of fixed Kerr-Schild time $\hat{t}$. On the other hand, it is usually simplest if the choice of hypersurface respects as many isometries as possible \cite{Papadopoulos:2020qik}. That means that it is in particular more convenient to parameterize the congruence on an initial condition hypersurface that is invariant under (spacetime) time translations. For Schwarzschild, the setup is simple enough that the choice makes little difference but it is for this isometry reason that we instead choose a constant radius $r = r_{\rm ref}$ hypersurface to parametrize the initial positions of the null geodesic congruence. With that choice the Kerr-Schild coordinates satisfy
\begin{equation}
\hat{t}|_{U=0} = \hat{t}_0 \, , \quad r|_{U=0} = r_{\rm ref} \, , \quad 	\theta|_{U=0} = \theta_0 \, , \quad \phi|_{U=0} = \phi_0 \, .
\end{equation}
By our choice $L_{\phi} = 0$, the initial azimuthal angle is preserved such that even in general $\phi = \phi_0$. In addition, the radial null geodesic motion at fixed energy and angular momentum doesn't depend on any of the other coordinates. Therefore, using the radial geodesic equation \eqref{eqn:Srgeod}, we can find $r(U)$ by inverting 
\begin{equation}\label{eqn:SrUapp}
U = \int^{r(U)}_{r_{\rm ref}} \frac{d\rho}{R'(\rho) -\frac{2M}{\rho} \left(\hat{E}+R'(\rho)\right)} \, .
\end{equation}
On the other hand, the solutions to the geodesic equations for $\theta$ and $\hat{t}$ do of course depend on $\theta_0$ and $\hat{t}_0$ in addition to radial motion, but simply as integration constants such that, from
\begin{equation}
u^{\theta} = \frac{d\theta}{dU} = \frac{L}{r(U)^2} \, , \quad  	u^{\hat{t}} = \frac{d\hat{t}}{dU} = \hat{E}+\frac{2M}{r(U)} \left(\hat{E}+R'(r)\right)\ \, ,
\end{equation}
with $R'(r)|_{r=r(U)}$, we find
\begin{equation}\label{eqn:StUapp}
\begin{aligned}
\theta(U, \hat{t}_0) &= \theta_0(V,\hat{t}_0)+ L\int^{r(U)}_{r_{\rm ref}} d\rho \, \frac{1}{\rho^2}\frac{1}{R'(\rho) -\frac{2M}{\rho} \left(\hat{E}+R'(\rho)\right)} \, ,\\
\hat{t}(U,\hat{t}_0) &= \hat{t}_0 + \int^{r(U)}_{r_{\rm ref}} d\rho \, \frac{\hat{E}+\frac{2M}{\rho} \left(\hat{E}+ R'(\rho)\right)}{R'(\rho) -\frac{2M}{\rho} \left(\hat{E}+R'(\rho)\right)} \, .
\end{aligned}
\end{equation}
Next, the action associated to the null geodesic congruence, from \eqref{eqn:Su}, is
\begin{equation}
V = \hat{E} \hat{t} - (R(r)-R(r_{\rm ref})) - L \theta \, ,
\end{equation}
which, when evaluated on \eqref{eqn:SrUapp} and \eqref{eqn:StUapp} gives
\begin{equation}
V = \hat{E} \hat{t}_0 - L \theta_0 \, .
\end{equation}
Specifically, as expected, $V$ does not depend on $U$ and can therefore be used as a coordinate for initial condition hypersurface. To see very explicitly how this happens, note that  using \eqref{eqn:uschwarzschild} 
\begin{equation}
-L^2  +\rho^2 \hat{E}^2-\rho^2 (R'(\rho))^2 +2M \rho \left(\hat{E}+R'(\rho)\right)R'(\rho) +2M \rho \hat{E} \left(\hat{E}+ R'(\rho)\right) = 0 \, .
\end{equation}
Our choice of adapted coordinates is to use $V$ instead of $\theta_0$
\begin{equation}\label{eqn:theta0app}
\theta_0(V,\hat{t}_0) = \frac{1}{L}\left(\hat{E}\hat{t}_0  - V\right) \, .
\end{equation}
In summary, \eqref{eqn:SrUapp}, \eqref{eqn:StUapp}, and  \eqref{eqn:theta0app} together are presented as \eqref{eqn:Schwarzschildadapted} in the main text. 

\subsection{The Kerr black hole}\label{app:Kerr}

In this section, for completeness, we provide some additional details related to the Kerr example in Section \ref{sec:Kerr}. First, using the frame \eqref{eqn:Kerrframe} and definitions \eqref{eqn:spincoeffssymbols}, the spin coefficients are given by 
\begin{alignat}{3}
\kappa &= 0 \, ,  \quad &\sigma& = 0 \, , \nonumber \\  \rho& = \frac{1}{r-iy} \, , \quad  &\tau& = -i\frac{(r+iy)\sqrt{a^2-y^2}}{\sqrt{2}(r^2+y^2)^{3/2}}  \, , \nonumber \\
\kappa' &= 0 \, ,  \quad &\sigma'& = 0  \, , \nonumber\\ 
\rho' &= -\frac{\Delta_r}{\sqrt{2} (r-iy)(r^2+y^2)} \, , \quad &\tau'& = -i\frac{(r+iy)\sqrt{a^2-y^2}}{\sqrt{2}(r^2+y^2)^{3/2}}  \, , \nonumber  \\
\epsilon &= \frac{i y}{2(r^2+y^2)} \, ,  \quad &\epsilon'& = \frac{r^3-(r+iy) \Delta_r+2(r-M)y^2-ra^2}{4(r^2+y^2)^2}  \, , \\ \beta& = -i\frac{(a^2-y^2)(r+iy)+i(r^2-a^2)y}{2 \sqrt{2} (r^2+y^2)^{3/2}\sqrt{a^2-y^2}} \, , \quad &\beta'& = -i\frac{(a^2-y^2)(r+iy)+i(r^2+a^2)y}{2 \sqrt{2} (r^2+y^2)^{3/2}\sqrt{a^2-y^2}}  \, . \label{eqn:Kspincoeffssymbols}
\end{alignat}
The principal null congruences are still geodesic and twist-free but now additionally have some twist. The timelike surface-element\footnote{The presence of the twists of the principal null congruences and the analogous quantities in $\tau$, $\tau'$ prevent having actual embeddings. In fact, the absence of these orthogonal embedded 2-surfaces is arguably the most relevant, qualitative difference from Schwarzschild.} spanned by the null direction now is not flat, even if $M = 0$ while its transverse spatial counterpart (the 2-sphere for Schwarzschild) still does not depend on the mass $M$, so it is identical for flat space with the corresponding Kerr-Schild frames,  but now also depends on $r$. Notable, embeddable 2-surfaces orthonormal or along the principal null-frame directions are the equator at $y^2 = a^2$ and the horizon at $\Delta_r = 0$.

The equations \eqref{eqn:S2killingeqn} for Schwarzschild, with the slightly more general ansatz,
\begin{equation}\label{eqn:Kchiansatz}
\chi_{AB} = 2 \chi_{01}(r,y) o_{(A} \iota_{B)} \, ,
\end{equation}
and allowing for the additional, non-vanishing spin coefficients in \eqref{eqn:Kspincoeffssymbols} compared to \eqref{eqn:Sspincoeffssymbols}, generalizes to
\begin{equation}\label{eqn:K2killingeqn}
\begin{aligned}
\left(\frac{d}{dr} - \rho\right)\chi_{01} &= 0  \, , \quad \left(\sqrt{\frac{a^2-y^2}{r^2+y^2}}\frac{d}{dy} + \sqrt{2} \tau'\right)\chi_{01} = 0 
\, , \\ \left(\frac{\Delta_r}{r^2+y^2}\frac{d}{dr} - 2\rho'\right)\chi_{01} &= 0   \, , \quad 
\left(\sqrt{\frac{a^2-y^2}{r^2+y^2}}\frac{d}{d\theta} + \sqrt{2} \tau\right)\chi_{01} = 0 \, .
\end{aligned}
\end{equation}
As a result, $\chi_{01} \propto r-iy = \zeta$ as desired. \\

Next, we will provide a first step in checking that the tangent two-spinor $\alpha^A$, given in \eqref{eqn:Kalpha}, is indeed parallel transported along the null geodesic with tangent given in \eqref{eqn:ukerr}.
\begin{equation}\label{eqn:Kalphaequations}
\begin{aligned}
-\iota_A	\alpha^B \bar{\alpha}^{B'} \nabla_{BB'}\alpha^A = \left\lbrack  |\alpha^o|^2  \left(-\frac{\partial \alpha^o}{\partial r} - \tau' \alpha^{\iota}+\epsilon \alpha^o \right) + |\alpha^{\iota}|^2 \left( \frac{\Delta_r}{2 (r^2+y^2)} \frac{\partial \alpha^{o}}{\partial r} - \epsilon' \alpha^{o} \right)\right \rbrack \\
+  \left\lbrack \alpha^o\bar{\alpha}^{\iota'} \left(-\sqrt{\frac{a^2-y^2}{2 (r^2+y^2)}}\frac{\partial \alpha^o}{\partial y} -\rho' \alpha^{\iota} + \beta \alpha^o \right)+  \alpha^{\iota}\bar{\alpha}^{o'} \left(-\sqrt{\frac{a^2-y^2}{2(r^2+y^2)}}\frac{\partial \alpha^o}{\partial y} -\beta' \alpha^o \right) \right\rbrack \, , \\
o_A	\alpha^B \bar{\alpha}^{B'} \nabla_{BB'}\alpha^A = \left\lbrack -|\alpha^o|^2  \left(\frac{\partial \alpha^{\iota}}{\partial r}+\epsilon \alpha^{\iota} \right) + |\alpha^{\iota}|^2 \left( \frac{\Delta_r}{2(r^2+y^2)}  \frac{\partial \alpha^{\iota}}{\partial r} + \epsilon' \alpha^{\iota} - \tau \alpha^{o}\right) \right \rbrack  \\
+ \left\lbrack \alpha^o\bar{\alpha}^{\iota'} \left(-\sqrt{\frac{a^2-y^2}{2(r^2+y^2)}}\frac{\partial \alpha^{\iota}}{\partial y} -\beta \alpha^{\iota}\right)+  \alpha^{\iota}\bar{\alpha}^{o'} \left(-\sqrt{\frac{a^2-y^2}{2(r^2+y^2)}}\frac{\partial \alpha^{\iota}}{\partial y} +\beta' \alpha^{\iota} - \rho \alpha^o \right) \right\rbrack \, .
\end{aligned}
\end{equation}
At this point, it serves no purpose to continue to write down the calculation, which is uninformative. However, we have verified that the equations \eqref{eqn:Kalphaequations} are satisfied by \eqref{eqn:Kalpha}.

\subsection{Vacuum Petrov type D spacetimes}\label{app:PlebanskiDemianski}

In this section, we provide some additional details related to the vacuum Petrov type D spacetimes example in Section \ref{sec:typeD}. From the build-up towards this general case, it is clear that for the general structure of the double copy and the Penrose limit, the main ingredient that we need is the existence of the Killing spinor $\chi_{AB}$. On the other hand, in practice, one would still need to be more explicitly about, say, the Kerr-Schild form and geodesics. We will therefore present here for completeness the (uncharged) Plebanski-Demianski line element \cite{Kinnersley:1969zza,Plebanski:1976gy}, which is the most general Petrov type D solution of the vacuum Einstein equations. However, see \cite{Griffiths:2005qp} for more details.

The (vacuum) Plebanski-Demianski line element, with vanishing cosmological constant, in complexified coordinates $\lbrace u,v,p,q \rbrace$ is given by
\begin{equation}\label{eqn:PDmetric}
(1-pq)^2 ds^2 = -2i (du+q^2 dv)dp +2(du-p^2 dv)dq - \frac{P(p)}{p^2+q^2}(du+q^2 dv)^2 +\frac{Q(q)}{p^2+q^2}(du-p^2 dv)^2 \, ,
\end{equation}
where the the auxiliary polynomials $P(p)$ and $Q(q)$ are given by
\begin{equation}
\begin{aligned}
P(p) &= \Gamma (1-p^4) +2 N p -\Xi p^2 + 2M p^3 \, ,  \\
Q(q) &= \Gamma (1-q^4) -2 M q + \Xi q^2 - 2N q^3 \, , 
\end{aligned}
\end{equation}
with free parameters $M$, $N$, $\Gamma$, and $\Xi$. We present the metric immediately in the complex form \eqref{eqn:PDmetric}, which is of the double Kerr-Schild form that was used in \cite{Luna:2018dpt} to connect to the single copy flat space.

The double Kerr-Schild form of \eqref{eqn:PDmetric} requires us to slightly modify parts of the discussion in the main text. First, instead of \eqref{eqn:KSmetric}, the metric is of the form
\begin{equation}\label{eqn:metricKSdouble}
g_{\mu \nu} = \eta_{\mu \nu} + \phi_k k_{\mu} k_{\nu} + \phi_m m_{\nu} m_{\mu} \, .
\end{equation}
for null, geodesic vector fields $k$ and $m$, which are moreover orthogonal to each other\footnote{The fact that
$k_{\mu}m^{\mu} = 0$ is our motivation to call the second null vector $m$ instead of $l$. Moreover, the requirement
to have a double Kerr-Schild metric, which involves say $m$, explains why it was necessary to complexify the
metric.}, as well as scalar functions $\phi_k$ and $\phi_m$. The mutual orthogonality of $m$ and $k$ implies first of all that one can unambiguously raise and lower their indices with either the flat or the curved metric. Moreover, the extension of the Kerr-Schild shift \eqref{eqn:frameshift} is given by
\begin{equation}\label{eqn:KSshiftdouble}
\begin{aligned}
n^{(g)}_{\mu} &= n^{(\eta)} + \frac{\phi_k}{2} k^{(\eta)}_{\mu} \, , \quad  k^{(g)}_{\mu} = k^{(\eta)}_{\mu} \\
\tilde{m}^{(g)}_{\mu} &= \tilde{m}^{(\eta)} + \frac{\phi_m}{2} m^{(\eta)}_{\mu} \, , \quad  m^{(g)}_{\mu} = m^{(\eta)}_{\mu} \, ,
\end{aligned}
\end{equation}
where, as before
\begin{equation}
\eta_{\mu \nu} = 2 k^{(\eta)}_{(\mu} n^{(\eta)}_{\nu)} -  2 m^{(\eta)}_{(\mu} \tilde{m}^{(\eta)}_{\nu)} \, , \quad 	g_{\mu \nu} = 2 k^{(g)}_{(\mu} n^{(g)}_{\nu)} -  2 m^{(g)}_{(\mu} \tilde{m}^{(g)}_{\nu)} \, ,
\end{equation}
but we use $\tilde{m}$ instead of $\bar{m}$ to emphasize that these are not necessarily complex conjugate for the metric in its complex form. Note also that the vector fields $k$ and $m$ are identical on both flat and curved spacetime, so again unambiguously $k_{\mu} = k_{\mu}^{(\eta)} = k_{\mu}^{(g)}$ and $m_{\mu} = m_{\mu}^{(\eta)} = m_{\mu}^{(g)}$. Similarly, the shift in Infeld-van der Waerden symbols \eqref{eqn:IvW} is readily extended to 
\begin{equation}\label{eqn:IvWdouble}
\sigma^{(g)}{}{}^{AA'}_{\mu}   = \sigma^{(\eta)}{}{}^{AA'}_{\mu} + \frac{\phi_k}{2} k_{\mu} \sigma^{(\eta)}{}{}^{AA'}_{\nu} k^{\nu} + \frac{\phi_m}{2} m_{\mu} \sigma^{(\eta)}{}{}^{AA'}_{\nu} m^{\nu} \, .
\end{equation}

A principal null frame for the metric \eqref{eqn:PDmetric}, that takes the form \eqref{eqn:KSshiftdouble}, is given by
\begin{alignat}{1}
k^{(g)}_{\mu}dx^{\mu} &= k^{(\eta)}_{\mu}dx^{\mu}  \, , \nonumber \\
n^{(g)}_{\mu}dx^{\mu} &= n^{(\eta)}_{\mu}dx^{\mu} - \frac{q (M+Nq^2)}{(1-pq)^2 (p^2+q^2)}   k^{(\eta)}_{\mu}dx^{\mu}  \, , \nonumber \\
m^{(g)}_{\mu}dx^{\mu} &= m^{(\eta)}_{\mu}dx^{\mu}  \, , \nonumber \\
\tilde{m}^{(g)}_{\mu}dx^{\mu} &= \tilde{m}^{(\eta)}_{\mu}dx^{\mu} + \frac{p (N+Mp^2)}{(1-pq)^2 (p^2+q^2)} m^{(\eta)}_{\mu}dx^{\mu} \, . \label{eqn:PDPNframe}
\end{alignat}
with
\begin{alignat}{1}
k^{(\eta)}_{\mu}dx^{\mu} &= du -p^2 dv  \, , \nonumber \\
n^{(\eta)}_{\mu}dx^{\mu} &= \frac{\Gamma(1-q^4)+ \Xi q^2}{2 (1-pq)^2(p^2+q^2)} du - p^2\frac{\Gamma(1-q^4) + \Xi q^2 }{2 (1-pq)^2(p^2+q^2)} dv +\frac{dq}{(1-pq)^2}\, , \nonumber \\
m^{(\eta)}_{\mu}dx^{\mu} &= du +q^2 dv\, , \nonumber \\
\tilde{m}^{(\eta)}_{\mu}dx^{\mu} &= \frac{\Gamma(1-p^4)+ \Xi p^2}{2 (1-pq)^2(p^2+q^2)} du + q^2\frac{\Gamma(1-p^4) - \Xi p^2 }{2 (1-pq)^2(p^2+q^2)} dv +\frac{i dp}{(1-pq)^2} \, . \label{eqn:PDPNframeflat}
\end{alignat}
The scalar functions $\phi_k$ and $\phi_m$, in \eqref{eqn:metricKSdouble} and \eqref{eqn:KSshiftdouble}, are thus given by
\begin{equation}
\phi_k = - \frac{q (M+Nq^2)}{(1-pq)^2 (p^2+q^2)}  \, , \quad \phi_m  = \frac{p (N+Mp^2)}{(1-pq)^2 (p^2+q^2)} \, .
\end{equation}
Using the frame \eqref{eqn:PDPNframe}, we find
\begin{alignat}{3}
\kappa &= 0 \, ,  \quad &\sigma& = 0 \, , \nonumber \\  
\rho& = -\frac{(i + p^2)(1-pq)}{p+iq} \, , \quad  &\tau& =  \frac{(-i + q^2)(1-pq)}{p+iq}  \, , \nonumber \\
\kappa' &= 0 \, ,  \quad &\sigma'& = 0  \, , \nonumber\\ 
\rho' &= -\frac{(i+p^2)(2Mq+2Nq^3-\Gamma(1-q^4)-\Xi q^2)}{2(p-iq)(p+iq)^2(1-pq)} \, , \\ \tau' &= -\frac{(-i+q^2)(2Np+2Np^3+\Gamma(1-p^4)-\Xi p^2)}{2(p-iq)(p+iq)^2(1-pq)}  \, ,  
\label{eqn:PDspincoeffssymbols}
\end{alignat}
where for brevity we omit the GHP gauge fields $\epsilon$, $\epsilon'$, $\beta$, and $\beta'$. In addition, as expected for a Petrov type D spacetime
\begin{equation}
\Psi_2 = i \frac{(M-i N)(1-pq)^3}{(p+iq)^3} \, ,\quad  \quad \Psi_0 = \Psi_1 = \Psi_3 = \Psi_4 = 0 \, .
\end{equation}
We should stress here that the above spin coefficients and Weyl scalars are potentially incomplete as written. Contrary to the previous examples, $m$ and $\tilde{m}$ are not complex conjugates. Therefore, there are additional scalars that are not necessarily related to the ones presented above. An interesting example is that $\tilde{\tau} \neq \bar{\tau}$. On the other hand, we do find
\begin{equation}
\tilde{\Psi}_2 = \bar{\Psi}_2 = -i \frac{(M+i N)(1-pq)^3}{(p-iq)^3} \, .
\end{equation}
which is the quantity of relevance for the discussion of the Weyl double copy in the main text.
As in the other Petrov type D examples, the Weyl spinor is given by
\begin{equation}
\Psi_{ABCD} = 6 i  \frac{(M-i N)(1-pq)^3}{(p+iq)^3}o_{(A} o_B \iota_C \iota_{D)} \, ,
\end{equation}
or
\begin{equation}
\Psi_{ABCD} = -  \frac{\tilde{M}^5}{\zeta^5} 3 \chi_{(AB}  \chi_{CD)} \, , \quad \chi_{AB} = \sqrt{2} \frac{\zeta}{\tilde{M}^2} o_{(A} \iota_{B)} \, .
\end{equation}
with
\begin{equation}
\tilde{M} = M-i N \, , \quad 	\zeta = \frac{q-ip}{1-pq} \, .
\end{equation}

The double Kerr-Schild prescription of \cite{Luna:2018dpt}, compatible with the Weyl double copy, is given by\footnote{One can (and would generically) change the constant charges $N^{(\eta)}$ and $M^{(\eta)}$ to which $N$ and $M$ are mapped, see \cite{Luna:2018dpt}.}
\begin{equation}\label{eqn:APBsc}
A = \frac{2 M^{(\eta)} q}{p^2 + q^2}k +  \frac{2 N^{(\eta)} p}{p^2 + q^2} m \, .  
\end{equation}
However, as was also pointed out by \cite{Luna:2018dpt},
\begin{equation}
\phi_k \not\propto  \frac{2 M^{(\eta)} q}{p^2 + q^2} \, , \quad \phi_m \not\propto \frac{2 N^{(\eta)} p}{p^2 + q^2} \, .
\end{equation}
Moreover, the prefactors of $k$ and $m$ in \eqref{eqn:APBsc} do not satisfy the flat space wave equation. However, a rescaling of the frames\footnote{The discussion of rescaling the frame vectors can also be usefully phrased in terms of the GHP-weight of the Kerr-Schild single and zeroth copy, which indeed have definite, non-zero weight. The requirement that the flat space field equations are satisfied generically fixes the GHP frame up to a constant.}
\begin{equation}
\left(l,n,m,\tilde{m}\right) \to  \left(\frac{l}{1-pq}, \left(1-pq \right)n,\frac{m}{1-pq}, \left(1-pq \right)\tilde{m}\right) \, ,
\end{equation}
together with
\begin{equation}\label{eqn:phishifts}
\phi_k \to \left(1-pq\right)^2\phi_k \, , \quad  	\phi_m \to \left(1-pq\right)^2\phi_m \, , 
\end{equation}
leaves invariant the form of the flat Kerr-Schild metric as well as of $\Psi_2$ but it changes \eqref{eqn:APBsc} into
\begin{equation}\label{eqn:APBscmod}
A =  \frac{2 M^{(\eta)} q \left(1-pq\right)}{p^2 + q^2}k +  \frac{2 N^{(\eta)} p \left(1-pq\right)}{p^2 + q^2} m \, .  
\end{equation}
while
\begin{equation}\label{eqn:KSphis}
\phi_k = - \frac{q (M+Nq^2)}{p^2+q^2}  \, , \quad \phi_m  = \frac{p (N+Mp^2)}{p^2+q^2} \, .
\end{equation}
Obviously, $\phi_k$ and $\phi_m$ in this frame still do not correspond to the prefactors of $k$ and $m$ in \eqref{eqn:APBscmod}. On the other hand, those prefactors, denoted for convenience and to distinguish from \eqref{eqn:KSphis} as
\begin{equation}\label{eqn:KSphisKS}
\phi^{(KS)}_k = \frac{2 M^{(\eta)} q \left(1-pq\right)}{p^2 + q^2} \, , \quad \phi^{(KS)}_m  = \frac{2 N^{(\eta)} p \left(1-pq\right)}{p^2 + q^2} \, .
\end{equation}
now at least also satisfy the flat space wave equation.

Even if \eqref{eqn:KSphisKS} and \eqref{eqn:KSphis} are not quite the same, we would like to understand the relation to the Weyl zeroth copy
\begin{equation}
S = -2 \frac{(\tilde{M}^{(\eta)})^2}{\tilde{M}} \frac{1-pq}{q-ip} \, ,
\end{equation}
which of course also satisfies the flat space wave equation. In the earlier, ordinary Kerr-Schild backgrounds, the relation was that the Kerr-Schild zeroth copy is, up to a constant, the real part of the Weyl zeroth copy, see  \eqref{eqn:WeyltoKSrelation}. Here, taking for simplicity $\tilde{M}^{(\eta)} = \tilde{M}$, that gives
\begin{equation}
S + S^* = -2 \frac{1-pq}{p^2 + q^2}(N^{(\eta)} p + M^{(\eta)} q) = \phi^{(KS)}_k + \phi^{(KS)}_m \, .
\end{equation}

\end{appendices}

\bibliographystyle{jhep}

\begin{thebibliography}{100}
	
	\bibitem{Bern:2019prr}
	Z.~Bern, J.J.~Carrasco, M.~Chiodaroli, H.~Johansson and R.~Roiban, \emph{{The
			Duality Between Color and Kinematics and its Applications}},
	\href{https://arxiv.org/abs/1909.01358}{{\ttfamily 1909.01358}}.
	
	\bibitem{Bern:2022wqg}
	Z.~Bern, J.J.~Carrasco, M.~Chiodaroli, H.~Johansson and R.~Roiban, \emph{{The
			SAGEX review on scattering amplitudes Chapter 2: An invitation to
			color-kinematics duality and the double copy}},
	\href{https://doi.org/10.1088/1751-8121/ac93cf}{\emph{J. Phys. A} {\bfseries
			55} (2022) 443003} [\href{https://arxiv.org/abs/2203.13013}{{\ttfamily
			2203.13013}}].
	
	\bibitem{Adamo:2022dcm}
	T.~Adamo, J.J.M.~Carrasco, M.~Carrillo-Gonz\'alez, M.~Chiodaroli, H.~Elvang,
	H.~Johansson et~al., \emph{{Snowmass White Paper: the Double Copy and its
			Applications}},  in \emph{{Snowmass 2021}}, 4, 2022
	[\href{https://arxiv.org/abs/2204.06547}{{\ttfamily 2204.06547}}].
	
	\bibitem{Bern:2019nnu}
	Z.~Bern, C.~Cheung, R.~Roiban, C.-H.~Shen, M.P.~Solon and M.~Zeng,
	\emph{{Scattering Amplitudes and the Conservative Hamiltonian for Binary
			Systems at Third Post-Minkowskian Order}},
	\href{https://doi.org/10.1103/PhysRevLett.122.201603}{\emph{Phys. Rev. Lett.}
		{\bfseries 122} (2019) 201603}
	[\href{https://arxiv.org/abs/1901.04424}{{\ttfamily 1901.04424}}].
	
	\bibitem{Bern:2019crd}
	Z.~Bern, C.~Cheung, R.~Roiban, C.-H.~Shen, M.P.~Solon and M.~Zeng, \emph{{Black
			Hole Binary Dynamics from the Double Copy and Effective Theory}},
	\href{https://doi.org/10.1007/JHEP10(2019)206}{\emph{JHEP} {\bfseries 10}
		(2019) 206} [\href{https://arxiv.org/abs/1908.01493}{{\ttfamily
			1908.01493}}].
	
	\bibitem{Bern:2020uwk}
	Z.~Bern, J.~Parra-Martinez, R.~Roiban, E.~Sawyer and C.-H.~Shen, \emph{{Leading
			Nonlinear Tidal Effects and Scattering Amplitudes}},
	\href{https://doi.org/10.1007/JHEP05(2021)188}{\emph{JHEP} {\bfseries 05}
		(2021) 188} [\href{https://arxiv.org/abs/2010.08559}{{\ttfamily
			2010.08559}}].
	
	\bibitem{Bern:2021yeh}
	Z.~Bern, J.~Parra-Martinez, R.~Roiban, M.S.~Ruf, C.-H.~Shen, M.P.~Solon et~al.,
	\emph{{Scattering Amplitudes, the Tail Effect, and Conservative Binary
			Dynamics at O(G4)}},
	\href{https://doi.org/10.1103/PhysRevLett.128.161103}{\emph{Phys. Rev. Lett.}
		{\bfseries 128} (2022) 161103}
	[\href{https://arxiv.org/abs/2112.10750}{{\ttfamily 2112.10750}}].
	
	\bibitem{Bern:2022jvn}
	Z.~Bern, J.~Parra-Martinez, R.~Roiban, M.S.~Ruf, C.-H.~Shen, M.P.~Solon et~al.,
	\emph{{Scattering amplitudes and conservative dynamics at the fourth
			post-Minkowskian order}},
	\href{https://doi.org/10.22323/1.416.0051}{\emph{PoS} {\bfseries LL2022}
		(2022) 051}.
	
	\bibitem{Kawai:1985xq}
	H.~Kawai, D.C.~Lewellen and S.H.H.~Tye, \emph{{A Relation Between Tree
			Amplitudes of Closed and Open Strings}},
	\href{https://doi.org/10.1016/0550-3213(86)90362-7}{\emph{Nucl. Phys. B}
		{\bfseries 269} (1986) 1}.
	
	\bibitem{Adamo:2017nia}
	T.~Adamo, E.~Casali, L.~Mason and S.~Nekovar, \emph{{Scattering on plane waves
			and the double copy}},
	\href{https://doi.org/10.1088/1361-6382/aa9961}{\emph{Class. Quant. Grav.}
		{\bfseries 35} (2018) 015004}
	[\href{https://arxiv.org/abs/1706.08925}{{\ttfamily 1706.08925}}].
	
	\bibitem{Adamo:2023fbj}
	T.~Adamo, G.~Bogna, L.~Mason and A.~Sharma, \emph{{Scattering on self-dual
			Taub-NUT}},  \href{https://arxiv.org/abs/2309.03834}{{\ttfamily 2309.03834}}.
	
	\bibitem{Sivaramakrishnan:2021srm}
	A.~Sivaramakrishnan, \emph{{Towards color-kinematics duality in generic
			spacetimes}}, \href{https://doi.org/10.1007/JHEP04(2022)036}{\emph{JHEP}
		{\bfseries 04} (2022) 036}
	[\href{https://arxiv.org/abs/2110.15356}{{\ttfamily 2110.15356}}].
	
	\bibitem{Herderschee:2022ntr}
	A.~Herderschee, R.~Roiban and F.~Teng, \emph{{On the differential
			representation and color-kinematics duality of AdS boundary correlators}},
	\href{https://doi.org/10.1007/JHEP05(2022)026}{\emph{JHEP} {\bfseries 05}
		(2022) 026} [\href{https://arxiv.org/abs/2201.05067}{{\ttfamily
			2201.05067}}].
	
	\bibitem{Cheung:2022pdk}
	C.~Cheung, J.~Parra-Martinez and A.~Sivaramakrishnan, \emph{{On-shell
			correlators and color-kinematics duality in curved symmetric spacetimes}},
	\href{https://doi.org/10.1007/JHEP05(2022)027}{\emph{JHEP} {\bfseries 05}
		(2022) 027} [\href{https://arxiv.org/abs/2201.05147}{{\ttfamily
			2201.05147}}].
	
	\bibitem{Ilderton:2024oly}
	A.~Ilderton and W.~Lindved, \emph{{Toward double copy on arbitrary
			backgrounds}},  \href{https://arxiv.org/abs/2405.10016}{{\ttfamily
			2405.10016}}.
	
	\bibitem{Adamo:2017qyl}
	T.~Adamo, \emph{{Lectures on twistor theory}},
	\href{https://doi.org/10.22323/1.323.0003}{\emph{PoS} {\bfseries Modave2017}
		(2018) 003} [\href{https://arxiv.org/abs/1712.02196}{{\ttfamily
			1712.02196}}].
	
	\bibitem{Monteiro:2014cda}
	R.~Monteiro, D.~O'Connell and C.D.~White, \emph{{Black holes and the double
			copy}}, \href{https://doi.org/10.1007/JHEP12(2014)056}{\emph{JHEP} {\bfseries
			12} (2014) 056} [\href{https://arxiv.org/abs/1410.0239}{{\ttfamily
			1410.0239}}].
	
	\bibitem{Luna:2016due}
	A.~Luna, R.~Monteiro, I.~Nicholson, D.~O'Connell and C.D.~White, \emph{{The
			double copy: Bremsstrahlung and accelerating black holes}},
	\href{https://doi.org/10.1007/JHEP06(2016)023}{\emph{JHEP} {\bfseries 06}
		(2016) 023} [\href{https://arxiv.org/abs/1603.05737}{{\ttfamily
			1603.05737}}].
	
	\bibitem{Luna:2015paa}
	A.~Luna, R.~Monteiro, D.~O'Connell and C.D.~White, \emph{{The classical double
			copy for Taub\textendash{}NUT spacetime}},
	\href{https://doi.org/10.1016/j.physletb.2015.09.021}{\emph{Phys. Lett. B}
		{\bfseries 750} (2015) 272}
	[\href{https://arxiv.org/abs/1507.01869}{{\ttfamily 1507.01869}}].
	
	\bibitem{Kim:2019jwm}
	K.~Kim, K.~Lee, R.~Monteiro, I.~Nicholson and D.~Peinador~Veiga, \emph{{The
			Classical Double Copy of a Point Charge}},
	\href{https://doi.org/10.1007/JHEP02(2020)046}{\emph{JHEP} {\bfseries 02}
		(2020) 046} [\href{https://arxiv.org/abs/1912.02177}{{\ttfamily
			1912.02177}}].
	
	\bibitem{Luna:2020adi}
	A.~Luna, S.~Nagy and C.~White, \emph{{The convolutional double copy: a case
			study with a point}},
	\href{https://doi.org/10.1007/JHEP09(2020)062}{\emph{JHEP} {\bfseries 09}
		(2020) 062} [\href{https://arxiv.org/abs/2004.11254}{{\ttfamily
			2004.11254}}].
	
	\bibitem{Luna:2018dpt}
	A.~Luna, R.~Monteiro, I.~Nicholson and D.~O'Connell, \emph{{Type D Spacetimes
			and the Weyl Double Copy}},
	\href{https://doi.org/10.1088/1361-6382/ab03e6}{\emph{Class. Quant. Grav.}
		{\bfseries 36} (2019) 065003}
	[\href{https://arxiv.org/abs/1810.08183}{{\ttfamily 1810.08183}}].
	
	\bibitem{Godazgar:2021iae}
	H.~Godazgar, M.~Godazgar, R.~Monteiro, D.~Peinador~Veiga and C.N.~Pope,
	\emph{{Asymptotic Weyl double copy}},
	\href{https://doi.org/10.1007/JHEP11(2021)126}{\emph{JHEP} {\bfseries 11}
		(2021) 126} [\href{https://arxiv.org/abs/2109.07866}{{\ttfamily
			2109.07866}}].
	
	\bibitem{Berman:2018hwd}
	D.S.~Berman, E.~Chac\'on, A.~Luna and C.D.~White, \emph{{The self-dual
			classical double copy, and the Eguchi-Hanson instanton}},
	\href{https://doi.org/10.1007/JHEP01(2019)107}{\emph{JHEP} {\bfseries 01}
		(2019) 107} [\href{https://arxiv.org/abs/1809.04063}{{\ttfamily
			1809.04063}}].
	
	\bibitem{Lee:2018gxc}
	K.~Lee, \emph{{Kerr-Schild Double Field Theory and Classical Double Copy}},
	\href{https://doi.org/10.1007/JHEP10(2018)027}{\emph{JHEP} {\bfseries 10}
		(2018) 027} [\href{https://arxiv.org/abs/1807.08443}{{\ttfamily
			1807.08443}}].
	
	\bibitem{Cho:2019ype}
	W.~Cho and K.~Lee, \emph{{Heterotic Kerr-Schild Double Field Theory and
			Classical Double Copy}},
	\href{https://doi.org/10.1007/JHEP07(2019)030}{\emph{JHEP} {\bfseries 07}
		(2019) 030} [\href{https://arxiv.org/abs/1904.11650}{{\ttfamily
			1904.11650}}].
	
	\bibitem{Cristofoli:2020hnk}
	A.~Cristofoli, \emph{{Gravitational shock waves and scattering amplitudes}},
	\href{https://doi.org/10.1007/JHEP11(2020)160}{\emph{JHEP} {\bfseries 11}
		(2020) 160} [\href{https://arxiv.org/abs/2006.08283}{{\ttfamily
			2006.08283}}].
	
	\bibitem{Easson:2021asd}
	D.A.~Easson, T.~Manton and A.~Svesko, \emph{{Sources in the Weyl Double Copy}},
	\href{https://doi.org/10.1103/PhysRevLett.127.271101}{\emph{Phys. Rev. Lett.}
		{\bfseries 127} (2021) 271101}
	[\href{https://arxiv.org/abs/2110.02293}{{\ttfamily 2110.02293}}].
	
	\bibitem{Easson:2022zoh}
	D.A.~Easson, T.~Manton and A.~Svesko, \emph{{Einstein-Maxwell theory and the
			Weyl double copy}},
	\href{https://doi.org/10.1103/PhysRevD.107.044063}{\emph{Phys. Rev. D}
		{\bfseries 107} (2023) 044063}
	[\href{https://arxiv.org/abs/2210.16339}{{\ttfamily 2210.16339}}].
	
	\bibitem{Easson:2020esh}
	D.A.~Easson, C.~Keeler and T.~Manton, \emph{{Classical double copy of
			nonsingular black holes}},
	\href{https://doi.org/10.1103/PhysRevD.102.086015}{\emph{Phys. Rev. D}
		{\bfseries 102} (2020) 086015}
	[\href{https://arxiv.org/abs/2007.16186}{{\ttfamily 2007.16186}}].
	
	\bibitem{Mkrtchyan:2022ulc}
	K.~Mkrtchyan and M.~Svazas, \emph{{Solutions in Nonlinear Electrodynamics and
			their double copy regular black holes}},
	\href{https://doi.org/10.1007/JHEP09(2022)012}{\emph{JHEP} {\bfseries 09}
		(2022) 012} [\href{https://arxiv.org/abs/2205.14187}{{\ttfamily
			2205.14187}}].
	
	\bibitem{Ilderton:2018lsf}
	A.~Ilderton, \emph{{Screw-symmetric gravitational waves: a double copy of the
			vortex}}, \href{https://doi.org/10.1016/j.physletb.2018.04.069}{\emph{Phys.
			Lett. B} {\bfseries 782} (2018) 22}
	[\href{https://arxiv.org/abs/1804.07290}{{\ttfamily 1804.07290}}].
	
	\bibitem{Barrientos:2024uuq}
	J.~Barrientos, A.~Cisterna, M.~Hassaine and J.~Oliva, \emph{{Revisiting
			Buchdahl transformations: New static and rotating black holes in vacuum,
			double copy, and hairy extensions}},
	\href{https://arxiv.org/abs/2404.12194}{{\ttfamily 2404.12194}}.
	
	\bibitem{Ortaggio:2023cdz}
	M.~Ortaggio, V.~Pravda and A.~Pravdova, \emph{{Kerr-Schild double copy for
			Kundt spacetimes of any dimension}},
	\href{https://doi.org/10.1007/JHEP02(2024)069}{\emph{JHEP} {\bfseries 2024}
		(2024) 069} [\href{https://arxiv.org/abs/2312.00706}{{\ttfamily
			2312.00706}}].
	
	\bibitem{Adamo:2022rob}
	T.~Adamo, A.~Cristofoli and P.~Tourkine, \emph{{The ultrarelativistic limit of
			Kerr}}, \href{https://doi.org/10.1007/JHEP02(2023)107}{\emph{JHEP} {\bfseries
			02} (2023) 107} [\href{https://arxiv.org/abs/2209.05730}{{\ttfamily
			2209.05730}}].
	
	\bibitem{Keeler:2020rcv}
	C.~Keeler, T.~Manton and N.~Monga, \emph{{From Navier-Stokes to Maxwell via
			Einstein}}, \href{https://doi.org/10.1007/JHEP08(2020)147}{\emph{JHEP}
		{\bfseries 08} (2020) 147}
	[\href{https://arxiv.org/abs/2005.04242}{{\ttfamily 2005.04242}}].
	
	\bibitem{Keeler:2024bdt}
	C.~Keeler and N.~Monga, \emph{{On type-II Spacetimes and the Double Copy for
			Fluids Metrics}},  \href{https://arxiv.org/abs/2404.03195}{{\ttfamily
			2404.03195}}.
	
	\bibitem{Godazgar:2020zbv}
	H.~Godazgar, M.~Godazgar, R.~Monteiro, D.~Peinador~Veiga and C.N.~Pope,
	\emph{{Weyl Double Copy for Gravitational Waves}},
	\href{https://doi.org/10.1103/PhysRevLett.126.101103}{\emph{Phys. Rev. Lett.}
		{\bfseries 126} (2021) 101103}
	[\href{https://arxiv.org/abs/2010.02925}{{\ttfamily 2010.02925}}].
	
	\bibitem{White:2020sfn}
	C.D.~White, \emph{{Twistorial Foundation for the Classical Double Copy}},
	\href{https://doi.org/10.1103/PhysRevLett.126.061602}{\emph{Phys. Rev. Lett.}
		{\bfseries 126} (2021) 061602}
	[\href{https://arxiv.org/abs/2012.02479}{{\ttfamily 2012.02479}}].
	
	\bibitem{Chacon:2021wbr}
	E.~Chac\'on, S.~Nagy and C.D.~White, \emph{{The Weyl double copy from twistor
			space}}, \href{https://doi.org/10.1007/JHEP05(2021)239}{\emph{JHEP}
		{\bfseries 05} (2021) 2239}
	[\href{https://arxiv.org/abs/2103.16441}{{\ttfamily 2103.16441}}].
	
	\bibitem{Chacon:2021lox}
	E.~Chac\'on, S.~Nagy and C.D.~White, \emph{{Alternative formulations of the
			twistor double copy}},
	\href{https://doi.org/10.1007/JHEP03(2022)180}{\emph{JHEP} {\bfseries 03}
		(2022) 180} [\href{https://arxiv.org/abs/2112.06764}{{\ttfamily
			2112.06764}}].
	
	\bibitem{Han:2022ubu}
	S.~Han, \emph{{Weyl double copy and massless free-fields in curved
			spacetimes}}, \href{https://doi.org/10.1088/1361-6382/ac96c2}{\emph{Class.
			Quant. Grav.} {\bfseries 39} (2022) 225009}
	[\href{https://arxiv.org/abs/2204.01907}{{\ttfamily 2204.01907}}].
	
	\bibitem{Alawadhi:2019urr}
	R.~Alawadhi, D.S.~Berman, B.~Spence and D.~Peinador~Veiga, \emph{{S-duality and
			the double copy}}, \href{https://doi.org/10.1007/JHEP03(2020)059}{\emph{JHEP}
		{\bfseries 03} (2020) 059}
	[\href{https://arxiv.org/abs/1911.06797}{{\ttfamily 1911.06797}}].
	
	\bibitem{Alawadhi:2020jrv}
	R.~Alawadhi, D.S.~Berman and B.~Spence, \emph{{Weyl doubling}},
	\href{https://doi.org/10.1007/JHEP09(2020)127}{\emph{JHEP} {\bfseries 09}
		(2020) 127} [\href{https://arxiv.org/abs/2007.03264}{{\ttfamily
			2007.03264}}].
	
	\bibitem{Alkac:2023glx}
	G.~Alkac, M.K.~Gumus, O.~Kasikci, M.A.~Olpak and M.~Tek, \emph{{Regularized
			Weyl double copy}},
	\href{https://doi.org/10.1103/PhysRevD.109.084047}{\emph{Phys. Rev. D}
		{\bfseries 109} (2024) 084047}
	[\href{https://arxiv.org/abs/2310.06048}{{\ttfamily 2310.06048}}].
	
	\bibitem{Armstrong-Williams:2023ssz}
	K.~Armstrong-Williams and C.D.~White, \emph{{A spinorial double copy for $
			\mathcal{N} $ = 0 supergravity}},
	\href{https://doi.org/10.1007/JHEP05(2023)047}{\emph{JHEP} {\bfseries 05}
		(2023) 047} [\href{https://arxiv.org/abs/2303.04631}{{\ttfamily
			2303.04631}}].
	
	\bibitem{Chawla:2022ogv}
	S.~Chawla and C.~Keeler, \emph{{Aligned fields double copy to Kerr-NUT-(A)dS}},
	\href{https://doi.org/10.1007/JHEP04(2023)005}{\emph{JHEP} {\bfseries 04}
		(2023) 005} [\href{https://arxiv.org/abs/2209.09275}{{\ttfamily
			2209.09275}}].
	
	\bibitem{Mao:2023yle}
	P.~Mao and W.~Zhao, \emph{{Asymptotic Weyl double copy in Newman-Penrose
			formalism}}, \href{https://doi.org/10.1007/JHEP02(2024)171}{\emph{JHEP}
		{\bfseries 02} (2024) 171}
	[\href{https://arxiv.org/abs/2312.17160}{{\ttfamily 2312.17160}}].
	
	\bibitem{Easson:2023dbk}
	D.A.~Easson, G.~Herczeg, T.~Manton and M.~Pezzelle, \emph{{Isometries and the
			double copy}}, \href{https://doi.org/10.1007/JHEP09(2023)162}{\emph{JHEP}
		{\bfseries 09} (2023) 162}
	[\href{https://arxiv.org/abs/2306.13687}{{\ttfamily 2306.13687}}].
	
	\bibitem{Luna:2022dxo}
	A.~Luna, N.~Moynihan and C.D.~White, \emph{{Why is the Weyl double copy local
			in position space?}},
	\href{https://doi.org/10.1007/JHEP12(2022)046}{\emph{JHEP} {\bfseries 12}
		(2022) 046} [\href{https://arxiv.org/abs/2208.08548}{{\ttfamily
			2208.08548}}].
	
	\bibitem{newman1968new}
	E.T.~Newman and R.~Penrose, \emph{New conservation laws for zero rest-mass
		fields in asymptotically flat space-time}, {\emph{Proceedings of the Royal
			Society of London. Series A. Mathematical and Physical Sciences} {\bfseries
			305} (1968) 175}.
	
	\bibitem{Adamo:2021dfg}
	T.~Adamo and U.~Kol, \emph{{Classical double copy at null infinity}},
	\href{https://doi.org/10.1088/1361-6382/ac635e}{\emph{Class. Quant. Grav.}
		{\bfseries 39} (2022) 105007}
	[\href{https://arxiv.org/abs/2109.07832}{{\ttfamily 2109.07832}}].
	
	\bibitem{penrose1976any}
	R.~Penrose, \emph{Any space-time has a plane wave as a limit},  in
	\emph{Differential geometry and relativity}, pp.~271--275, Springer (1976).
	
	\bibitem{penrose1984spinors}
	R.~Penrose and W.~Rindler, \emph{Spinors and space-time: Volume 1, Two-spinor
		calculus and relativistic fields}, vol.~1, Cambridge University Press (1984).
	
	\bibitem{Penrose:1986ca}
	R.~Penrose and W.~Rindler, \emph{{SPINORS AND SPACE-TIME. VOL. 2: SPINOR AND
			TWISTOR METHODS IN SPACE-TIME GEOMETRY}}, Cambridge Monographs on
	Mathematical Physics, Cambridge University Press (4, 1988),
	\href{https://doi.org/10.1017/CBO9780511524486}{10.1017/CBO9780511524486}.
	
	\bibitem{Kosower:2022yvp}
	D.A.~Kosower, R.~Monteiro and D.~O'Connell, \emph{{The SAGEX review on
			scattering amplitudes Chapter 14: Classical gravity from scattering
			amplitudes}}, \href{https://doi.org/10.1088/1751-8121/ac8846}{\emph{J. Phys.
			A} {\bfseries 55} (2022) 443015}
	[\href{https://arxiv.org/abs/2203.13025}{{\ttfamily 2203.13025}}].
	
	\bibitem{blau2011plane}
	M.~Blau, \emph{Plane waves and penrose limits}, {\emph{Lecture Notes for the
			ICTP School on Mathematics in String and Field Theory (June 2-13 2003)}
		(2011) }.
	
	\bibitem{Kerr:1965vyg}
	R.P.~Kerr and A.~Schild, \emph{{Republication of: A new class of vacuum
			solutions of the Einstein field equations}},
	\href{https://doi.org/10.1007/s10714-009-0857-z}{\emph{Gen. Rel. Grav.}
		{\bfseries 41} (2009) 2485}.
	
	\bibitem{Didenko:2022qxq}
	V.E.~Didenko and N.K.~Dosmanbetov, \emph{{Classical Double Copy and Higher-Spin
			Fields}}, \href{https://doi.org/10.1103/PhysRevLett.130.071603}{\emph{Phys.
			Rev. Lett.} {\bfseries 130} (2023) 071603}
	[\href{https://arxiv.org/abs/2210.04704}{{\ttfamily 2210.04704}}].
	
	\bibitem{Liang:2023zxo}
	Q.~Liang and S.~Nagy, \emph{{Convolutional double copy in (anti) de Sitter
			space}}, \href{https://doi.org/10.1007/JHEP04(2024)139}{\emph{JHEP}
		{\bfseries 04} (2024) 139}
	[\href{https://arxiv.org/abs/2311.14319}{{\ttfamily 2311.14319}}].
	
	\bibitem{Farnsworth:2023mff}
	K.~Farnsworth, M.L.~Graesser and G.~Herczeg, \emph{{Double Kerr-Schild
			spacetimes and the Newman-Penrose map}},
	\href{https://doi.org/10.1007/JHEP10(2023)010}{\emph{JHEP} {\bfseries 10}
		(2023) 010} [\href{https://arxiv.org/abs/2306.16445}{{\ttfamily
			2306.16445}}].
	
	\bibitem{Han:2022mze}
	S.~Han, \emph{{The Weyl double copy in vacuum spacetimes with a cosmological
			constant}}, \href{https://doi.org/10.1007/JHEP09(2022)238}{\emph{JHEP}
		{\bfseries 09} (2022) 238}
	[\href{https://arxiv.org/abs/2205.08654}{{\ttfamily 2205.08654}}].
	
	\bibitem{Alkac:2021bav}
	G.~Alkac, M.K.~Gumus and M.~Tek, \emph{{The Kerr-Schild Double Copy in Lifshitz
			Spacetime}}, \href{https://doi.org/10.1007/JHEP05(2021)214}{\emph{JHEP}
		{\bfseries 05} (2021) 214}
	[\href{https://arxiv.org/abs/2103.06986}{{\ttfamily 2103.06986}}].
	
	\bibitem{Gurses:2018ckx}
	M.~Gurses and B.~Tekin, \emph{{Classical Double Copy: Kerr-Schild-Kundt metrics
			from Yang-Mills Theory}},
	\href{https://doi.org/10.1103/PhysRevD.98.126017}{\emph{Phys. Rev. D}
		{\bfseries 98} (2018) 126017}
	[\href{https://arxiv.org/abs/1810.03411}{{\ttfamily 1810.03411}}].
	
	\bibitem{Prabhu:2020avf}
	S.G.~Prabhu, \emph{{The classical double copy in curved spacetimes:
			perturbative Yang-Mills from the bi-adjoint scalar}},
	\href{https://doi.org/10.1007/JHEP05(2024)117}{\emph{JHEP} {\bfseries 05}
		(2024) 117} [\href{https://arxiv.org/abs/2011.06588}{{\ttfamily
			2011.06588}}].
	
	\bibitem{Carrillo-Gonzalez:2017iyj}
	M.~Carrillo-Gonz\'alez, R.~Penco and M.~Trodden, \emph{{The classical double
			copy in maximally symmetric spacetimes}},
	\href{https://doi.org/10.1007/JHEP04(2018)028}{\emph{JHEP} {\bfseries 04}
		(2018) 028} [\href{https://arxiv.org/abs/1711.01296}{{\ttfamily
			1711.01296}}].
	
	\bibitem{Bahjat-Abbas:2017htu}
	N.~Bahjat-Abbas, A.~Luna and C.D.~White, \emph{{The Kerr-Schild double copy in
			curved spacetime}},
	\href{https://doi.org/10.1007/JHEP12(2017)004}{\emph{JHEP} {\bfseries 12}
		(2017) 004} [\href{https://arxiv.org/abs/1710.01953}{{\ttfamily
			1710.01953}}].
	
	\bibitem{Hawking:1973uf}
	S.W.~Hawking and G.F.R.~Ellis, \emph{{The Large Scale Structure of
			Space-Time}}, Cambridge Monographs on Mathematical Physics, Cambridge
	University Press (2, 2023),
	\href{https://doi.org/10.1017/9781009253161}{10.1017/9781009253161}.
	
	\bibitem{penrose1972techniques}
	R.~Penrose, \emph{Techniques in differential topology in relativity}, SIAM
	(1972).
	
	\bibitem{rosen1937plane}
	N.~Rosen, \emph{Plane polarized waves in the general theory of relativity},
	{\emph{Phys. Z. Sowjetunion} {\bfseries 12} (1937) 366}.
	
	\bibitem{robinson1956report}
	I.~Robinson, \emph{Report to the eddington group},  1956.
	
	\bibitem{penrose1965remarkable}
	R.~Penrose, \emph{A remarkable property of plane waves in general relativity},
	{\emph{Reviews of Modern Physics} {\bfseries 37} (1965) 215}.
	
	\bibitem{brinkmann1925einstein}
	H.~Brinkmann, \emph{Einstein spaces which are mapped conformally on each
		other}, {\emph{Mathematische Annalen} {\bfseries 94} (1925) 119}.
	
	\bibitem{Blau:2006ar}
	M.~Blau, D.~Frank and S.~Weiss, \emph{{Fermi coordinates and Penrose limits}},
	\href{https://doi.org/10.1088/0264-9381/23/11/020}{\emph{Class. Quant. Grav.}
		{\bfseries 23} (2006) 3993}
	[\href{https://arxiv.org/abs/hep-th/0603109}{{\ttfamily hep-th/0603109}}].
	
	\bibitem{Tod:2019urw}
	P.~Tod, \emph{{Spacetimes with all Penrose limits diagonalisable}},
	\href{https://doi.org/10.1088/1361-6382/ab738a}{\emph{Class. Quant. Grav.}
		{\bfseries 37} (2020) 075021}
	[\href{https://arxiv.org/abs/1909.07756}{{\ttfamily 1909.07756}}].
	
	\bibitem{Freedman:2012zz}
	D.Z.~Freedman and A.~Van~Proeyen, \emph{{Supergravity}}, Cambridge Univ. Press,
	Cambridge, UK (5, 2012),
	\href{https://doi.org/10.1017/CBO9781139026833}{10.1017/CBO9781139026833}.
	
	\bibitem{Gueven:2000ru}
	R.~Gueven, \emph{{Plane wave limits and T duality}},
	\href{https://doi.org/10.1016/S0370-2693(00)00517-7}{\emph{Phys. Lett. B}
		{\bfseries 482} (2000) 255}
	[\href{https://arxiv.org/abs/hep-th/0005061}{{\ttfamily hep-th/0005061}}].
	
	\bibitem{Biggs:2023sqw}
	A.~Biggs and J.~Maldacena, \emph{{Scaling similarities and quasinormal modes of
			D0 black hole solutions}},
	\href{https://doi.org/10.1007/JHEP11(2023)155}{\emph{JHEP} {\bfseries 11}
		(2023) 155} [\href{https://arxiv.org/abs/2303.09974}{{\ttfamily
			2303.09974}}].
	
	\bibitem{Duval:2017els}
	C.~Duval, G.W.~Gibbons, P.A.~Horvathy and P.M.~Zhang, \emph{{Carroll symmetry
			of plane gravitational waves}},
	\href{https://doi.org/10.1088/1361-6382/aa7f62}{\emph{Class. Quant. Grav.}
		{\bfseries 34} (2017) 175003}
	[\href{https://arxiv.org/abs/1702.08284}{{\ttfamily 1702.08284}}].
	
	\bibitem{Ciambelli:2023mir}
	L.~Ciambelli, L.~Freidel and R.G.~Leigh, \emph{{Null Raychaudhuri: canonical
			structure and the dressing time}},
	\href{https://doi.org/10.1007/JHEP01(2024)166}{\emph{JHEP} {\bfseries 01}
		(2024) 166} [\href{https://arxiv.org/abs/2309.03932}{{\ttfamily
			2309.03932}}].
	
	\bibitem{Horowitz:1989bv}
	G.T.~Horowitz and A.R.~Steif, \emph{{Space-Time Singularities in String
			Theory}}, \href{https://doi.org/10.1103/PhysRevLett.64.260}{\emph{Phys. Rev.
			Lett.} {\bfseries 64} (1990) 260}.
	
	\bibitem{Horowitz:1990sr}
	G.T.~Horowitz and A.R.~Steif, \emph{{Strings in Strong Gravitational Fields}},
	\href{https://doi.org/10.1103/PhysRevD.42.1950}{\emph{Phys. Rev. D}
		{\bfseries 42} (1990) 1950}.
	
	\bibitem{Kiritsis:1993jk}
	E.~Kiritsis and C.~Kounnas, \emph{{String propagation in gravitational wave
			backgrounds}},
	\href{https://doi.org/10.1016/0370-2693(94)90655-6}{\emph{Phys. Lett. B}
		{\bfseries 320} (1994) 264}
	[\href{https://arxiv.org/abs/hep-th/9310202}{{\ttfamily hep-th/9310202}}].
	
	\bibitem{Russo:2002rq}
	J.G.~Russo and A.A.~Tseytlin, \emph{{On solvable models of type 2B superstring
			in NS NS and RR plane wave backgrounds}},
	\href{https://doi.org/10.1088/1126-6708/2002/04/021}{\emph{JHEP} {\bfseries
			04} (2002) 021} [\href{https://arxiv.org/abs/hep-th/0202179}{{\ttfamily
			hep-th/0202179}}].
	
	\bibitem{Papadopoulos:2002bg}
	G.~Papadopoulos, J.G.~Russo and A.A.~Tseytlin, \emph{{Solvable model of strings
			in a time dependent plane wave background}},
	\href{https://doi.org/10.1088/0264-9381/20/5/313}{\emph{Class. Quant. Grav.}
		{\bfseries 20} (2003) 969}
	[\href{https://arxiv.org/abs/hep-th/0211289}{{\ttfamily hep-th/0211289}}].
	
	\bibitem{Berenstein:2002jq}
	D.E.~Berenstein, J.M.~Maldacena and H.S.~Nastase, \emph{{Strings in flat space
			and pp waves from N=4 superYang-Mills}},
	\href{https://doi.org/10.1088/1126-6708/2002/04/013}{\emph{JHEP} {\bfseries
			04} (2002) 013} [\href{https://arxiv.org/abs/hep-th/0202021}{{\ttfamily
			hep-th/0202021}}].
	
	\bibitem{Eberhardt:2018exh}
	L.~Eberhardt and K.~Ferreira, \emph{{The plane-wave spectrum from the
			worldsheet}}, \href{https://doi.org/10.1007/JHEP10(2018)109}{\emph{JHEP}
		{\bfseries 10} (2018) 109}
	[\href{https://arxiv.org/abs/1805.12155}{{\ttfamily 1805.12155}}].
	
	\bibitem{Adamo:2018mpq}
	T.~Adamo, E.~Casali, L.~Mason and S.~Nekovar, \emph{{Plane wave backgrounds and
			colour-kinematics duality}},
	\href{https://doi.org/10.1007/JHEP02(2019)198}{\emph{JHEP} {\bfseries 02}
		(2019) 198} [\href{https://arxiv.org/abs/1810.05115}{{\ttfamily
			1810.05115}}].
	
	\bibitem{Adamo:2020qru}
	T.~Adamo and A.~Ilderton, \emph{{Classical and quantum double copy of
			back-reaction}}, \href{https://doi.org/10.1007/JHEP09(2020)200}{\emph{JHEP}
		{\bfseries 09} (2020) 200}
	[\href{https://arxiv.org/abs/2005.05807}{{\ttfamily 2005.05807}}].
	
	\bibitem{Lobo:2017bfh}
	I.P.~Lobo and G.G.~Carvalho, \emph{{The geometry of null-like disformal
			transformations}},
	\href{https://doi.org/10.1142/S0219887819501809}{\emph{Int. J. Geom. Meth.
			Mod. Phys.} {\bfseries 16} (2019) 1950180}
	[\href{https://arxiv.org/abs/1707.01784}{{\ttfamily 1707.01784}}].
	
	\bibitem{Gonzo:2021drq}
	R.~Gonzo and C.~Shi, \emph{{Geodesics from classical double copy}},
	\href{https://doi.org/10.1103/PhysRevD.104.105012}{\emph{Phys. Rev. D}
		{\bfseries 104} (2021) 105012}
	[\href{https://arxiv.org/abs/2109.01072}{{\ttfamily 2109.01072}}].
	
	\bibitem{Ball:2023xnr}
	A.~Ball, A.~Bencke, Y.~Chen and A.~Volovich, \emph{{Hidden symmetry in the
			double copy}}, \href{https://doi.org/10.1007/JHEP10(2023)085}{\emph{JHEP}
		{\bfseries 10} (2023) 085}
	[\href{https://arxiv.org/abs/2307.01338}{{\ttfamily 2307.01338}}].
	
	\bibitem{Chawla:2023bsu}
	S.~Chawla and C.~Keeler, \emph{{Black hole horizons from the double copy}},
	\href{https://doi.org/10.1088/1361-6382/acfe57}{\emph{Class. Quant. Grav.}
		{\bfseries 40} (2023) 225004}
	[\href{https://arxiv.org/abs/2306.02417}{{\ttfamily 2306.02417}}].
	
	\bibitem{He:2023iew}
	J.-L.~He and J.-H.~Huang, \emph{{Cosmological horizons from classical double
			copy}}, \href{https://doi.org/10.1016/j.physletb.2024.138579}{\emph{Phys.
			Lett. B} {\bfseries 851} (2024) 138579}
	[\href{https://arxiv.org/abs/2312.00972}{{\ttfamily 2312.00972}}].
	
	\bibitem{Walker:1970un}
	M.~Walker and R.~Penrose, \emph{{On quadratic first integrals of the geodesic
			equations for type [22] spacetimes}},
	\href{https://doi.org/10.1007/BF01649445}{\emph{Commun. Math. Phys.}
		{\bfseries 18} (1970) 265}.
	
	\bibitem{Patricot:2003dh}
	C.~Patricot, \emph{{Kaigorodov spaces and their Penrose limits}},
	\href{https://doi.org/10.1088/0264-9381/20/11/310}{\emph{Class. Quant. Grav.}
		{\bfseries 20} (2003) 2087}
	[\href{https://arxiv.org/abs/hep-th/0302073}{{\ttfamily hep-th/0302073}}].
	
	\bibitem{Bialynicki-Birula:2004bvr}
	I.~Bialynicki-Birula, \emph{{Particle beams guided by electromagnetic vortices:
			New solutions of the Lorentz, Schrodinger, Klein-Gordon and Dirac
			equations}}, \href{https://doi.org/10.1103/PhysRevLett.93.020402}{\emph{Phys.
			Rev. Lett.} {\bfseries 93} (2004) 020402}
	[\href{https://arxiv.org/abs/physics/0403078}{{\ttfamily physics/0403078}}].
	
	\bibitem{boyer1967maximal}
	R.H.~Boyer and R.W.~Lindquist, \emph{Maximal analytic extension of the kerr
		metric}, {\emph{Journal of mathematical physics} {\bfseries 8} (1967) 265}.
	
	\bibitem{Frolov:2017kze}
	V.P.~Frolov, P.~Krtous and D.~Kubiznak, \emph{{Black holes, hidden symmetries,
			and complete integrability}},
	\href{https://doi.org/10.1007/s41114-017-0009-9}{\emph{Living Rev. Rel.}
		{\bfseries 20} (2017) 6} [\href{https://arxiv.org/abs/1705.05482}{{\ttfamily
			1705.05482}}].
	
	\bibitem{Newman:1965tw}
	E.T.~Newman and A.I.~Janis, \emph{{Note on the Kerr spinning particle metric}},
	\href{https://doi.org/10.1063/1.1704350}{\emph{J. Math. Phys.} {\bfseries 6}
		(1965) 915}.
	
	\bibitem{Arkani-Hamed:2019ymq}
	N.~Arkani-Hamed, Y.-t.~Huang and D.~O'Connell, \emph{{Kerr black holes as
			elementary particles}},
	\href{https://doi.org/10.1007/JHEP01(2020)046}{\emph{JHEP} {\bfseries 01}
		(2020) 046} [\href{https://arxiv.org/abs/1906.10100}{{\ttfamily
			1906.10100}}].
	
	\bibitem{Guevara:2020xjx}
	A.~Guevara, B.~Maybee, A.~Ochirov, D.~O'connell and J.~Vines, \emph{{A
			worldsheet for Kerr}},
	\href{https://doi.org/10.1007/JHEP03(2021)201}{\emph{JHEP} {\bfseries 03}
		(2021) 201} [\href{https://arxiv.org/abs/2012.11570}{{\ttfamily
			2012.11570}}].
	
	\bibitem{Carter:1968rr}
	B.~Carter, \emph{{Global structure of the Kerr family of gravitational
			fields}}, \href{https://doi.org/10.1103/PhysRev.174.1559}{\emph{Phys. Rev.}
		{\bfseries 174} (1968) 1559}.
	
	\bibitem{Bardeen:1973tla}
	J.M.~Bardeen, \emph{{Timelike and null geodesics in the Kerr metric}},
	{\emph{Proceedings, Ecole d'Et\'e de Physique Th\'eorique: Les Astres Occlus
			: Les Houches, France, August, 1972, 215-240} (1973) 215}.
	
	\bibitem{Chandrasekhar:1985kt}
	S.~Chandrasekhar, \emph{{The mathematical theory of black holes}} (1985).
	
	\bibitem{Gralla:2019ceu}
	S.E.~Gralla and A.~Lupsasca, \emph{{Null geodesics of the Kerr exterior}},
	\href{https://doi.org/10.1103/PhysRevD.101.044032}{\emph{Phys. Rev. D}
		{\bfseries 101} (2020) 044032}
	[\href{https://arxiv.org/abs/1910.12881}{{\ttfamily 1910.12881}}].
	
	\bibitem{Compere:2021bkk}
	G.~Comp\`ere, Y.~Liu and J.~Long, \emph{{Classification of radial Kerr geodesic
			motion}}, \href{https://doi.org/10.1103/PhysRevD.105.024075}{\emph{Phys. Rev.
			D} {\bfseries 105} (2022) 024075}
	[\href{https://arxiv.org/abs/2106.03141}{{\ttfamily 2106.03141}}].
	
	\bibitem{Cieslik:2023qdc}
	A.~Cie\'slik, E.~Hackmann and P.~Mach, \emph{{Kerr geodesics in terms of
			Weierstrass elliptic functions}},
	\href{https://doi.org/10.1103/PhysRevD.108.024056}{\emph{Phys. Rev. D}
		{\bfseries 108} (2023) 024056}
	[\href{https://arxiv.org/abs/2305.07771}{{\ttfamily 2305.07771}}].
	
	\bibitem{Igata:2019pgb}
	T.~Igata, H.~Ishihara and Y.~Yasunishi, \emph{{Observability of spherical
			photon orbits in near-extremal Kerr black holes}},
	\href{https://doi.org/10.1103/PhysRevD.100.044058}{\emph{Phys. Rev. D}
		{\bfseries 100} (2019) 044058}
	[\href{https://arxiv.org/abs/1904.00271}{{\ttfamily 1904.00271}}].
	
	\bibitem{Fransen:2023eqj}
	K.~Fransen, \emph{{Quasinormal modes from Penrose limits}},
	\href{https://doi.org/10.1088/1361-6382/acf26d}{\emph{Class. Quant. Grav.}
		{\bfseries 40} (2023) 205004}
	[\href{https://arxiv.org/abs/2301.06999}{{\ttfamily 2301.06999}}].
	
	\bibitem{Hollowood:2009qz}
	T.J.~Hollowood, G.M.~Shore and R.J.~Stanley, \emph{{The Refractive Index of
			Curved Spacetime II: QED, Penrose Limits and Black Holes}},
	\href{https://doi.org/10.1088/1126-6708/2009/08/089}{\emph{JHEP} {\bfseries
			08} (2009) 089} [\href{https://arxiv.org/abs/0905.0771}{{\ttfamily
			0905.0771}}].
	
	\bibitem{Papadopoulos:2020qik}
	G.~Papadopoulos, \emph{{Separability, plane wave limits and rotating black
			holes}}, \href{https://doi.org/10.1088/1361-6382/ac1cf8}{\emph{Class. Quant.
			Grav.} {\bfseries 38} (2021) 195018}
	[\href{https://arxiv.org/abs/2007.04702}{{\ttfamily 2007.04702}}].
	
	\bibitem{Kubiznak:2008zs}
	D.~Kubiznak, V.P.~Frolov, P.~Krtous and P.~Connell, \emph{{Parallel-propagated
			frame along null geodesics in higher-dimensional black hole spacetimes}},
	\href{https://doi.org/10.1103/PhysRevD.79.024018}{\emph{Phys. Rev. D}
		{\bfseries 79} (2009) 024018}
	[\href{https://arxiv.org/abs/0811.0012}{{\ttfamily 0811.0012}}].
	
	\bibitem{Cariglia:2018erv}
	M.~Cariglia, T.~Houri, P.~Krtous and D.~Kubiznak, \emph{{On Integrability of
			the Geodesic Deviation Equation}},
	\href{https://doi.org/10.1140/epjc/s10052-018-6133-1}{\emph{Eur. Phys. J. C}
		{\bfseries 78} (2018) 661}
	[\href{https://arxiv.org/abs/1805.07677}{{\ttfamily 1805.07677}}].
	
	\bibitem{penrose19952}
	R.~Penrose, \emph{In $\{$2 2$\}$ vacuums, the null-datum on null rays is
		exactly r- 5}, {\emph{General Relativity and Gravitation} {\bfseries 27}
		(1995) 1323}.
	
	\bibitem{penrose1980golden}
	R.~Penrose, \emph{Golden oldie: null hypersurface initial data for classical
		fields of arbitrary spin and for general relativity}, {\emph{General
			Relativity and Gravitation} {\bfseries 12} (1980) 225}.
	
	\bibitem{Aksteiner:2014zyp}
	S.~Aksteiner, \emph{{Geometry and analysis on black hole spacetimes}}, Ph.D.
	thesis, Leibniz U., Hannover, 2014.
	\newblock 10.15488/8214.
	
	\bibitem{geroch1973space}
	R.~Geroch, A.~Held and R.~Penrose, \emph{A space-time calculus based on pairs
		of null directions}, {\emph{Journal of Mathematical Physics} {\bfseries 14}
		(1973) 874}.
	
	\bibitem{Kunze:2004qd}
	K.E.~Kunze, \emph{{Behavior of curvature and matter in the Penrose limit}},
	\href{https://doi.org/10.1103/PhysRevD.71.063518}{\emph{Phys. Rev. D}
		{\bfseries 71} (2005) 063518}
	[\href{https://arxiv.org/abs/gr-qc/0411115}{{\ttfamily gr-qc/0411115}}].
	
	\bibitem{Monteiro:2018xev}
	R.~Monteiro, I.~Nicholson and D.~O'Connell, \emph{{Spinor-helicity and the
			algebraic classification of higher-dimensional spacetimes}},
	\href{https://doi.org/10.1088/1361-6382/ab03df}{\emph{Class. Quant. Grav.}
		{\bfseries 36} (2019) 065006}
	[\href{https://arxiv.org/abs/1809.03906}{{\ttfamily 1809.03906}}].
	
	\bibitem{Hollowood:2007ku}
	T.J.~Hollowood and G.M.~Shore, \emph{{The Refractive index of curved spacetime:
			The Fate of causality in QED}},
	\href{https://doi.org/10.1016/j.nuclphysb.2007.11.034}{\emph{Nucl. Phys. B}
		{\bfseries 795} (2008) 138}
	[\href{https://arxiv.org/abs/0707.2303}{{\ttfamily 0707.2303}}].
	
	\bibitem{Harte:2012jg}
	A.I.~Harte, \emph{{Strong lensing, plane gravitational waves and transient
			flashes}}, \href{https://doi.org/10.1088/0264-9381/30/7/075011}{\emph{Class.
			Quant. Grav.} {\bfseries 30} (2013) 075011}
	[\href{https://arxiv.org/abs/1210.1449}{{\ttfamily 1210.1449}}].
	
	\bibitem{Harte:2015ila}
	A.I.~Harte, \emph{{Optics in a nonlinear gravitational plane wave}},
	\href{https://doi.org/10.1088/0264-9381/32/17/175017}{\emph{Class. Quant.
			Grav.} {\bfseries 32} (2015) 175017}
	[\href{https://arxiv.org/abs/1502.03658}{{\ttfamily 1502.03658}}].
	
	\bibitem{harte2013tails}
	A.I.~Harte, \emph{Tails of plane wave spacetimes: Wave-wave scattering in
		general relativity}, {\emph{Physical Review D} {\bfseries 88} (2013) 084059}.
	
	\bibitem{Zhang:2017rno}
	P.M.~Zhang, C.~Duval, G.W.~Gibbons and P.A.~Horvathy, \emph{{The Memory Effect
			for Plane Gravitational Waves}},
	\href{https://doi.org/10.1016/j.physletb.2017.07.050}{\emph{Phys. Lett. B}
		{\bfseries 772} (2017) 743}
	[\href{https://arxiv.org/abs/1704.05997}{{\ttfamily 1704.05997}}].
	
	\bibitem{Zhang:2017geq}
	P.M.~Zhang, C.~Duval, G.W.~Gibbons and P.A.~Horvathy, \emph{{Soft gravitons and
			the memory effect for plane gravitational waves}},
	\href{https://doi.org/10.1103/PhysRevD.96.064013}{\emph{Phys. Rev. D}
		{\bfseries 96} (2017) 064013}
	[\href{https://arxiv.org/abs/1705.01378}{{\ttfamily 1705.01378}}].
	
	\bibitem{Flanagan:2019ezo}
	E.E.~Flanagan, A.M.~Grant, A.I.~Harte and D.A.~Nichols, \emph{{Persistent
			gravitational wave observables: Nonlinear plane wave spacetimes}},
	\href{https://doi.org/10.1103/PhysRevD.101.104033}{\emph{Phys. Rev. D}
		{\bfseries 101} (2020) 104033}
	[\href{https://arxiv.org/abs/1912.13449}{{\ttfamily 1912.13449}}].
	
	\bibitem{Harte:2012uw}
	A.I.~Harte and T.D.~Drivas, \emph{{Caustics and wave propagation in curved
			spacetimes}}, \href{https://doi.org/10.1103/PhysRevD.85.124039}{\emph{Phys.
			Rev. D} {\bfseries 85} (2012) 124039}
	[\href{https://arxiv.org/abs/1202.0540}{{\ttfamily 1202.0540}}].
	
	\bibitem{Perlman:2014cwa}
	E.S.~Perlman, S.A.~Rappaport, W.A.~Christiansen, Y.J.~Ng, J.~DeVore and
	D.~Pooley, \emph{{New Constraints on Quantum Gravity from X-ray and Gamma-Ray
			Observations}},
	\href{https://doi.org/10.1088/0004-637X/805/1/10}{\emph{Astrophys. J.}
		{\bfseries 805} (2015) 10} [\href{https://arxiv.org/abs/1411.7262}{{\ttfamily
			1411.7262}}].
	
	\bibitem{Cardoso:2016rao}
	V.~Cardoso, E.~Franzin and P.~Pani, \emph{{Is the gravitational-wave ringdown a
			probe of the event horizon?}},
	\href{https://doi.org/10.1103/PhysRevLett.116.171101}{\emph{Phys. Rev. Lett.}
		{\bfseries 116} (2016) 171101}
	[\href{https://arxiv.org/abs/1602.07309}{{\ttfamily 1602.07309}}].
	
	\bibitem{Baker:2017hug}
	T.~Baker, E.~Bellini, P.G.~Ferreira, M.~Lagos, J.~Noller and I.~Sawicki,
	\emph{{Strong constraints on cosmological gravity from GW170817 and GRB
			170817A}}, \href{https://doi.org/10.1103/PhysRevLett.119.251301}{\emph{Phys.
			Rev. Lett.} {\bfseries 119} (2017) 251301}
	[\href{https://arxiv.org/abs/1710.06394}{{\ttfamily 1710.06394}}].
	
	\bibitem{Cunha:2018acu}
	P.V.P.~Cunha and C.A.R.~Herdeiro, \emph{{Shadows and strong gravitational
			lensing: a brief review}},
	\href{https://doi.org/10.1007/s10714-018-2361-9}{\emph{Gen. Rel. Grav.}
		{\bfseries 50} (2018) 42} [\href{https://arxiv.org/abs/1801.00860}{{\ttfamily
			1801.00860}}].
	
	\bibitem{Johnson:2019ljv}
	M.D.~Johnson et~al., \emph{{Universal interferometric signatures of a black
			hole\textquoteright{}s photon ring}},
	\href{https://doi.org/10.1126/sciadv.aaz1310}{\emph{Sci. Adv.} {\bfseries 6}
		(2020) eaaz1310} [\href{https://arxiv.org/abs/1907.04329}{{\ttfamily
			1907.04329}}].
	
	\bibitem{Hadar:2022xag}
	S.~Hadar, D.~Kapec, A.~Lupsasca and A.~Strominger, \emph{{Holography of the
			photon ring}}, \href{https://doi.org/10.1088/1361-6382/ac8d43}{\emph{Class.
			Quant. Grav.} {\bfseries 39} (2022) 215001}
	[\href{https://arxiv.org/abs/2205.05064}{{\ttfamily 2205.05064}}].
	
	\bibitem{Paugnat:2022qzy}
	H.~Paugnat, A.~Lupsasca, F.~Vincent and M.~Wielgus, \emph{{Photon ring test of
			the Kerr hypothesis: Variation in the ring shape}},
	\href{https://doi.org/10.1051/0004-6361/202244216}{\emph{Astron. Astrophys.}
		{\bfseries 668} (2022) A11}
	[\href{https://arxiv.org/abs/2206.02781}{{\ttfamily 2206.02781}}].
	
	\bibitem{Cardenas-Avendano:2023dzo}
	A.~C\'ardenas-Avenda\~no and A.~Lupsasca, \emph{{Prediction for the
			interferometric shape of the first black hole photon ring}},
	\href{https://doi.org/10.1103/PhysRevD.108.064043}{\emph{Phys. Rev. D}
		{\bfseries 108} (2023) 064043}
	[\href{https://arxiv.org/abs/2305.12956}{{\ttfamily 2305.12956}}].
	
	\bibitem{Lee:2023kry}
	V.S.H.~Lee, K.M.~Zurek and Y.~Chen, \emph{{Astronomical image blurring from
			transversely correlated quantum gravity fluctuations}},
	\href{https://doi.org/10.1103/PhysRevD.109.084005}{\emph{Phys. Rev. D}
		{\bfseries 109} (2024) 084005}
	[\href{https://arxiv.org/abs/2312.06757}{{\ttfamily 2312.06757}}].
	
	\bibitem{Jia:2024mlb}
	H.~Jia, E.~Quataert, A.~Lupsasca and G.N.~Wong, \emph{{Photon Ring
			Interferometric Signatures Beyond The Universal Regime}},
	\href{https://arxiv.org/abs/2405.08804}{{\ttfamily 2405.08804}}.
	
	\bibitem{Gueven:1987ad}
	R.~Gueven, \emph{{Plane Waves in Effective Field Theories of Superstrings}},
	\href{https://doi.org/10.1016/0370-2693(87)90254-1}{\emph{Phys. Lett. B}
		{\bfseries 191} (1987) 275}.
	
	\bibitem{Amati:1988sa}
	D.~Amati and C.~Klimcik, \emph{{Nonperturbative Computation of the Weyl Anomaly
			for a Class of Nontrivial Backgrounds}},
	\href{https://doi.org/10.1016/0370-2693(89)91092-7}{\emph{Phys. Lett. B}
		{\bfseries 219} (1989) 443}.
	
	\bibitem{Blau:2003dz}
	M.~Blau, M.~Borunda, M.~O'Loughlin and G.~Papadopoulos, \emph{{Penrose limits
			and space-time singularities}},
	\href{https://doi.org/10.1088/0264-9381/21/7/L02}{\emph{Class. Quant. Grav.}
		{\bfseries 21} (2004) L43}
	[\href{https://arxiv.org/abs/hep-th/0312029}{{\ttfamily hep-th/0312029}}].
	
	\bibitem{Blau:2004yi}
	M.~Blau, M.~Borunda, M.~O'Loughlin and G.~Papadopoulos, \emph{{The Universality
			of Penrose limits near space-time singularities}},
	\href{https://doi.org/10.1088/1126-6708/2004/07/068}{\emph{JHEP} {\bfseries
			07} (2004) 068} [\href{https://arxiv.org/abs/hep-th/0403252}{{\ttfamily
			hep-th/0403252}}].
	
	\bibitem{Giddings:2007bw}
	S.B.~Giddings, D.J.~Gross and A.~Maharana, \emph{{Gravitational effects in
			ultrahigh-energy string scattering}},
	\href{https://doi.org/10.1103/PhysRevD.77.046001}{\emph{Phys. Rev. D}
		{\bfseries 77} (2008) 046001}
	[\href{https://arxiv.org/abs/0705.1816}{{\ttfamily 0705.1816}}].
	
	\bibitem{Craps:2008bv}
	B.~Craps, F.~De~Roo and O.~Evnin, \emph{{Can free strings propagate across
			plane wave singularities?}},
	\href{https://doi.org/10.1088/1126-6708/2009/03/105}{\emph{JHEP} {\bfseries
			03} (2009) 105} [\href{https://arxiv.org/abs/0812.2900}{{\ttfamily
			0812.2900}}].
	
	\bibitem{Hollowood:2011yh}
	T.J.~Hollowood and G.M.~Shore, \emph{{The Effect of Gravitational Tidal Forces
			on Renormalized Quantum Fields}},
	\href{https://doi.org/10.1007/JHEP02(2012)120}{\emph{JHEP} {\bfseries 02}
		(2012) 120} [\href{https://arxiv.org/abs/1111.3174}{{\ttfamily 1111.3174}}].
	
	\bibitem{Martinec:2020cml}
	E.J.~Martinec and N.P.~Warner, \emph{{The Harder They Fall, the Bigger They
			Become: Tidal Trapping of Strings by Microstate Geometries}},
	\href{https://doi.org/10.1007/JHEP04(2021)259}{\emph{JHEP} {\bfseries 04}
		(2021) 259} [\href{https://arxiv.org/abs/2009.07847}{{\ttfamily
			2009.07847}}].
	
	\bibitem{Bena:2020iyw}
	I.~Bena, A.~Houppe and N.P.~Warner, \emph{{Delaying the Inevitable: Tidal
			Disruption in Microstate Geometries}},
	\href{https://doi.org/10.1007/JHEP02(2021)103}{\emph{JHEP} {\bfseries 02}
		(2021) 103} [\href{https://arxiv.org/abs/2006.13939}{{\ttfamily
			2006.13939}}].
	
	\bibitem{Dodelson:2020lal}
	M.~Dodelson and H.~Ooguri, \emph{{Singularities of thermal correlators at
			strong coupling}},
	\href{https://doi.org/10.1103/PhysRevD.103.066018}{\emph{Phys. Rev. D}
		{\bfseries 103} (2021) 066018}
	[\href{https://arxiv.org/abs/2010.09734}{{\ttfamily 2010.09734}}].
	
	\bibitem{Nishii:2021ylb}
	K.~Nishii and D.~Yoshida, \emph{{String excitation by initial singularity of
			inflation}}, \href{https://doi.org/10.1007/JHEP10(2021)025}{\emph{JHEP}
		{\bfseries 10} (2021) 025}
	[\href{https://arxiv.org/abs/2105.12339}{{\ttfamily 2105.12339}}].
	
	\bibitem{Balivada:2023akk}
	A.~Balivada, P.R.~Padhi and A.~Virmani, \emph{{Tidal forces in Kerr-AdS and
			Grey galaxies}}, \href{https://doi.org/10.1088/1361-6382/ad494b}{\emph{Class.
			Quant. Grav.} {\bfseries 41} (2024) 125008}
	[\href{https://arxiv.org/abs/2309.14672}{{\ttfamily 2309.14672}}].
	
	\bibitem{Horowitz:2024dch}
	G.T.~Horowitz, M.~Kolanowski, G.N.~Remmen and J.E.~Santos, \emph{{Sudden
			breakdown of effective field theory near cool Kerr-Newman black holes}},
	\href{https://doi.org/10.1007/JHEP05(2024)122}{\emph{JHEP} {\bfseries 05}
		(2024) 122} [\href{https://arxiv.org/abs/2403.00051}{{\ttfamily
			2403.00051}}].
	
	\bibitem{Guo:2024pvv}
	B.~Guo, S.D.~Hampton and N.P.~Warner, \emph{{Inscribing geodesic circles on the
			face of the superstratum}},
	\href{https://doi.org/10.1007/JHEP05(2024)224}{\emph{JHEP} {\bfseries 05}
		(2024) 224} [\href{https://arxiv.org/abs/2401.17366}{{\ttfamily
			2401.17366}}].
	
	\bibitem{Maldacena:1997re}
	J.M.~Maldacena, \emph{{The Large N limit of superconformal field theories and
			supergravity}}, \href{https://doi.org/10.4310/ATMP.1998.v2.n2.a1}{\emph{Adv.
			Theor. Math. Phys.} {\bfseries 2} (1998) 231}
	[\href{https://arxiv.org/abs/hep-th/9711200}{{\ttfamily hep-th/9711200}}].
	
	\bibitem{Witten:1998qj}
	E.~Witten, \emph{{Anti-de Sitter space and holography}},
	\href{https://doi.org/10.4310/ATMP.1998.v2.n2.a2}{\emph{Adv. Theor. Math.
			Phys.} {\bfseries 2} (1998) 253}
	[\href{https://arxiv.org/abs/hep-th/9802150}{{\ttfamily hep-th/9802150}}].
	
	\bibitem{Aharony:1999ti}
	O.~Aharony, S.S.~Gubser, J.M.~Maldacena, H.~Ooguri and Y.~Oz, \emph{{Large N
			field theories, string theory and gravity}},
	\href{https://doi.org/10.1016/S0370-1573(99)00083-6}{\emph{Phys. Rept.}
		{\bfseries 323} (2000) 183}
	[\href{https://arxiv.org/abs/hep-th/9905111}{{\ttfamily hep-th/9905111}}].
	
	\bibitem{Plefka:2003nb}
	J.C.~Plefka, \emph{{Lectures on the plane wave string / gauge theory duality}},
	\href{https://doi.org/10.1002/prop.200310121}{\emph{Fortsch. Phys.}
		{\bfseries 52} (2004) 264}
	[\href{https://arxiv.org/abs/hep-th/0307101}{{\ttfamily hep-th/0307101}}].
	
	\bibitem{Sadri:2003pr}
	D.~Sadri and M.M.~Sheikh-Jabbari, \emph{{The Plane wave / superYang-Mills
			duality}}, \href{https://doi.org/10.1103/RevModPhys.76.853}{\emph{Rev. Mod.
			Phys.} {\bfseries 76} (2004) 853}
	[\href{https://arxiv.org/abs/hep-th/0310119}{{\ttfamily hep-th/0310119}}].
	
	\bibitem{Pound:2021qin}
	A.~Pound and B.~Wardell, \emph{{Black hole perturbation theory and
			gravitational self-force}},
	\href{https://arxiv.org/abs/2101.04592}{{\ttfamily 2101.04592}}.
	
	\bibitem{Kinnersley:1969zza}
	W.~Kinnersley, \emph{{Type D Vacuum Metrics}},
	\href{https://doi.org/10.1063/1.1664958}{\emph{J. Math. Phys.} {\bfseries 10}
		(1969) 1195}.
	
	\bibitem{Plebanski:1976gy}
	J.F.~Plebanski and M.~Demianski, \emph{{Rotating, charged, and uniformly
			accelerating mass in general relativity}},
	\href{https://doi.org/10.1016/0003-4916(76)90240-2}{\emph{Annals Phys.}
		{\bfseries 98} (1976) 98}.
	
	\bibitem{Griffiths:2005qp}
	J.B.~Griffiths and J.~Podolsky, \emph{{A New look at the Plebanski-Demianski
			family of solutions}},
	\href{https://doi.org/10.1142/S0218271806007742}{\emph{Int. J. Mod. Phys. D}
		{\bfseries 15} (2006) 335}
	[\href{https://arxiv.org/abs/gr-qc/0511091}{{\ttfamily gr-qc/0511091}}].
	
\end{thebibliography}

 \newcommand{\noop}[1]{}

\providecommand{\href}[2]{#2}\begingroup\raggedright\endgroup

\end{document}